\documentclass[reprint, superscriptaddress, amsmath, amssymb, aps, prx,
]{revtex4-2}
\usepackage{graphicx}
\usepackage{dcolumn}
\usepackage{bm}
\usepackage{braket}
\usepackage[colorlinks]{hyperref}
\usepackage{amsmath}
\usepackage{booktabs}
\usepackage{tabularx}
\usepackage{makecell}

\usepackage{subfigure}

\hypersetup{colorlinks=true, citecolor=blue, urlcolor=blue, linkcolor=blue}

\begin{document}

\title{Experimental benchmarking of quantum state overlap estimation strategies with photonic systems}

\author{Hao Zhan}
\author{Ben Wang}
\author{Minghao Mi}
\author{Jie Xie}
\author{Liang Xu}
\affiliation{National Laboratory of Solid State Microstructures, Key Laboratory of Intelligent Optical Sensing and Manipulation, College of Engineering and Applied Sciences, Jiangsu Physical Science Research Center, and Collaborative Innovation Center of Advanced Microstructures, Nanjing University, Nanjing 210093, China}
\author{Aonan Zhang}
\email[Previous email: ]{a.zhang@imperial.ac.uk}
\email[\\Current email: ]{aonan.zhang@physics.ox.ac.uk}
\affiliation{National Laboratory of Solid State Microstructures, Key Laboratory of Intelligent Optical Sensing and Manipulation, College of Engineering and Applied Sciences, Jiangsu Physical Science Research Center, and Collaborative Innovation Center of Advanced Microstructures, Nanjing University, Nanjing 210093, China}
\affiliation{Department of Physics, Imperial College London, Prince Consort Road, London SW7 2AZ, United Kingdom}
\author{Lijian Zhang}
\email[]{lijian.zhang@nju.edu.cn}
\affiliation{National Laboratory of Solid State Microstructures, Key Laboratory of Intelligent Optical Sensing and Manipulation, College of Engineering and Applied Sciences, Jiangsu Physical Science Research Center, and Collaborative Innovation Center of Advanced Microstructures, Nanjing University, Nanjing 210093, China}


\date{\today}

\begin{abstract}
    Accurately estimating the overlap between quantum states is a fundamental task in quantum information processing. While various strategies using distinct quantum measurements have been proposed for overlap estimation, the lack of experimental benchmarks on estimation precision limits strategy selection in different situations. Here we compare the performance of four practical strategies for overlap estimation, including tomography-tomography, tomography-projection, Schur collective measurement and optical swap test using photonic quantum systems. We encode the quantum states on the polarization and path degrees of freedom of single photons. The corresponding measurements are performed by photon detection on certain modes following single-photon mode transformation or two-photon interference. We further propose an adaptive strategy with optimized precision in full-range overlap estimation. Our results shed new light on extracting the parameter of interest from quantum systems, prompting the design of efficient quantum protocols.
\end{abstract}


\maketitle

\section{Introduction}
Quantum information processing tasks are normally accomplished by estimating specific parameters encoded in the output states instead of the full knowledge of the states. The estimation of the overlap $c = |\braket{\psi|\phi}|^2$ between two unknown quantum states $\ket{\psi}$ and $\ket{\phi}$ is a quintessential example underlying various applications, including relative quantum information~\cite{PhysRevA.70.032321, PhysRevA.73.022341, RevModPhys.79.555, PhysRevX.10.031023,doi:10.1126/sciadv.adj4249}, entanglement estimation~\cite{ PhysRevLett.88.217901, PhysRevLett.95.260502,PhysRevA.75.032338,PhysRevA.106.062413}, cross-platform verification~\cite{PhysRevLett.124.010504} and quantum algorithms~\cite{J_2022, RN66}. In particular, state overlap estimation plays a pivotal role in various quantum machine learning algorithms~\cite{biamonte2017quantum,e25020287}, such as quantum neural network training~\cite{s41534-017-0032-4,s43588-021-00084-1,s41467-023-39785-8, RN67,PhysRevA.106.042431}, quantum support vector machine~\cite{PhysRevLett.113.130503, havlivcek2019supervised,PhysRevLett.122.040504,RN80,PhysRevX.9.041029,PhysRevA.107.022402} and variational quantum learning~\cite{doi:10.1126/sciadv.aav2761,RN81,PhysRevA.101.032323}, in which the state overlaps serve as cost functions or kernel functions. However, to date, these applications usually assume the ability of ideal and precise state overlap estimation without considering the precision and imperfections in realistic experiments, which may limit the performance of their actual implementations.
\par
The most intuitive way to estimate the overlap is performing full tomography to reconstruct both quantum states and then calculate the overlap directly. This strategy can be modified by only performing tomography of one state $\ket{\phi}$ and projecting another state $\ket{\psi}$ onto the estimate $\ket{\tilde{\phi}}$, the success probability of which gives the overlap between the two states. On the other hand, a widely used strategy in many quantum protocols~\cite{PhysRevA.106.042431,havlivcek2019supervised, PhysRevLett.87.167902,RN70,agliardi2023,white2024robust} is a joint measurement on $\ket{\psi}\ket{\phi}$ called the swap test~\cite{PhysRevLett.87.167902}. Swap test has been realized in various quantum systems~\cite{nguyen2021experimental,PhysRevApplied.18.014047}, for example, with the Hong-Ou-Mandel interference (HOMI) of photons~\cite{PhysRevLett.59.2044,PhysRevA.87.052330,s41377-021-00608-4}. There have been efforts to further improve the implementation of the swap test through variational quantum approaches to find shorter-depth algorithms~\cite{Cincio_2018}, as well as to estimate both the amplitude and phase of the inner product $\braket{\psi|\phi}$~\cite{Huggins_2020}. Recently, optimal strategy for overlap estimation has been proposed, achieving ultimate precision among all possible strategies~\cite{PhysRevLett.124.060503}. Yet the optimal strategy involves formidable experimental costs requiring joint measurements on all copies of the quantum states. This gap between theoretical proposals and experimental capabilities, which restricts the practical implementation of many quantum protocols, necessitates benchmarking the attainable precision of overlap estimation strategies feasible with current technologies.
\par
To bridge this gap, here we experimentally evaluate the precision of overlap estimation strategies on the photonic platform. Photonics has emerged as a promising platform for various quantum information applications including quantum machine learning~\cite{s41534-019-0174-7,qute.202200125}, benefited from the development of photonic quantum circuits that have already matured in the implementation of optical neural networks~\cite{s41586-020-2973-6, 10.1117/1.AP.4.2.026004,doi:10.1126/sciadv.adg7904}. The advantages in high-dimensional encoding and programmable operations using linear optics can be readily generalized to implement quantum-optical neural networks at the single-photon level~\cite{s41534-017-0032-4, Chabaud2021quantummachine}. Developing tailored overlap estimation strategies with optimized precision and efficiency is therefore crucial for the development of photonic quantum machine learning. Moreover, these tailored strategies can be adapted to diverse quantum technology platforms, broadening their application scope.
\par
\begin{figure*}[htbp]
    \centering
    \includegraphics[width= 0.95 \textwidth]{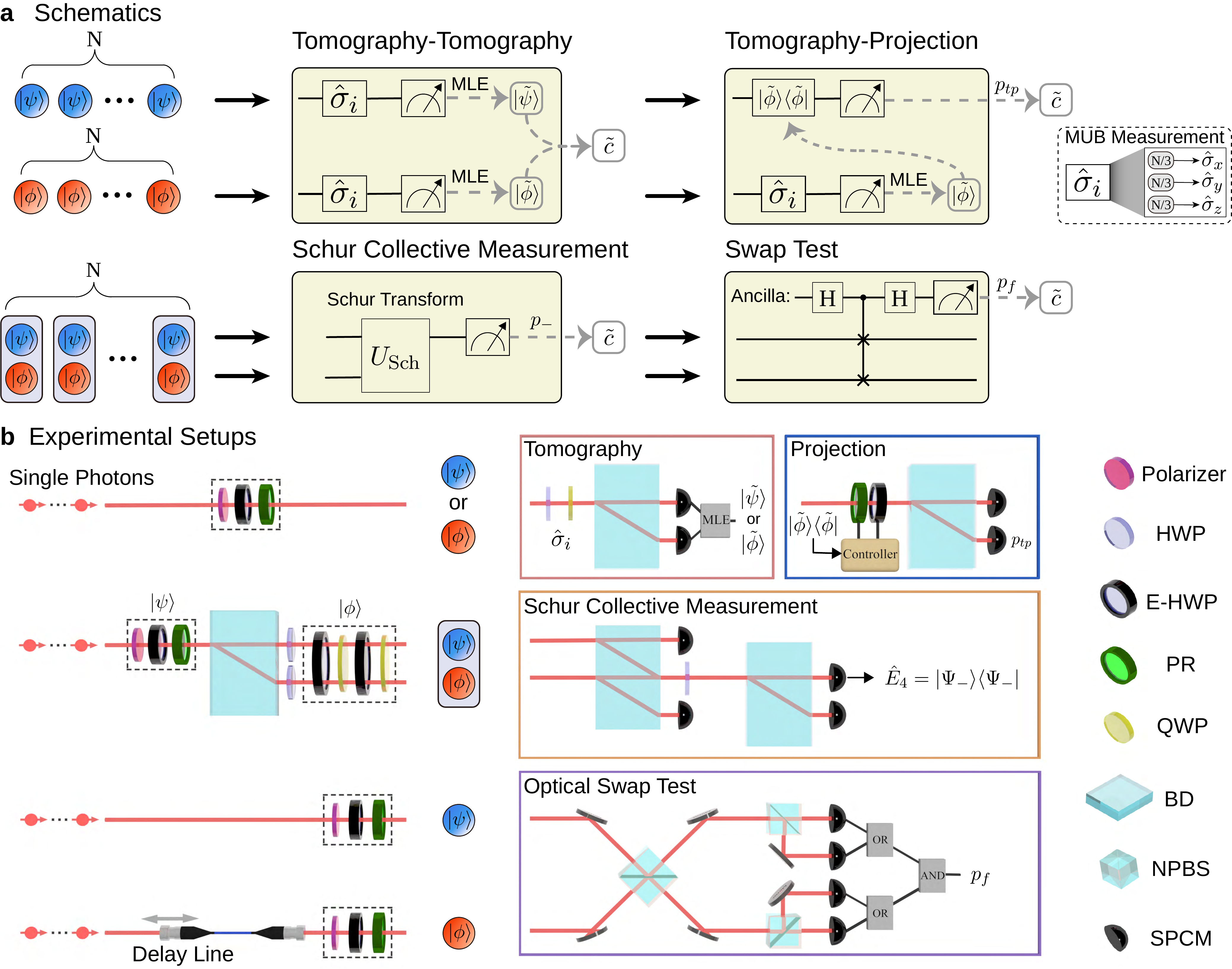}
    \caption{\textbf{Schematics and experimental setups of overlap estimation strategies.} \textbf{a} Schematics. Tomography-Tomography (TT): Perform quantum state tomography to reconstruct both states using mutually unbiased bases (MUB) and calculate the overlap between $\ket{\tilde{\psi}}$ and $\ket{\tilde{\phi}}$. MLE: maximum likelihood estimation. Tomography-Projection (TP): $\ket{\phi}$ is reconstructed by tomography, and $\ket{\psi}$ is projected onto the estimate $\ket{\tilde{\phi}}$. Schur Collective Measurement (SCM): Schur transform~\cite{PhysRevLett.97.170502} with computational basis measurements with is applied to project the joint state $\ket{\psi}\ket{\phi}$ on the Schur bases. Swap Test: Swap test is conducted on the state $\ket{\psi}\ket{\phi}$, where we realize HOMI as the optical swap test (OST). \textbf{b} Experimental setups. Single photons are generated via spontaneous parametric down conversion. Qubits $\ket{\psi}$ and $\ket{\phi}$ are encoded in the polarization DoF of different single photons in TT, TP and OST. In SCM, the two-qubit state $\ket{\psi}\ket{\phi}$ is encoded in the polarization and path DoF of the single photon. Tomography module (red frame) with wave-plates and a beam displacer performs the measurements of Pauli operators $\hat{\sigma}_i$ in TT and TP strategies. Projection module (blue frame) registers the successful projections onto $\ket{\tilde{\phi}}$ to estimate $p_{tp}$ in TP strategy. SCM module (orange frame) conducts Schur basis projective measurements on $\ket{\psi}\ket{\phi}$. OST module (purple frame) utilizes HOMI to obtain the ``fail" probability $p_f$ of the swap test. (E-)HWP (electronically controlled) half wave-plate, PR liquid crystal phase retarder, QWP quarter wave-plate, BD beam displacer, NPBS non-polarizing beam-splitter, SPCM single photon counting module.}
    \label{Fig1_Schematic-Exp}
\end{figure*}

In this work, we benchmark four practical overlap estimation strategies suitable for current photonic technologies: tomography-tomography (TT), tomography-projection (TP), Schur collective measurement (SCM), and optical swap test (OST), as illustrated in Fig.~\ref{Fig1_Schematic-Exp}. By encoding qubit states into various degrees of freedom (DoF) of photons, we experimentally perform the corresponding measurements with linear optics and quantify the estimation precision as a function of the true overlap. Our results demonstrate that different strategies yield varying overlap-dependent precision. By comparing performance across different overlap ranges, we develop an adaptive strategy that combines TP and SCM strategies to achieve optimized precision across the full overlap interval. Furthermore, we quantify the contributions of tomography errors or specific measurement outcome statistics to the final precision for each strategy, elucidating key performance factors. Extending this analysis to higher-dimensional states, we discuss the scaling of the performance of each strategy with respect to state dimension, highlighting the dimension-independence of SCM and OST, and analyzing TT and TP performance under different tomographic measurement schemes including joint and local measurements. These findings provide insights into the analysis of overlap estimation precision and help to design the practical strategy with optimized performance.
\section{Results}
\subsection{Overlap estimation strategy performance assessment}
Given $N$ pairs of two unknown pure qubit states $\ket{\psi}$ and $\ket{\phi}$, without loss of generality, these states can be expressed as $\ket{\psi}=U\ket{0}$ and $\ket{\phi}=U(\sqrt{c}\ket{0}+e^{i\varphi}\sqrt{1-c}\ket{1})$, where $U\in SU(2)$ and $c$ represents their overlap. Here, we primarily focus on estimating the overlap between two qubits, with the high-dimensional cases discussed later.  An overlap estimation strategy denoted by $s$ involves a general positive operator-valued measure (POVM) $\{\hat{E}_k^{(s)}\}$ on all copies of the quantum states and the estimation of the overlap as $\tilde{c}_s(k)$ based on the outcome $k$. Under a specific choice of $U$ and $\varphi$, the mean squared error of overlap is given by $v_{s}(c,N|U,\varphi) =  \sum_{k}[\tilde{c}_{s}(k)-c]^2 \text{Tr}\left[\hat{E}_k^{(s)}\ket{\Phi}\bra{\Phi}\right]$, where $\ket{\Phi}=(\ket{\psi}\ket{\phi})^{\otimes N}$. Notably, when the estimator $\tilde{c}_{s}(k)$ is (asymptotically) unbiased, $v_{s}(c,N|U,\varphi)$ is equivalent to the variance of $\tilde{c}_{s}(k)$. To compare the average performance of strategy $s$ over all possible quantum states, we consider randomly sampled qubit pairs with a fixed overlap $c$, where $U$ is Haar-distributed in $SU(2)$ and $\varphi$ is a uniformly distributed phase between $0$ and $2\pi$. More details can be found in the Supplementary Information (SI). The precision of strategy $s$ can be quantified by the average variance
\begin{equation}
    v_{s}(c,N)=\frac{1}{2\pi}\int_U\int_{0}^{2\pi}v_{s}(c,N|U,\varphi)dUd{\varphi},
    \label{specific-var-eq}
\end{equation}
where $dU$ is the Haar measure. We observe that the average variance $v_{s}(c,N)$ for each strategy exhibits a scaling behavior of $O(1/N)$. To offset the influence of the copy number $N$, we introduce the scaled average variance $Nv_{s}(c)$, which only depends on $c$ at large $N$, as the performance assessment metric of the overlap estimation strategy $s$.
Figure~\hyperref[Fig1_Schematic-Exp]{1}a illustrates the four practical strategies for overlap estimation:
\begin{enumerate}
    \item \textit{Tomography-Tomography} (TT). Reconstruct the two quantum states through quantum state tomography based on mutually unbiased bases (MUB)~\cite{doi:10.1142/S0219749910006502}, i.e., measuring the Pauli operators $(\hat{\sigma}_x,\hat{\sigma}_y,\hat{\sigma}_z)$ on $N/3$ copies of $\ket{\phi}$ and $\ket{\psi}$ respectively. The estimate states, $\ket{\tilde{\phi}}$ and $\ket{\tilde{\psi}}$, yield an overlap estimator as $\tilde{c}_{tt} = |\braket{\tilde{\psi}|\tilde{\phi}}|^2$.
    \item \textit{Tomography-Projection} (TP). Reconstruct $|\phi\rangle$ with the same quantum state tomography procedure in TT, project the $N$ copies of $|\psi\rangle$ onto the estimate $\ket{\tilde{\phi}}$ and record the number of successful projections $k$. The overlap estimator is $\tilde{c}_{tp} = k/N$ with the expectation value $p_{tp} = |\braket{\psi|\tilde{\phi}}|^2$.
    \item \textit{Schur Collective Measurement} (SCM). Perform collective measurement on each of the $N$ pairs of the qubits $\ket{\psi}\ket{\phi}$ and record the number of successful projections $k$ onto the singlet state $\ket{\Psi_-} = (\ket{01}-\ket{10})/\sqrt{2}$, where the probability of successful projection is $p_-=(1-|\braket{\psi|\phi}|^2)/2$. The estimator of overlap is given by $\tilde{c}_{scm}=1-2k/N$.
    \item \textit{Optical Swap Test} (OST). Implement a multi-mode HOMI between each of the $N$ photon pairs encoding the state $\ket{\psi}\ket{\phi}$ with the pseudo photon-number-resolving detectors (PPNRD). The states will ``fail'' or ``pass'' the test and we register $k$ ``fail'' outcomes out of $N$ measurements (see the SI for definitions). The overlap estimator is $\tilde{c}_{ost}=(1-2k/N)/\Gamma$, where $\Gamma$ is the indistinguishability between the internal modes of two photons (explained later).
\end{enumerate}
\par
We derive the average variances $v_s(c,N)$ for all the four strategies (see the SI for derivations). The summary of these strategies is presented in Table~\ref{ove_table}.
\begin{table*}[htbp]
\centering
\caption{\textbf{Summary of the four overlap estimation strategies.} The average variances $v(c,N)$ are derived in the asymptotic limit ($N\to \infty$). $\{\hat{\sigma}_i\}$: three Pauli operators. $k$: Measurement outcome statistic in corresponding strategy. $\kappa$: scaled average infidelity in the pure qubit tomography based on MUB  (see Materials and methods). HOMI: Hong-Ou-Mandel interference. PPNRD: pseudo photon-number-resolving detectors. $\Gamma$: indistinguishability between the internal modes of two photons in HOMI.}
\renewcommand\arraystretch{1.1}
\setlength{\tabcolsep}{5.3mm}
{
\begin{tabular}{lcccc}
\toprule
\toprule
\  & TT & TP & SCM &  OST \\
\midrule
Measurement & $\hat{\sigma}_i$, $\hat{\sigma}_i$ & \makecell{$\hat{\sigma}_i$, $\ket{\tilde{\phi}}\bra{\tilde{\phi}}$ }  & \makecell{$\hat{E}_- = \ket{\Psi_-}\bra{\Psi_-}$} & \makecell{HOMI, PPNRD}\\
Estimator $\tilde{c}$ & $|\braket{\tilde{\psi}|\tilde{\phi}}|^2$ & $k/N$ & $1-2k/N$ & $(1-2k/N)/\Gamma$ \\
$v(c,N)$ & $4\kappa c(1-c)/N$ & $(2\kappa+1)c(1-c)/N$ & $(1-c^2)/N$ & $(3-\Gamma c)(1-\Gamma^2 c^2)/2N\Gamma^2$ \\
\bottomrule
\end{tabular}
}
\label{ove_table}
\end{table*}
\subsection{Photonic implementation of estimation strategies}
We experimentally benchmark the aforementioned overlap estimation strategies using photonic systems. The experimental setups, depicted in Fig.~\hyperref[Fig1_Schematic-Exp]{1}b, consist of state preparation modules and four measurement modules. Different combination of the state preparation module and the measurement module forms the corresponding strategy.
\par
In the TT and TP strategies, we encode the qubit states $\ket{\psi}$ or $\ket{\phi}$ on the polarization DoF of the heralded single photons generated through the spontaneous parametric down-conversion process. The horizontal $\ket{H}$ and vertical $\ket{V}$ polarizations of the photon represent the computational bases $\ket{0}$ and $\ket{1}$, respectively. In both strategies, measurements of Pauli operators to perform the state tomography are implemented with one half-wave plate, one quarter-wave plate, and one beam displacer (BD). The wave-plates are set into three configurations to implement the three bases in MUB, followed by two single photon counting modules (SPCMs) to register the measurement outcomes. In the TP strategy, a set of electronically-controlled wave-plates enables the projection of $\ket{\psi}$ onto the reconstructed state $\ket{\tilde{\phi}}$ from the state tomography result of $\ket{\phi}$. The clicks of the corresponding SPCM are registered as the successful projections.
\par
In the SCM strategy, we encode the first qubit $\ket{\psi}$ on the path DoF of a single photon, while the second qubit $\ket{\phi}$ is encoded on the polarization DoF of the photon~\cite{RN71}. The encoding basis is $\ket{00}=\ket{s_0}\ket{H},\ket{01}=\ket{s_0}\ket{V},\ket{10}=\ket{s_1}\ket{H},\ket{11}=\ket{s_1}\ket{V}$, with $s_0$ (down) and $s_1$ (up) denoting two path modes of the photon. The POVM in the SCM strategy involves four projectors which realize the projections on the Schur bases~\cite{PhysRevLett.97.170502}: $\hat{E}_1=\ket{00}\bra{00},\hat{E}_2=\ket{11}\bra{11},\hat{E}_+=\ket{\Psi_+}\bra{\Psi_+},\hat{E}_-=\ket{\Psi_-}\bra{\Psi_-}$ with $\ket{\Psi_+} =(\ket{01}+\ket{10})/\sqrt{2} $ and $\ket{\Psi_-} =(\ket{01}-\ket{10})/\sqrt{2}$. It is noteworthy that we only need the outcome probability of $\hat{E}_-$ while the other three are need for the normalization condition (see Materials and methods).
To realize these projectors, as illustrated at the SCM module in Fig.~\hyperref[Fig1_Schematic-Exp]{1}b, the first BD splits the horizontal and vertical polarization modes of the two path modes. The horizontal polarization of the $s_0$ path and the vertical polarization of the $s_1$ path are detected by two single-photon counting modules (SPCMs), which realize projectors $\hat{E}_1$ and $\hat{E}_2$. The half-wave plate and another BD, followed by two SPCMs, implement the projectors $\hat{E}_+$ and $\hat{E}_-$ (see the SI for the details).
\begin{figure}[!t]
    \centering
    \includegraphics[width=0.5\textwidth]{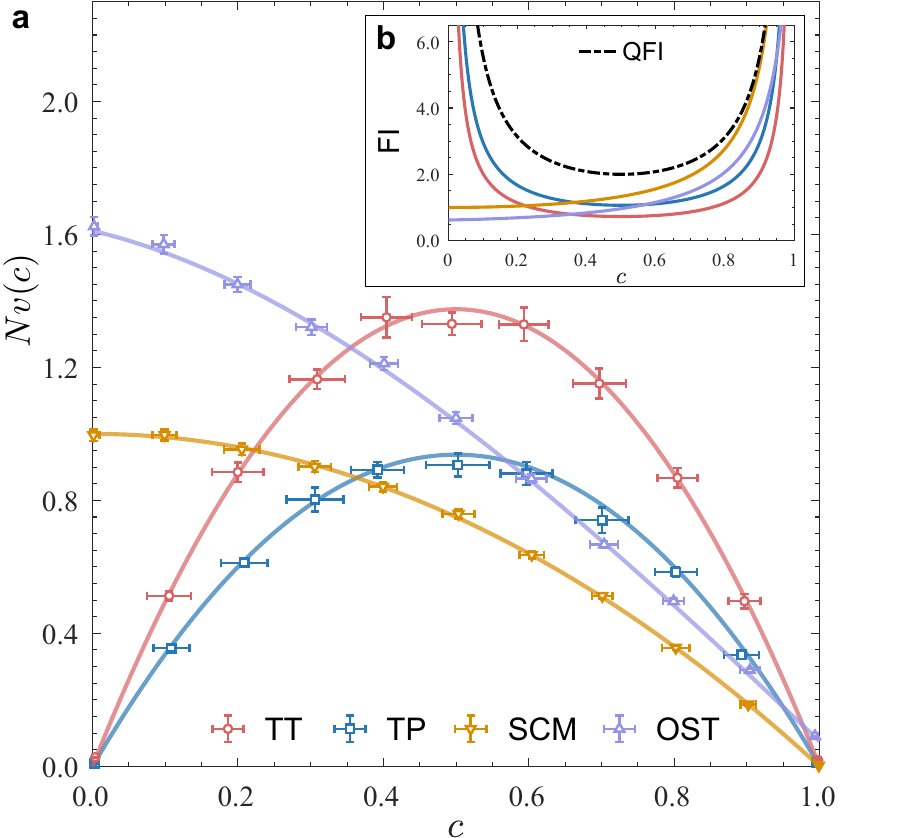}
    \caption{\textbf{Experimentally measured scaled average variances $Nv(c)$ and the corresponding Fisher information vs the value of overlap $c$.} \textbf{a} Experimentally determined (markers) and theoretical (solid lines) $Nv(c)$ for four overlap estimation strategies with the copy number $N = 900$. Vertical error bars represent the uncertainties of $Nv(c)$ over 10 runs of the experiments. Horizontal error bars denote the standard deviation of the exact overlap values for different qubit pairs generated in the experiments (see Materials and methods). \textbf{b} Normalized Fisher information (FI) per state pair at large $N$ for each strategy. 
    The black dashed line indicates the quantum Fisher information (QFI).}
    \label{Fig2_data_ove}
\end{figure}
In the OST strategy, we encode $\ket{\psi}$ and $\ket{\phi}$ on the polarization DOF of two different photons, where we regard other modes of the photon as internal modes. The optical swap test is implemented via a multi-mode HOMI~\cite{PhysRevA.87.052330} for each pair of the two photons at a balanced non-polarizing beam splitter (NPBS). After the interference, a combination of a balanced NPBS followed by two SPCMs is placed at each output port of the NPBS to function as a PPNRD, the ``pass'' outcome of the OST corresponds to the event that both photons exit the same port of the first NPBS, while the ``fail" outcome corresponds to the coincidence events that two photons are detected in different output ports of the first NPBS. Due to experimental imperfections, the two photons from the SPDC source exhibit reduced indistinguishability even when they encode the same qubit state, due to the mismatch of their internal modes, primarily the spectral mode~\cite{PhysRevA.56.1627}. We quantify this indistinguishability as $\Gamma = 0.965$, which is estimated by the maximum visibility of HOMI. In the SI, we derive the unbiased overlap estimator and its associated variance in the presence of $\Gamma$. Our analysis confirms the feasibility of performing overlap estimation using the OST strategy even in the presence of practical experimental imperfections, though the precision is reduced.
\par
\subsection{Overlap-dependent precision of strategies}
\begin{figure*}[!ht]
\centering
    \includegraphics[width=1\textwidth]{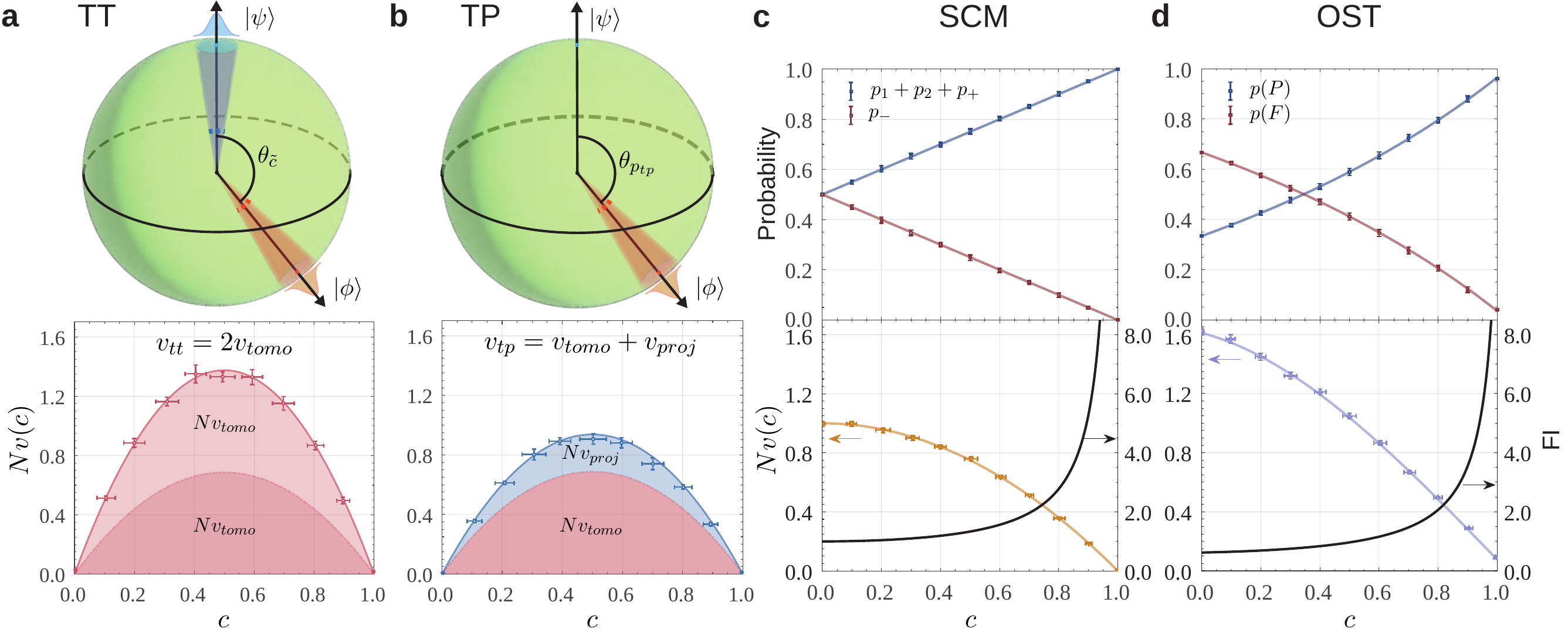}
    \caption{\textbf{Detailed analysis of the contributions to the average variance of different overlap estimation strategies.} Representations of the tomography error distributions in the Bloch sphere and their overlap-dependent contributions to the average variance in \textbf{a} TT and \textbf{b} TP strategies. Projection error accounts for the remaining portion of the average variance in TP strategy. Outcome probabilities, scaled average variances $Nv(c)$ and normalized Fisher information (FI) are presented in \textbf{c} for SCM, and \textbf{d} for OST.  In \textbf{c}, ``$p_{1}+p_{2}+p_{+}$'' represents the sum of the probabilities of the first three projectors, while $p_-$ corresponds to the probability of the last projector $\hat{E}_-$. In \textbf{d}, $p(P)$ and $p(F)$ indicate the raw detection probability of ``pass'' and ``fail'' outcomes under the PPNRD setup, respectively. Markers and lines denote experimental and theoretical results, respectively. Experimentally measured probabilities are estimated from total datasets comprising 180,000 measurements per qubit pair, and the error bars represent the standard deviation of the probabilities obtained from different qubit pairs.}
    \label{Fig3_total}
\end{figure*}
To provide a fair comparison for different overlap estimation strategies, we employ the same number of quantum states for each strategy. Specifically, we perform the experiments for $11$ overlap values equally spaced in the range $[0,1]$. For each overlap $c$, we uniformly and randomly sample $M=100$ qubit pairs $\ket{\psi_m(c)}$ and $\ket{\phi_m(c)}$, with $|\braket{\psi_m(c)|\phi_m(c)}|^2=c$ and $m\in\{1,2,...,M\}$. For each qubit pair, we collect the measurement outcomes for $N=900$ copies to obtain an estimated overlap $\tilde{c}_m$. This data collection and estimation process is repeated $n=20$ times to give the estimated variance $\tilde{v}_m(c)$. By averaging over $M$ sampled qubit pairs, which is approximately equivalent to integrate with $SU(2)$ in Eq.~\hyperref[specific-var-eq]{(1)}, we obtain the measured average variance for the strategy. To further determine the uncertainties of the estimated variance, $R=10$ independent experiments are conducted, producing a total data set $\{\{\{\tilde{c}_m^{j,r}\}_{j=1}^{n}\}_{m=1}^{M}\}_{r=1}^{R}$ of $100 \times 20 \times 10$ estimations for each overlap value of a strategy (see Materials and methods for details of data processing).
\par
Figure~\hyperref[Fig2_data_ove]{2}a shows the experimentally measured average variances scaled by the number of copies $Nv(c)$ for the four strategies. The results exhibit a clear overlap-dependent performance for all strategies, aligning well with theoretical predictions. The average variances of the two local measurement strategies, TT and TP, show symmetric behaviors in the entire overlap range. Both strategies achieve higher precision near $c=0$ and $c=1$ but lower precision for intermediate overlaps around $0.5$. TP outperforms TT for all values of $c$, due to the fact that the tailored projective measurement in TP provides more overlap information compared with tomography. In contrast, the two joint measurement strategies, SCM and OST, exhibit monotonic behaviors, i.e., they achieve lower precision for small $c$ but higher precision for large $c$ in comparison with TT and TP. Notably, SCM and OST are two different experimental realizations of the destructive swap test~\cite{PhysRevA.87.052330}. Therefore, they are expected to exhibit the same performance. Yet, the actual performance of OST in our experiment worse than that of SCM. We attribute this performance gap to experimental imperfections in the OST setup, detailed further in the SI. The first factor is the use of PPNRDs, which introduce extra photon loss and alter the outcome probability distribution, which is especially detrimental for small overlaps. The second factor is the limited HOMI visibility of two photons, leading to a constant reduction in precision over the whole range of overlaps. These two factors together contribute to the reduced OST precision observed in the experiment.
Furthermore, we evaluate the overlap estimator in each strategy by calculating the normalized Fisher information (FI) per state pair, as shown in Fig.~\hyperref[Fig2_data_ove]{2}b. The Cramér-Rao bound, defined as the inverse of the FI, provides a lower bound on the variance of any unbiased estimator for a parameter~\cite{cramer1999mathematical}. In the large $N$ limit, the normalized FI is equivalent to the inverse of corresponding $Nv(c)$ for each strategy, indicating that their overlap estimators saturate the Cramér-Rao bound. Notably, when considering large overlaps, the FI for the SCM strategy converges towards the quantum Fisher information~\cite{PhysRevLett.124.060503,PhysRevLett.72.3439}, which is the upper bound of FI for all possible measurement strategies, indicating the SCM strategy achieves ultimate precision for large overlaps. It reveals that the collective measurements involving more copies of states cannot outperform the SCM only involving a pair of states when the overlap approaches unity. 
\par
In fact, the distinct behaviors for the four estimation strategies arise from the characteristics of their measurements and estimators. Both separable measurement strategies can be separated into two separable measurement and estimation processes. Therefore, the average variance can be decomposed into the contribution either from two state tomography processes (TT), or one state tomography and the projective measurement (TP). The contribution of each part is given as (see the SI for derivations)
\begin{align}
    v_{tomo}(c,N) = &  \frac{2\kappa c(1-c)}{N}, \label{eq:tomo}\\
    v_{proj}(c,N) = & \frac{ c(1-c)}{N}, \label{eq:proj}
\end{align}
respectively, where $\kappa$ denotes the scaled average infidelity in the tomography process (see Materials and methods). It is noteworthy that the inherent error in tomography process is independent of the overlap $c$, whereas its contribution to the overlap estimation variance is overlap-dependent. Different combinations of the variances in Eqs.~\hyperref[eq:tomo]{(2-3)} lead to the overall average variance of the two strategies, as illustrated in Fig.~\hyperref[Fig3_total]{3}a and Fig.~\hyperref[Fig3_total]{3}b. For the joint measurement strategies, the overlap is estimated directly from outcomes of the joint measurements on the qubit pair. The average variance is directly related to the Fisher information from the probability distribution of measurement outcomes, as shown in Fig.~\hyperref[Fig3_total]{3}c and Fig.~\hyperref[Fig3_total]{3}d. 
\par
The aforementioned variance results allow us to determine the number of copies of states required to achieve a desired precision in overlap estimation. Applying Chebyshev’s inequality, an overlap can be estimated with an error bounded by $|\tilde{c}_s-c|\leq\varepsilon$ and ensure a probability exceeding $1-\eta$ by using approximately $N\sim f_s(c)/\eta\varepsilon^2$ copies of states (see the SI for details). Here, $f_s(c)$ denotes the scaled average variance for strategy $s$. Consequently, the overlap estimation error $\varepsilon$ scales as $O(1/\sqrt{N})$. The fact that different strategies exhibit same scaling behavior for $N$, justifies the efforts on developing practical strategies to reduce the scaled average variance.
\begin{figure*}[!ht]
    \centering
    \includegraphics[width=1\textwidth]{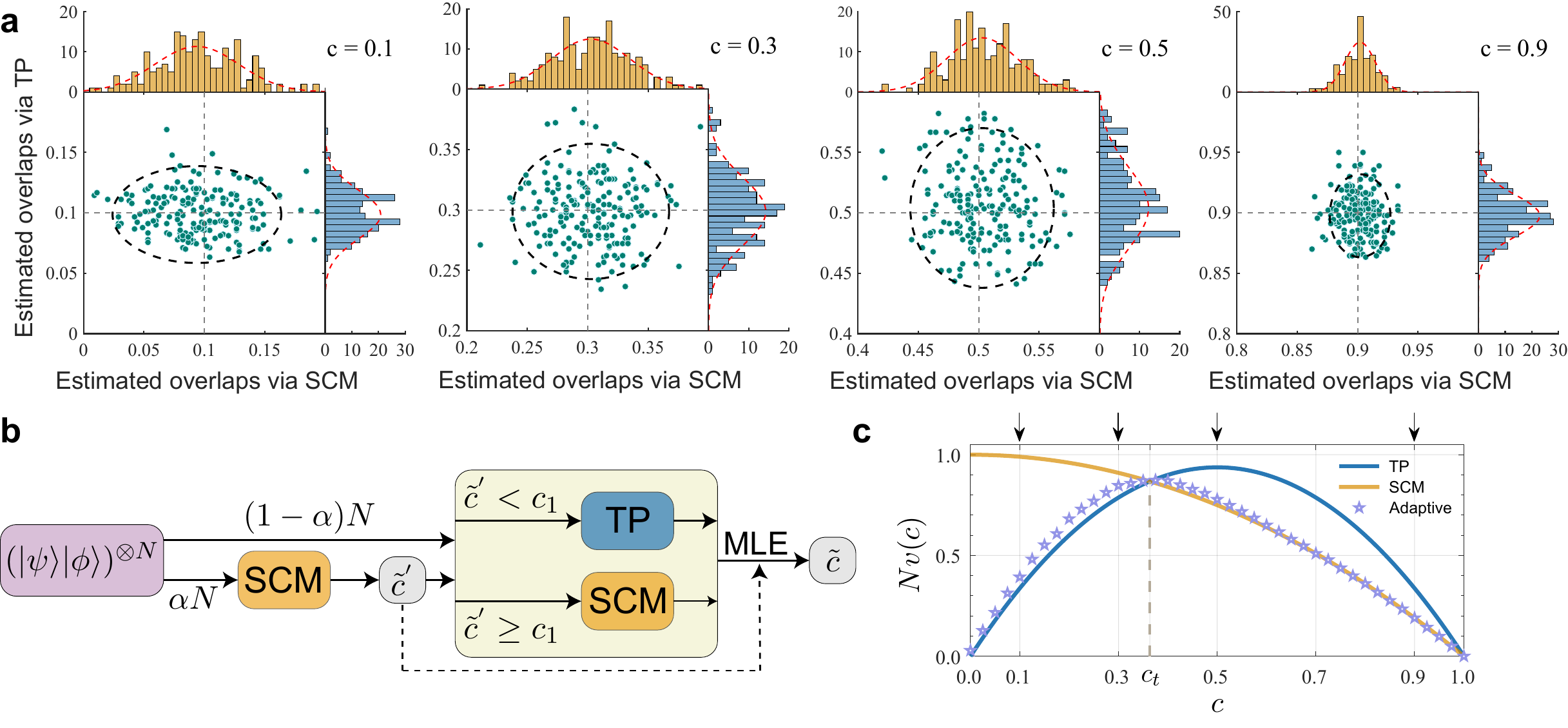}
    \caption{\textbf{Adaptive overlap estimation strategy based on TP and SCM. } \textbf{a} Comparison between the TP strategy and the SCM strategy for four exemplary overlaps indicated in \textbf{c}. Markers represent the experimentally estimated overlaps obtained using both strategies with $N=900$ copies of the sampled qubit pair. The histograms along the horizontal (SCM) and vertical (TP) axes depict the marginal distributions of 200 estimated overlaps. The red dashed lines in the histograms represent the corresponding Gaussian function fittings. The black dashed line ellipse in each panel, with principal axes proportional to the standard deviations of the marginal distributions, demonstrates the relative precision of the two strategies. \textbf{b} Schematic of the two-step adaptive strategy for overlap estimation. In the first step, an initial estimation of overlap $\tilde{c}^{'}$ is obtained using SCM with $\alpha N$ ($0<\alpha<1$) copies of $\ket{\psi}\ket{\phi}$. Based on the comparison between $\tilde{c}^{'}$ and $c_t$, one of the TP and SCM strategies is chosen for the second step with $(1-\alpha)N$ copies of $\ket{\psi}\ket{\phi}$. The final estimation $\tilde{c}$ is obtained through the maximum likelihood estimation (MLE) using the two-step overlap estimation results.  \textbf{c} Simulation results of the adaptive strategy. $N=900$ and $\alpha = 1/30$ are adopted in the simulation.}
    \label{Fig4_compare_adaptive}
\end{figure*}
\subsection{Adaptive overlap estimation strategy}
From the above experiments we can conclude that the optimal strategy among the four investigated ones varies with overlap value. As a detailed comparison, Figure~\hyperref[Fig4_compare_adaptive]{4}a compares the experimentally estimated overlaps $\tilde{c}$ with TP and SCM for different $c$. From this comparison and the average variances in Fig.~\hyperref[Fig2_data_ove]{2}a, we identify that the average variances of TP and SCM intersect at overlap $c_t=4/11$. In other words, the most efficient strategy among the four strategies is TP when the overlap $ c<c_t $ and SCM when $ c\geq c_t $. Leveraging this observation, we propose a two-step adaptive strategy that combines TP and SCM strategies, as illustrated in Fig.~\hyperref[Fig4_compare_adaptive]{4}b. Our simulation results $Nv(c)$ for the adaptive strategy are shown in Fig.~\hyperref[Fig4_compare_adaptive]{4}c. In the first step of the adaptive strategy, the SCM strategy is employed on $\alpha N$ pairs of states to get a rough estimation $\tilde{c}^{'}$, which then determines the strategy used in the second step. Notably, the copies of states used in the first step are not used in tomography process when the second step involves the TP strategy. Although the estimation variance of the adaptive strategy slightly deviates from that of the TP strategy when $c<c_t$ due to the resource consumption in the first step, our adaptive strategy still achieves nearly optimal estimation precision across the full range of overlap values compared with the four static strategies.
\par
\subsection{Overlap estimation of high-dimensional states}
The preceding analysis of the average variance of overlap estimation strategies can be generalized to high-dimensional and multi-qubit quantum states. Consider estimating the overlap between two $d$-dimensional states in a sufficient-copy scenario ($N\gg d$). For separable measurement strategies, the contribution of tomography errors to the average variance in this high-dimensional case is (see the SI for derivations)
\begin{equation}
    v_{tomo}(c,N)=\frac{2\kappa c(1-c)}{(d-1)N},
    \label{eq_vd_tomo}
\end{equation}
where $\kappa$ is the scaled average infidelity of the underlying pure state tomography approach. For $d=2$, this recovers the result in Eq.~\hyperref[eq:tomo]{(2)}. The factor $1/(d-1)$ in Eq.~\hyperref[eq_vd_tomo]{(4)} arises because, in high-dimensional state spaces, the ratio of the tomography errors projected onto the subspace spanned by $\ket{\psi}$ and $\ket{\phi}$ diminishes, thereby reducing their impact on overlap estimation. The scaling of $\kappa$ with respect to $d$ depends on the specific tomography measurements employed. When joint measurements across all copies are allowed, or when arbitrary independent measurements on each copy are permitted, $\kappa$ scales as $O(d)$~\cite{PhysRevA.72.032325,KUENG201788,10.1145/2897518.2897585}, resulting in a dimension-independent average variance of $O(c(1-c)/N)$ for both TT and TP strategies. This dimension-independence holds specifically under the sufficient-copy condition ($N \gg d$). When the high-dimensional states are $n$-qubit states ($d=2^n$), restricting tomography to local, single-qubit measurements leads to $\kappa$ scaling as $O(d^2\log d)=O(4^n n)$~\cite{Flammia_2012}, and a dimension-dependent average variance of $O(4^n nc(1-c)/N)$ for TT and TP.
\par
For two joint measurement strategies, both SCM and OST can be extended to higher dimensions while maintaining precision independent of $d$ (see the SI for details). In the sufficient-copy scenario, allowing joint measurements for tomography yields comparable performance across all four strategies. However, SCM and OST significantly outperform TT and TP for multi-qubit states when the latter are restricted to local measurements on each qubit for tomography.
\par
In the limited-copy scenario ($N\sim d$), tomography yields highly inaccurate estimations due to information incompleteness and substantial statistical errors. Therefore, the errors of TT and TP strategies deviate significantly from the average variances derived in the sufficient-copy scenario, as the bias becomes non-negligible. Both $v_{tt}$ and $v_{tp}$ are then dominated by a constant error scale as $O(1)$ (see the SI for derivations). This problem is exacerbated by increasing qubit number $n$, leading to exponential growth in $d$ and rendering quantum state tomography infeasible. In these situations, SCM and OST offer a significant advantage due to their inherent dimension independence.
\par
\section{Discussion}
In this work, we present a comprehensive investigation of four representative strategies for estimating the overlap of two unknown quantum states using a photonic setup. We compare the performance in terms of the average estimation variance of the separable measurement strategies including tomography-tomography (TT) and tomography-projection (TP) with that of the joint measurements strategies including Schur collective measurement (SCM) and optical swap test (OST). Our experimental results demonstrate the superior performance of the TP strategy over the TT strategy for all overlap values considered. Moreover, although in principle the OST strategy matches the performance of the SCM strategy, it exhibits poorer performance in the presence of experimental imperfections when compared to the SCM strategy, which indicates that high-dimension encoded single-photon systems are more robust against experimental errors. These results reveal that the optimal strategy among the four varies depends on the overlap values. To approach the optimal performance across the full range of overlaps, we further design an adaptive strategy combining TP and SCM strategies. Our experiments with single qubits show that separable measurements involving tomography achieve precision comparable to joint measurements performed on pairs of states. Yet, for quantum states with higher dimensions and multiple qubits, theoretical analysis reveals that SCM and OST benefit from dimension-independence, providing a significant advantage, whereas TT and TP become highly dimension-dependent when the number of copies is limited or only single-qubit measurements are available for tomography. By elucidating the overlap-dependent precision with practical setups, our work provides new insights into designing measurement strategies for extracting parameters of interest from quantum states, a vital task in quantum information applications~\cite{PhysRevA.70.032321, PhysRevA.73.022341, RevModPhys.79.555, PhysRevX.10.031023, PhysRevLett.88.217901,PhysRevLett.95.260502,PhysRevA.75.032338,PhysRevA.106.062413,s41467-023-39785-8,RN67,PhysRevA.106.042431,PhysRevLett.113.130503,havlivcek2019supervised,PhysRevA.107.022402,RN80,PhysRevX.9.041029,PhysRevA.101.032323,PhysRevLett.87.167902,RN70,agliardi2023,PhysRevLett.122.040504,RN81,doi:10.1126/sciadv.aav2761,doi:10.1126/sciadv.adj4249,PhysRevLett.124.010504}.
\par
Several avenues exist for future research to enhance the strategies presented here. The separable measurement strategies, TT and TP, can benefit from adaptive quantum state tomography techniques~\cite{PhysRevLett.89.277904,PhysRevLett.111.183601,RN91}. The SCM strategy can be further improved by incorporating collective measurements involving more than one pair of states, surpassing the performance of the ideal swap test, as discussed in~\cite{PhysRevLett.124.060503}. Moreover, the SCM and OST strategies can be generalized and applied to higher dimensional or multi-qubit quantum systems~\cite{s41377-021-00608-4,PhysRevLett.131.150803}. In practice, the efficiency of all strategies can be improved by utilizing faster optical systems~\cite{Zhang:22,PhysRevA.107.032608}.
\par
Our work provides an example of striking a balance between optimized performance and experimental complexity, aiming to minimize the gap between theoretical proposals and experimental attainable performance in overlap estimation strategies. Given the prevalence of overlap estimation in quantum machine learning algorithms~\cite{s41467-023-39785-8,RN67,PhysRevA.106.042431,PhysRevLett.113.130503,havlivcek2019supervised,PhysRevA.107.022402,PhysRevX.9.041029,PhysRevA.101.032323,PhysRevLett.87.167902,RN70,agliardi2023,PhysRevLett.122.040504,RN80,RN81,doi:10.1126/sciadv.aav2761}, the optimized estimation strategies can find immediate applications in quantum algorithms involving readout of state overlaps as the cost function or quantum kernel function~\cite{PhysRevLett.113.130503,havlivcek2019supervised,RN67,PhysRevA.106.042431,PhysRevLett.122.040504,RN80}. State overlaps quantify the similarity between data points mapped into the quantum feature space in quantum kernel methods, which have wide-ranging applications from data classification to training quantum models~\cite{schuld2021supervised,Bowie_2024,PhysRevA.107.032428, PhysRevA.98.062324,PhysRevA.105.042432,roncallo2024quantum}. We anticipate that the strategies explored here, along with the understanding of their corresponding precision, can be applied to construct quantum kernels, learn quantum systems and train quantum neural networks, resulting in improved training efficiency and overall performance.

\section*{Materials and methods}
\subsection{Precision of separable measurement strategies}
In TT and TP strategies, reconstructing the qubit states relies on the tomography based on MUB measurements with the prior knowledge that the state is pure. The tomography fidelity can be quantified by $F = |\braket{\psi|\tilde{\psi}}|^2$, defined as the overlap between the true state $\ket{\psi}$ and the reconstructed state $\ket{\tilde{\psi}}$. We consider the average fidelity $\overline{F}$, averaged over the distribution of the reconstructed state and the unitary $U$ where $\ket{\psi} = U\ket{0}$. At the asymptotic limit (the number of copies $N\to\infty$), the average fidelity is derived as $\overline{F} = 1-\kappa/N$, and $\kappa = N(1-\overline{F})$ is defined as the scaled average infidelity, with the analytical value $\kappa=11/8$ for MUB measurements. Through the error analysis in the tomography process, we can represent the reconstructed states $\ket{\tilde{\psi}}$ as 
\begin{equation}
\begin{split}
        \ket{\tilde{\psi}} = \cos{\chi}\ket{\psi}+\sin{\chi}e^{i\zeta}\ket{\psi_{\perp}},\\
\end{split}
\label{eq-psi-error}
\end{equation}
where $\ket{\psi_{\perp}} = U\ket{1}$, $\chi$ and $\zeta$ are two error parameters introduced by the tomography. Furthermore, we derive the average values for the functions of $\chi$ and $\zeta$ as: $ \overline{\braket{(\sin{\chi}\cos{\zeta})^2}} \approx \overline{\braket{(\sin{\chi}\sin{\zeta})^2}}\approx\overline{\left\langle \chi^2 \right\rangle}/2\approx 2\kappa/N$, where $\braket{\cdot}$ and the overline denote the average over the conditional probability distribution $p(\chi,\zeta|U)$ and the unitary $U$, respectively. 
\par
In the TT strategy, the other reconstructed state $\ket{\tilde{\phi}}$ has a similar form to Eq.~\hyperref[eq-psi-error]{(5)}, and the variance for TT can be expressed with the error parameters $\chi$ and $\zeta$. The average variance can then be shown as 
\begin{equation}
    v_{tt}(c,N) = 2\overline{\braket{\chi^2}}c(1-c) +O\left(\frac{1}{N^2}\right) \approx \frac{4\kappa c(1-c)}{N},
\end{equation}
here we keep the leading term.
\par
In the TP strategy, we only need to consider the tomography error of the one state $\ket{\phi}$, but together with an additional error introduced by the projection procedure. The successful projection probability $p_{tp} = |\braket{\tilde{\phi}|\psi}|^2$ can be shown as the function of parameters $\chi$ and $\zeta$ in $\ket{\tilde{\phi}}$. The average variance for TP strategy is derived as $v_{tp}(c,N) = \overline{\left\langle(p_{tp}-c)^2\right\rangle}+\overline{\left\langle p_{tp}(1-p_{tp})/N\right\rangle}$, where $\overline{\left\langle(p_{tp}-c)^2\right\rangle}$ implies the tomography error and $\overline{\left\langle p_{tp}(1-p_{tp})/N\right\rangle}$ denotes the projection error. The final result for $v_{tp}$ is given by
\begin{equation}
\begin{split}
        v_{tp}(c,N) & = \overline{\braket{\chi^2}}c(1-c)+\frac{c(1-c)}{N} +O\left(\frac{1}{N^2}\right) \\
        & \approx \frac{(2\kappa+1)c(1-c)}{N}.
\end{split}
\end{equation}
\par
Generalizing to $d$-dimensional states, the average variances for TT and TP can be shown as
\begin{equation}
    \begin{split}
        v_{tt}(c,N) & \approx \frac{4\kappa c(1-c)}{(d-1)N}, \\
        v_{tp}(c,N) & \approx \left(\frac{2\kappa }{d-1}+1\right)\frac{c(1-c)}{N},
    \end{split}
\end{equation}
where $d$ denotes the dimension of a single-copy state. The scaled average infidelity $\kappa$ for high-dimensional state tomography depends on $d$ and varies with the tomography approach. In the SI, we provide detailed derivations of the average variances and their high-dimensional generalizations, and also demonstrates that the estimators used in TT and TP are asymptotically unbiased.
\subsection{Precision of joint measurement strategies}
In a joint measurement strategy, the overlap information is extracted directly by performing a POVM $\{\hat{E}_i\}$ on the joint state $\ket{\Phi_0} = \ket{\psi}\ket{\phi}$, with each element $\hat{E}_i$ associated with a measurement outcome $i$. According to Born's rule, the probability of obtaining the outcome $i$ is $p_i =\text{Tr}\left(\hat{E}_i\ket{\Phi_0}\bra{\Phi_0}\right)$, which depends on the overlap $c$. Noting that the joint measurement is static, the precision of the overlap estimation is bounded by the Fisher information (FI) derived from the probability distribution as $I(c) = \sum_i p_i(d\log{p_i}/dc)^2$. In the SCM strategy, the measurements are described by four projectors $\{\hat{E}_1,\hat{E}_2,\hat{E}_+,\hat{E}_-\}$, and the corresponding probability distribution is given by
\begin{equation}
    p_1+p_2+p_+ = \frac{1+c}{2},\ p_- = \frac{1-c}{2},
\end{equation}
where $p_i = \bra{\Phi_0}\hat{E}_i\ket{\Phi_0}$. By combining the first three outcomes into one, we obtain binary outcomes where the probabilities solely rely on the overlap $c$. The FI per state pair is given by $I_{scm} = 1/(1-c^2)$. The overlap estimator $\tilde{c}_{scm} = 1-2k/N$, where $k$ is the number of occurrences of outcome $\hat{E}_-$ in $N$ measurements, saturates the Cramér-Rao bound with the variance
\begin{equation}
    v_{scm}(c,N) = \frac{1-c^2}{N}.
\end{equation}
\par
In the OST strategy, the ideal OST yields a binary outcome of either ``pass'' or ``fail'' with the probability of ``fail'' outcome given by $p_{f} = (1-c)/2$. However, due to experimental imperfections, the outcome probability distribution deviates from the ideal case. In our experiments, with the internal mode indistinguishability $\Gamma$ between two photons in HOMI and the PPNRD setup, the outcome probability distribution is given by
\begin{equation}
    p(P)   = \frac{1+\Gamma c}{3-\Gamma c},\ p(F)  = \frac{2-2\Gamma c}{3-\Gamma c},
\end{equation}
where $p(P)$ and $p(F)$ denote the probabilities that the PPNRD response the ``pass'' and ``fail'' outcomes, respectively. It is worth noting that the PPNRD introduces photon loss, which must be take into account in precision comparison. On average, for $N$ state pairs, only $N^{\prime} = (3-\Gamma c)N/4$ events are detected. To ensure a fair comparison, we calculate that effective FI per state pair as $I^{e}_{ost} = 2\Gamma^2/(3-\Gamma c)(1-\Gamma^2 c^2)$ to bound the precision of OST (see the SI for detailed derivation). Using the estimator $\tilde{c}_{ost} = (1-2k_f/N)/\Gamma$, the variance for the OST strategy is given by
\begin{equation}
    v_{ost}(c,N) = \frac{(3-\Gamma c)(1-\Gamma^2 c^2)}{2N\Gamma^2}.
\end{equation}
\par
In these two joint measurement strategies, the outcome probabilities depend solely on the overlap between the two states, rather than the specific states themselves. Therefore, the variance mentioned above is equal to the average variance in SCM and OST.
\subsection{Photon source}
Frequency-doubled light pulses ($\sim$150 fs duration, 415 nm central wavelength) originating from a Ti:Sapphire laser (76 MHz repetition rate; Coherent Mira-HP) pump a beta barium borate ($\beta$-BBO) crystal phase-matched for type-II beamlike spontaneous parametric down conversion (SPDC) to produce degenerate photon pairs (830 nm central wavelength). The photon pairs undergo spectral filtering with 3 nm full-width at half-maximum and are collected into single-mode fibers. The pump power is set to $\sim$100 mW to ensure a low probability of emitting two-photon pairs. In TT, TP, and SCM experiments, one of the photon pair is detected by a SPCM (Excelitas Technologies), while the other serves as a heralded single photon. In the OST experiment, both photons undergo HOMI. Despite the presence of systemic errors and interference drift, an average maximum HOMI visibility of $0.965\pm0.008$ is observed. 
\subsection{State preparation}
In the TT, TP and OST strategy experiments, a combination of an electronically controlled half wave-plate (E-HWP) and a liquid crystal phase retarder (LCPR, Thorlabs, LCC1113-B), prepares the horizontal polarized photon to the state
\begin{equation}
    \ket{\psi}\text{or}\ket{\phi}=\cos{2\theta}\ket{H}+e^{i\alpha}\sin{2\theta}\ket{V},
\end{equation}
where $\theta$ and $\alpha$ denote the E-HWP angle and the relative phase between two polarizations added by the LCPR, respectively. In the SCM strategy experiment, we firstly encode $\ket{\psi}$ on the polarization DoF of the single photon and use a BD and HWPs to transfer the polarization-encoded qubit to a path-encoded qubit. The second qubit $\ket{\phi}$ is then encoded on the polarization DoF of the photon through a E-HWP and a QHQ (QWP-HWP-QWP) wave-plate group, resulting in a two-qubit joint state
\begin{equation}
\begin{split}
       \ket{\psi}\otimes\ket{\phi} = &\ \left(\cos2\theta_1\ket{s_0}+e^{i\alpha_1}\sin{2\theta_1}\ket{s_1}\right)\\
       &\otimes\left(\cos{2\theta_2}\ket{H}+e^{i\alpha_2}\sin{2\theta_2}\ket{V}\right).
\end{split}
\end{equation}
Here, $\theta_1$ and $\alpha_1$ denote the E-HWP angle and the relative phase from the LCPR used to prepare $\ket{\psi}$, and $\theta_2$ and $\alpha_2$ denote the E-HWP angle and the relative phase from the QHQ group used to prepare $\ket{\phi}$.
\subsection{Data processing and uncertainty quantification}
For each chosen overlap $c$ in our experiments, we have a total data set of estimated overlaps $\{\{\{\tilde{c}_m^{j,r}\}_{j=1}^{n}\}_{m=1}^{M}\}_{r=1}^{R}$. Here, $R$ groups of data are collected by repetitive runs of the experiments for the TT, OST and SCM strategies, while in the TP strategy, the data is generated using the Bootstrap method from a single group to reduce data acquisition time. To obtain the estimated average variances $\{\tilde{v}^r\}_{r=1}^R$, we process the overlap data as
\begin{equation}
\begin{split}
        \tilde{v}^{r} = \frac{1}{M}\sum_{m=1}^M\tilde{v}^{r}_m,\ 
        \tilde{v}^{r}_m = \frac{1}{n-1}\sum_{j=1}^n(\tilde{c}_m^{j,r}-\frac{1}{n}\sum_{j=1}^n\tilde{c}_m^{j,r})^2.
\end{split}
\end{equation}
The mean and the standard deviation for the average variance are then calculated as 
\begin{equation}
    \tilde{v}=\frac{1}{R}\sum_r\tilde{v}^{r},\ \delta \tilde{v} = \sqrt{\frac{\sum_r(\tilde{v}^{r}-\tilde{v})^2}{R-1}}.
\end{equation}
Scaling the results by the copy number $N$, $N\tilde{v}$ and $N\delta\tilde{v}$ correspond to the scaled average variance $Nv(c)$ and the vertical uncertainty in Fig.~\hyperref[Fig2_data_ove]{2}a. Considering the systematic errors in state preparation and measurements, the exact overlaps being measured, between different qubit pairs in state preparation, may deviate from the target overlap $c$. To quantify this uncertainty, we estimate the exact overlaps from the data for the same pairs of states in large number of copies to obtain the exact overlap data set $\{\bar{c}_m\}_{m=1}^{M}$, where $\bar{c}_m=\sum_{j,r}\tilde{c}_m^{j,r}/nR$. The average exact overlap and corresponding standard deviation are given by
\begin{equation}
    \bar{c} = \frac{1}{M}\sum_m\bar{c}_m,\ \delta\bar{c} = \sqrt{\frac{\sum_m(\bar{c}_m-\bar{c})^2}{M-1}}.
\end{equation}
Here, $\bar{c}$ and $\delta\bar{c}$ indicate the overlap $c$ (markers) and the corresponding uncertainty (horizontal error bars) in Fig.~\hyperref[Fig2_data_ove]{2}a.

\section*{Acknowledgments}
The authors thank Nengkun Yu and Penghui Yao for helpful discussions and Zhenghao Yin for valuable comments on quantum machine learning. This work was supported by National Natural Science Foundation   of   China (Grants No. U24A2017, No. 12347104 and No. 12461160276), the National Key Research and Development Program of China (Grants No. 2023YFC2205802), Natural Science Foundation of Jiangsu Province (Grants No. BK20243060 and No. BK20233001), in part by State Key Laboratory of Advanced Optical Communication Systems and Networks, China.
\section*{Author contributions}
L.Z., A.Z. and H.Z. conceived the project. H.Z. and A.Z. developed the theoretical analysis, numerical calculation and experimental design. H.Z. performed the experiments, with contributions from B.W., M.M. and J.X.. H.Z., B.W., L.X., A.Z. and L.Z. analyzed the data. H.Z., A.Z. and L.Z. wrote the paper with input from all authors. 
\section*{Data availability} 
All data needed to evaluate the conclusions in the paper are present in the paper and/or the Supplementary Information.
\section*{Competing interests}
All authors declare no competing interests. 
\bibliography{OVE.bib}

\begin{thebibliography}{76}%
\makeatletter
\providecommand \@ifxundefined [1]{%
 \@ifx{#1\undefined}
}%
\providecommand \@ifnum [1]{%
 \ifnum #1\expandafter \@firstoftwo
 \else \expandafter \@secondoftwo
 \fi
}%
\providecommand \@ifx [1]{%
 \ifx #1\expandafter \@firstoftwo
 \else \expandafter \@secondoftwo
 \fi
}%
\providecommand \natexlab [1]{#1}%
\providecommand \enquote  [1]{``#1''}%
\providecommand \bibnamefont  [1]{#1}%
\providecommand \bibfnamefont [1]{#1}%
\providecommand \citenamefont [1]{#1}%
\providecommand \href@noop [0]{\@secondoftwo}%
\providecommand \href [0]{\begingroup \@sanitize@url \@href}%
\providecommand \@href[1]{\@@startlink{#1}\@@href}%
\providecommand \@@href[1]{\endgroup#1\@@endlink}%
\providecommand \@sanitize@url [0]{\catcode `\\12\catcode `\$12\catcode
  `\&12\catcode `\#12\catcode `\^12\catcode `\_12\catcode `\%12\relax}%
\providecommand \@@startlink[1]{}%
\providecommand \@@endlink[0]{}%
\providecommand \url  [0]{\begingroup\@sanitize@url \@url }%
\providecommand \@url [1]{\endgroup\@href {#1}{\urlprefix }}%
\providecommand \urlprefix  [0]{URL }%
\providecommand \Eprint [0]{\href }%
\providecommand \doibase [0]{http://dx.doi.org/}%
\providecommand \selectlanguage [0]{\@gobble}%
\providecommand \bibinfo  [0]{\@secondoftwo}%
\providecommand \bibfield  [0]{\@secondoftwo}%
\providecommand \translation [1]{[#1]}%
\providecommand \BibitemOpen [0]{}%
\providecommand \bibitemStop [0]{}%
\providecommand \bibitemNoStop [0]{.\EOS\space}%
\providecommand \EOS [0]{\spacefactor3000\relax}%
\providecommand \BibitemShut  [1]{\csname bibitem#1\endcsname}%
\let\auto@bib@innerbib\@empty
\bibitem [{\citenamefont {Bartlett}\ \emph {et~al.}(2004)\citenamefont
  {Bartlett}, \citenamefont {Rudolph},\ and\ \citenamefont
  {Spekkens}}]{PhysRevA.70.032321}%
  \BibitemOpen
  \bibfield  {author} {\bibinfo {author} {\bibfnamefont {S.~D.}\ \bibnamefont
  {Bartlett}}, \bibinfo {author} {\bibfnamefont {T.}~\bibnamefont {Rudolph}}, \
  and\ \bibinfo {author} {\bibfnamefont {R.~W.}\ \bibnamefont {Spekkens}},\
  }\href {\doibase 10.1103/PhysRevA.70.032321} {\bibfield  {journal} {\bibinfo
  {journal} {Phys. Rev. A}\ }\textbf {\bibinfo {volume} {70}},\ \bibinfo
  {pages} {032321} (\bibinfo {year} {2004})}\BibitemShut {NoStop}%
\bibitem [{\citenamefont {Bagan}\ \emph {et~al.}(2006)\citenamefont {Bagan},
  \citenamefont {Iblisdir},\ and\ \citenamefont {Mu\~noz
  Tapia}}]{PhysRevA.73.022341}%
  \BibitemOpen
  \bibfield  {author} {\bibinfo {author} {\bibfnamefont {E.}~\bibnamefont
  {Bagan}}, \bibinfo {author} {\bibfnamefont {S.}~\bibnamefont {Iblisdir}}, \
  and\ \bibinfo {author} {\bibfnamefont {R.}~\bibnamefont {Mu\~noz Tapia}},\
  }\href {\doibase 10.1103/PhysRevA.73.022341} {\bibfield  {journal} {\bibinfo
  {journal} {Phys. Rev. A}\ }\textbf {\bibinfo {volume} {73}},\ \bibinfo
  {pages} {022341} (\bibinfo {year} {2006})}\BibitemShut {NoStop}%
\bibitem [{\citenamefont {Bartlett}\ \emph {et~al.}(2007)\citenamefont
  {Bartlett}, \citenamefont {Rudolph},\ and\ \citenamefont
  {Spekkens}}]{RevModPhys.79.555}%
  \BibitemOpen
  \bibfield  {author} {\bibinfo {author} {\bibfnamefont {S.~D.}\ \bibnamefont
  {Bartlett}}, \bibinfo {author} {\bibfnamefont {T.}~\bibnamefont {Rudolph}}, \
  and\ \bibinfo {author} {\bibfnamefont {R.~W.}\ \bibnamefont {Spekkens}},\
  }\href {\doibase 10.1103/RevModPhys.79.555} {\bibfield  {journal} {\bibinfo
  {journal} {Rev. Mod. Phys.}\ }\textbf {\bibinfo {volume} {79}},\ \bibinfo
  {pages} {555} (\bibinfo {year} {2007})}\BibitemShut {NoStop}%
\bibitem [{\citenamefont {Tsang}\ \emph {et~al.}(2020)\citenamefont {Tsang},
  \citenamefont {Albarelli},\ and\ \citenamefont {Datta}}]{PhysRevX.10.031023}%
  \BibitemOpen
  \bibfield  {author} {\bibinfo {author} {\bibfnamefont {M.}~\bibnamefont
  {Tsang}}, \bibinfo {author} {\bibfnamefont {F.}~\bibnamefont {Albarelli}}, \
  and\ \bibinfo {author} {\bibfnamefont {A.}~\bibnamefont {Datta}},\ }\href
  {\doibase 10.1103/PhysRevX.10.031023} {\bibfield  {journal} {\bibinfo
  {journal} {Phys. Rev. X}\ }\textbf {\bibinfo {volume} {10}},\ \bibinfo
  {pages} {031023} (\bibinfo {year} {2020})}\BibitemShut {NoStop}%
\bibitem [{\citenamefont {Giordani}\ \emph {et~al.}(2023)\citenamefont
  {Giordani}, \citenamefont {Wagner}, \citenamefont {Esposito}, \citenamefont
  {Camillini}, \citenamefont {Hoch}, \citenamefont {Carvacho}, \citenamefont
  {Pentangelo}, \citenamefont {Ceccarelli}, \citenamefont {Piacentini},
  \citenamefont {Crespi}, \citenamefont {Spagnolo}, \citenamefont {Osellame},
  \citenamefont {Galvão},\ and\ \citenamefont
  {Sciarrino}}]{doi:10.1126/sciadv.adj4249}%
  \BibitemOpen
  \bibfield  {author} {\bibinfo {author} {\bibfnamefont {T.}~\bibnamefont
  {Giordani}}, \bibinfo {author} {\bibfnamefont {R.}~\bibnamefont {Wagner}},
  \bibinfo {author} {\bibfnamefont {C.}~\bibnamefont {Esposito}}, \bibinfo
  {author} {\bibfnamefont {A.}~\bibnamefont {Camillini}}, \bibinfo {author}
  {\bibfnamefont {F.}~\bibnamefont {Hoch}}, \bibinfo {author} {\bibfnamefont
  {G.}~\bibnamefont {Carvacho}}, \bibinfo {author} {\bibfnamefont
  {C.}~\bibnamefont {Pentangelo}}, \bibinfo {author} {\bibfnamefont
  {F.}~\bibnamefont {Ceccarelli}}, \bibinfo {author} {\bibfnamefont
  {S.}~\bibnamefont {Piacentini}}, \bibinfo {author} {\bibfnamefont
  {A.}~\bibnamefont {Crespi}}, \bibinfo {author} {\bibfnamefont
  {N.}~\bibnamefont {Spagnolo}}, \bibinfo {author} {\bibfnamefont
  {R.}~\bibnamefont {Osellame}}, \bibinfo {author} {\bibfnamefont {E.~F.}\
  \bibnamefont {Galvão}}, \ and\ \bibinfo {author} {\bibfnamefont
  {F.}~\bibnamefont {Sciarrino}},\ }\href {\doibase 10.1126/sciadv.adj4249}
  {\bibfield  {journal} {\bibinfo  {journal} {Science Advances}\ }\textbf
  {\bibinfo {volume} {9}},\ \bibinfo {pages} {eadj4249} (\bibinfo {year}
  {2023})}\BibitemShut {NoStop}%
\bibitem [{\citenamefont {Ekert}\ \emph {et~al.}(2002)\citenamefont {Ekert},
  \citenamefont {Alves}, \citenamefont {Oi}, \citenamefont {Horodecki},
  \citenamefont {Horodecki},\ and\ \citenamefont
  {Kwek}}]{PhysRevLett.88.217901}%
  \BibitemOpen
  \bibfield  {author} {\bibinfo {author} {\bibfnamefont {A.~K.}\ \bibnamefont
  {Ekert}}, \bibinfo {author} {\bibfnamefont {C.~M.}\ \bibnamefont {Alves}},
  \bibinfo {author} {\bibfnamefont {D.~K.~L.}\ \bibnamefont {Oi}}, \bibinfo
  {author} {\bibfnamefont {M.}~\bibnamefont {Horodecki}}, \bibinfo {author}
  {\bibfnamefont {P.}~\bibnamefont {Horodecki}}, \ and\ \bibinfo {author}
  {\bibfnamefont {L.~C.}\ \bibnamefont {Kwek}},\ }\href {\doibase
  10.1103/PhysRevLett.88.217901} {\bibfield  {journal} {\bibinfo  {journal}
  {Phys. Rev. Lett.}\ }\textbf {\bibinfo {volume} {88}},\ \bibinfo {pages}
  {217901} (\bibinfo {year} {2002})}\BibitemShut {NoStop}%
\bibitem [{\citenamefont {Mintert}\ \emph {et~al.}(2005)\citenamefont
  {Mintert}, \citenamefont {Kuś},\ and\ \citenamefont
  {Buchleitner}}]{PhysRevLett.95.260502}%
  \BibitemOpen
  \bibfield  {author} {\bibinfo {author} {\bibfnamefont {F.}~\bibnamefont
  {Mintert}}, \bibinfo {author} {\bibfnamefont {M.}~\bibnamefont {Kuś}}, \
  and\ \bibinfo {author} {\bibfnamefont {A.}~\bibnamefont {Buchleitner}},\
  }\href {\doibase 10.1103/PhysRevLett.95.260502} {\bibfield  {journal}
  {\bibinfo  {journal} {Phys. Rev. Lett.}\ }\textbf {\bibinfo {volume} {95}},\
  \bibinfo {pages} {260502} (\bibinfo {year} {2005})}\BibitemShut {NoStop}%
\bibitem [{\citenamefont {Walborn}\ \emph {et~al.}(2007)\citenamefont
  {Walborn}, \citenamefont {Ribeiro}, \citenamefont {Davidovich}, \citenamefont
  {Mintert},\ and\ \citenamefont {Buchleitner}}]{PhysRevA.75.032338}%
  \BibitemOpen
  \bibfield  {author} {\bibinfo {author} {\bibfnamefont {S.~P.}\ \bibnamefont
  {Walborn}}, \bibinfo {author} {\bibfnamefont {P.~H.~S.}\ \bibnamefont
  {Ribeiro}}, \bibinfo {author} {\bibfnamefont {L.}~\bibnamefont {Davidovich}},
  \bibinfo {author} {\bibfnamefont {F.}~\bibnamefont {Mintert}}, \ and\
  \bibinfo {author} {\bibfnamefont {A.}~\bibnamefont {Buchleitner}},\ }\href
  {\doibase 10.1103/PhysRevA.75.032338} {\bibfield  {journal} {\bibinfo
  {journal} {Phys. Rev. A}\ }\textbf {\bibinfo {volume} {75}},\ \bibinfo
  {pages} {032338} (\bibinfo {year} {2007})}\BibitemShut {NoStop}%
\bibitem [{\citenamefont {Consiglio}\ \emph {et~al.}(2022)\citenamefont
  {Consiglio}, \citenamefont {Apollaro},\ and\ \citenamefont
  {Wieśniak}}]{PhysRevA.106.062413}%
  \BibitemOpen
  \bibfield  {author} {\bibinfo {author} {\bibfnamefont {M.}~\bibnamefont
  {Consiglio}}, \bibinfo {author} {\bibfnamefont {T.~J.~G.}\ \bibnamefont
  {Apollaro}}, \ and\ \bibinfo {author} {\bibfnamefont {M.}~\bibnamefont
  {Wieśniak}},\ }\href {\doibase 10.1103/PhysRevA.106.062413} {\bibfield
  {journal} {\bibinfo  {journal} {Phys. Rev. A}\ }\textbf {\bibinfo {volume}
  {106}},\ \bibinfo {pages} {062413} (\bibinfo {year} {2022})}\BibitemShut
  {NoStop}%
\bibitem [{\citenamefont {Elben}\ \emph {et~al.}(2020)\citenamefont {Elben},
  \citenamefont {Vermersch}, \citenamefont {van Bijnen}, \citenamefont
  {Kokail}, \citenamefont {Brydges}, \citenamefont {Maier}, \citenamefont
  {Joshi}, \citenamefont {Blatt}, \citenamefont {Roos},\ and\ \citenamefont
  {Zoller}}]{PhysRevLett.124.010504}%
  \BibitemOpen
  \bibfield  {author} {\bibinfo {author} {\bibfnamefont {A.}~\bibnamefont
  {Elben}}, \bibinfo {author} {\bibfnamefont {B.}~\bibnamefont {Vermersch}},
  \bibinfo {author} {\bibfnamefont {R.}~\bibnamefont {van Bijnen}}, \bibinfo
  {author} {\bibfnamefont {C.}~\bibnamefont {Kokail}}, \bibinfo {author}
  {\bibfnamefont {T.}~\bibnamefont {Brydges}}, \bibinfo {author} {\bibfnamefont
  {C.}~\bibnamefont {Maier}}, \bibinfo {author} {\bibfnamefont {M.~K.}\
  \bibnamefont {Joshi}}, \bibinfo {author} {\bibfnamefont {R.}~\bibnamefont
  {Blatt}}, \bibinfo {author} {\bibfnamefont {C.~F.}\ \bibnamefont {Roos}}, \
  and\ \bibinfo {author} {\bibfnamefont {P.}~\bibnamefont {Zoller}},\ }\href
  {\doibase 10.1103/PhysRevLett.124.010504} {\bibfield  {journal} {\bibinfo
  {journal} {Phys. Rev. Lett.}\ }\textbf {\bibinfo {volume} {124}},\ \bibinfo
  {pages} {010504} (\bibinfo {year} {2020})}\BibitemShut {NoStop}%
\bibitem [{\citenamefont {Abhijith}\ \emph {et~al.}(2022)\citenamefont
  {Abhijith}, \citenamefont {Adetokunbo}, \citenamefont {John} \emph
  {et~al.}}]{J_2022}%
  \BibitemOpen
  \bibfield  {author} {\bibinfo {author} {\bibfnamefont {J.}~\bibnamefont
  {Abhijith}}, \bibinfo {author} {\bibfnamefont {A.}~\bibnamefont
  {Adetokunbo}}, \bibinfo {author} {\bibfnamefont {A.}~\bibnamefont {John}},
  \emph {et~al.},\ }\href {\doibase 10.1145/3517340} {\bibfield  {journal}
  {\bibinfo  {journal} {{ACM} Transactions on Quantum Computing}\ }\textbf
  {\bibinfo {volume} {3}},\ \bibinfo {pages} {1} (\bibinfo {year}
  {2022})}\BibitemShut {NoStop}%
\bibitem [{\citenamefont {Cerezo}\ \emph {et~al.}(2021)\citenamefont {Cerezo},
  \citenamefont {Arrasmith}, \citenamefont {Babbush}, \citenamefont {Benjamin},
  \citenamefont {Endo}, \citenamefont {Fujii}, \citenamefont {McClean},
  \citenamefont {Mitarai}, \citenamefont {Yuan}, \citenamefont {Cincio},\ and\
  \citenamefont {Coles}}]{RN66}%
  \BibitemOpen
  \bibfield  {author} {\bibinfo {author} {\bibfnamefont {M.}~\bibnamefont
  {Cerezo}}, \bibinfo {author} {\bibfnamefont {A.}~\bibnamefont {Arrasmith}},
  \bibinfo {author} {\bibfnamefont {R.}~\bibnamefont {Babbush}}, \bibinfo
  {author} {\bibfnamefont {S.~C.}\ \bibnamefont {Benjamin}}, \bibinfo {author}
  {\bibfnamefont {S.}~\bibnamefont {Endo}}, \bibinfo {author} {\bibfnamefont
  {K.}~\bibnamefont {Fujii}}, \bibinfo {author} {\bibfnamefont {J.~R.}\
  \bibnamefont {McClean}}, \bibinfo {author} {\bibfnamefont {K.}~\bibnamefont
  {Mitarai}}, \bibinfo {author} {\bibfnamefont {X.}~\bibnamefont {Yuan}},
  \bibinfo {author} {\bibfnamefont {L.}~\bibnamefont {Cincio}}, \ and\ \bibinfo
  {author} {\bibfnamefont {P.~J.}\ \bibnamefont {Coles}},\ }\href {\doibase
  10.1038/s42254-021-00348-9} {\bibfield  {journal} {\bibinfo  {journal}
  {Nature Reviews Physics}\ }\textbf {\bibinfo {volume} {3}},\ \bibinfo {pages}
  {625} (\bibinfo {year} {2021})}\BibitemShut {NoStop}%
\bibitem [{\citenamefont {Biamonte}\ \emph {et~al.}(2017)\citenamefont
  {Biamonte}, \citenamefont {Wittek}, \citenamefont {Pancotti}, \citenamefont
  {Rebentrost}, \citenamefont {Wiebe},\ and\ \citenamefont
  {Lloyd}}]{biamonte2017quantum}%
  \BibitemOpen
  \bibfield  {author} {\bibinfo {author} {\bibfnamefont {J.}~\bibnamefont
  {Biamonte}}, \bibinfo {author} {\bibfnamefont {P.}~\bibnamefont {Wittek}},
  \bibinfo {author} {\bibfnamefont {N.}~\bibnamefont {Pancotti}}, \bibinfo
  {author} {\bibfnamefont {P.}~\bibnamefont {Rebentrost}}, \bibinfo {author}
  {\bibfnamefont {N.}~\bibnamefont {Wiebe}}, \ and\ \bibinfo {author}
  {\bibfnamefont {S.}~\bibnamefont {Lloyd}},\ }\href {\doibase
  10.1038/nature23474} {\bibfield  {journal} {\bibinfo  {journal} {Nature}\
  }\textbf {\bibinfo {volume} {549}},\ \bibinfo {pages} {195} (\bibinfo {year}
  {2017})}\BibitemShut {NoStop}%
\bibitem [{\citenamefont {Zeguendry}\ \emph {et~al.}(2023)\citenamefont
  {Zeguendry}, \citenamefont {Jarir},\ and\ \citenamefont
  {Quafafou}}]{e25020287}%
  \BibitemOpen
  \bibfield  {author} {\bibinfo {author} {\bibfnamefont {A.}~\bibnamefont
  {Zeguendry}}, \bibinfo {author} {\bibfnamefont {Z.}~\bibnamefont {Jarir}}, \
  and\ \bibinfo {author} {\bibfnamefont {M.}~\bibnamefont {Quafafou}},\ }\href
  {\doibase 10.3390/e25020287} {\bibfield  {journal} {\bibinfo  {journal}
  {Entropy}\ }\textbf {\bibinfo {volume} {25}},\ \bibinfo {pages} {287}
  (\bibinfo {year} {2023})}\BibitemShut {NoStop}%
\bibitem [{\citenamefont {Wan}\ \emph {et~al.}(2017)\citenamefont {Wan},
  \citenamefont {Dahlsten}, \citenamefont {Kristjánsson}, \citenamefont
  {Gardner},\ and\ \citenamefont {Kim}}]{s41534-017-0032-4}%
  \BibitemOpen
  \bibfield  {author} {\bibinfo {author} {\bibfnamefont {K.~H.}\ \bibnamefont
  {Wan}}, \bibinfo {author} {\bibfnamefont {O.}~\bibnamefont {Dahlsten}},
  \bibinfo {author} {\bibfnamefont {H.}~\bibnamefont {Kristjánsson}}, \bibinfo
  {author} {\bibfnamefont {R.}~\bibnamefont {Gardner}}, \ and\ \bibinfo
  {author} {\bibfnamefont {M.~S.}\ \bibnamefont {Kim}},\ }\href {\doibase
  10.1038/s41534-017-0032-4} {\bibfield  {journal} {\bibinfo  {journal} {npj
  Quantum Information}\ }\textbf {\bibinfo {volume} {3}},\ \bibinfo {pages}
  {36} (\bibinfo {year} {2017})}\BibitemShut {NoStop}%
\bibitem [{\citenamefont {Abbas}\ \emph {et~al.}(2021)\citenamefont {Abbas},
  \citenamefont {Sutter}, \citenamefont {Zoufal}, \citenamefont {Lucchi},
  \citenamefont {Figalli},\ and\ \citenamefont {Woerner}}]{s43588-021-00084-1}%
  \BibitemOpen
  \bibfield  {author} {\bibinfo {author} {\bibfnamefont {A.}~\bibnamefont
  {Abbas}}, \bibinfo {author} {\bibfnamefont {D.}~\bibnamefont {Sutter}},
  \bibinfo {author} {\bibfnamefont {C.}~\bibnamefont {Zoufal}}, \bibinfo
  {author} {\bibfnamefont {A.}~\bibnamefont {Lucchi}}, \bibinfo {author}
  {\bibfnamefont {A.}~\bibnamefont {Figalli}}, \ and\ \bibinfo {author}
  {\bibfnamefont {S.}~\bibnamefont {Woerner}},\ }\href {\doibase
  10.1038/s43588-021-00084-1} {\bibfield  {journal} {\bibinfo  {journal}
  {Nature Computational Science}\ }\textbf {\bibinfo {volume} {1}},\ \bibinfo
  {pages} {403} (\bibinfo {year} {2021})}\BibitemShut {NoStop}%
\bibitem [{\citenamefont {Pan}\ \emph {et~al.}(2023)\citenamefont {Pan},
  \citenamefont {Lu}, \citenamefont {Wang}, \citenamefont {Hua}, \citenamefont
  {Xu}, \citenamefont {Li}, \citenamefont {Cai}, \citenamefont {Li},
  \citenamefont {Wang}, \citenamefont {Song}, \citenamefont {Zou},
  \citenamefont {Deng},\ and\ \citenamefont {Sun}}]{s41467-023-39785-8}%
  \BibitemOpen
  \bibfield  {author} {\bibinfo {author} {\bibfnamefont {X.}~\bibnamefont
  {Pan}}, \bibinfo {author} {\bibfnamefont {Z.}~\bibnamefont {Lu}}, \bibinfo
  {author} {\bibfnamefont {W.}~\bibnamefont {Wang}}, \bibinfo {author}
  {\bibfnamefont {Z.}~\bibnamefont {Hua}}, \bibinfo {author} {\bibfnamefont
  {Y.}~\bibnamefont {Xu}}, \bibinfo {author} {\bibfnamefont {W.}~\bibnamefont
  {Li}}, \bibinfo {author} {\bibfnamefont {W.}~\bibnamefont {Cai}}, \bibinfo
  {author} {\bibfnamefont {X.}~\bibnamefont {Li}}, \bibinfo {author}
  {\bibfnamefont {H.}~\bibnamefont {Wang}}, \bibinfo {author} {\bibfnamefont
  {Y.-P.}\ \bibnamefont {Song}}, \bibinfo {author} {\bibfnamefont {C.-L.}\
  \bibnamefont {Zou}}, \bibinfo {author} {\bibfnamefont {D.-L.}\ \bibnamefont
  {Deng}}, \ and\ \bibinfo {author} {\bibfnamefont {L.}~\bibnamefont {Sun}},\
  }\href {\doibase 10.1038/s41467-023-39785-8} {\bibfield  {journal} {\bibinfo
  {journal} {Nature Communications}\ }\textbf {\bibinfo {volume} {14}},\
  \bibinfo {pages} {4006} (\bibinfo {year} {2023})}\BibitemShut {NoStop}%
\bibitem [{\citenamefont {Beer}\ \emph {et~al.}(2020)\citenamefont {Beer},
  \citenamefont {Bondarenko}, \citenamefont {Farrelly}, \citenamefont
  {Osborne}, \citenamefont {Salzmann}, \citenamefont {Scheiermann},\ and\
  \citenamefont {Wolf}}]{RN67}%
  \BibitemOpen
  \bibfield  {author} {\bibinfo {author} {\bibfnamefont {K.}~\bibnamefont
  {Beer}}, \bibinfo {author} {\bibfnamefont {D.}~\bibnamefont {Bondarenko}},
  \bibinfo {author} {\bibfnamefont {T.}~\bibnamefont {Farrelly}}, \bibinfo
  {author} {\bibfnamefont {T.~J.}\ \bibnamefont {Osborne}}, \bibinfo {author}
  {\bibfnamefont {R.}~\bibnamefont {Salzmann}}, \bibinfo {author}
  {\bibfnamefont {D.}~\bibnamefont {Scheiermann}}, \ and\ \bibinfo {author}
  {\bibfnamefont {R.}~\bibnamefont {Wolf}},\ }\href {\doibase
  10.1038/s41467-020-14454-2} {\bibfield  {journal} {\bibinfo  {journal}
  {Nature Communications}\ }\textbf {\bibinfo {volume} {11}},\ \bibinfo {pages}
  {808} (\bibinfo {year} {2020})}\BibitemShut {NoStop}%
\bibitem [{\citenamefont {Hubregtsen}\ \emph {et~al.}(2022)\citenamefont
  {Hubregtsen}, \citenamefont {Wierichs}, \citenamefont {Gil-Fuster},
  \citenamefont {Derks}, \citenamefont {Faehrmann},\ and\ \citenamefont
  {Meyer}}]{PhysRevA.106.042431}%
  \BibitemOpen
  \bibfield  {author} {\bibinfo {author} {\bibfnamefont {T.}~\bibnamefont
  {Hubregtsen}}, \bibinfo {author} {\bibfnamefont {D.}~\bibnamefont
  {Wierichs}}, \bibinfo {author} {\bibfnamefont {E.}~\bibnamefont
  {Gil-Fuster}}, \bibinfo {author} {\bibfnamefont {P.-J. H.~S.}\ \bibnamefont
  {Derks}}, \bibinfo {author} {\bibfnamefont {P.~K.}\ \bibnamefont
  {Faehrmann}}, \ and\ \bibinfo {author} {\bibfnamefont {J.~J.}\ \bibnamefont
  {Meyer}},\ }\href {\doibase 10.1103/PhysRevA.106.042431} {\bibfield
  {journal} {\bibinfo  {journal} {Phys. Rev. A}\ }\textbf {\bibinfo {volume}
  {106}},\ \bibinfo {pages} {042431} (\bibinfo {year} {2022})}\BibitemShut
  {NoStop}%
\bibitem [{\citenamefont {Rebentrost}\ \emph {et~al.}(2014)\citenamefont
  {Rebentrost}, \citenamefont {Mohseni},\ and\ \citenamefont
  {Lloyd}}]{PhysRevLett.113.130503}%
  \BibitemOpen
  \bibfield  {author} {\bibinfo {author} {\bibfnamefont {P.}~\bibnamefont
  {Rebentrost}}, \bibinfo {author} {\bibfnamefont {M.}~\bibnamefont {Mohseni}},
  \ and\ \bibinfo {author} {\bibfnamefont {S.}~\bibnamefont {Lloyd}},\ }\href
  {\doibase 10.1103/PhysRevLett.113.130503} {\bibfield  {journal} {\bibinfo
  {journal} {Phys. Rev. Lett.}\ }\textbf {\bibinfo {volume} {113}},\ \bibinfo
  {pages} {130503} (\bibinfo {year} {2014})}\BibitemShut {NoStop}%
\bibitem [{\citenamefont {Havl{\'\i}{\v{c}}ek}\ \emph
  {et~al.}(2019)\citenamefont {Havl{\'\i}{\v{c}}ek}, \citenamefont
  {C{\'o}rcoles}, \citenamefont {Temme}, \citenamefont {Harrow}, \citenamefont
  {Kandala}, \citenamefont {Chow},\ and\ \citenamefont
  {Gambetta}}]{havlivcek2019supervised}%
  \BibitemOpen
  \bibfield  {author} {\bibinfo {author} {\bibfnamefont {V.}~\bibnamefont
  {Havl{\'\i}{\v{c}}ek}}, \bibinfo {author} {\bibfnamefont {A.~D.}\
  \bibnamefont {C{\'o}rcoles}}, \bibinfo {author} {\bibfnamefont
  {K.}~\bibnamefont {Temme}}, \bibinfo {author} {\bibfnamefont {A.~W.}\
  \bibnamefont {Harrow}}, \bibinfo {author} {\bibfnamefont {A.}~\bibnamefont
  {Kandala}}, \bibinfo {author} {\bibfnamefont {J.~M.}\ \bibnamefont {Chow}}, \
  and\ \bibinfo {author} {\bibfnamefont {J.~M.}\ \bibnamefont {Gambetta}},\
  }\href {\doibase 10.1038/s41586-019-0980-2} {\bibfield  {journal} {\bibinfo
  {journal} {Nature}\ }\textbf {\bibinfo {volume} {567}},\ \bibinfo {pages}
  {209} (\bibinfo {year} {2019})}\BibitemShut {NoStop}%
\bibitem [{\citenamefont {Schuld}\ and\ \citenamefont
  {Killoran}(2019)}]{PhysRevLett.122.040504}%
  \BibitemOpen
  \bibfield  {author} {\bibinfo {author} {\bibfnamefont {M.}~\bibnamefont
  {Schuld}}\ and\ \bibinfo {author} {\bibfnamefont {N.}~\bibnamefont
  {Killoran}},\ }\href {\doibase 10.1103/PhysRevLett.122.040504} {\bibfield
  {journal} {\bibinfo  {journal} {Phys. Rev. Lett.}\ }\textbf {\bibinfo
  {volume} {122}},\ \bibinfo {pages} {040504} (\bibinfo {year}
  {2019})}\BibitemShut {NoStop}%
\bibitem [{\citenamefont {Liu}\ \emph {et~al.}(2021)\citenamefont {Liu},
  \citenamefont {Arunachalam},\ and\ \citenamefont {Temme}}]{RN80}%
  \BibitemOpen
  \bibfield  {author} {\bibinfo {author} {\bibfnamefont {Y.}~\bibnamefont
  {Liu}}, \bibinfo {author} {\bibfnamefont {S.}~\bibnamefont {Arunachalam}}, \
  and\ \bibinfo {author} {\bibfnamefont {K.}~\bibnamefont {Temme}},\ }\href
  {\doibase 10.1038/s41567-021-01287-z} {\bibfield  {journal} {\bibinfo
  {journal} {Nature Physics}\ }\textbf {\bibinfo {volume} {17}},\ \bibinfo
  {pages} {1013} (\bibinfo {year} {2021})}\BibitemShut {NoStop}%
\bibitem [{\citenamefont {Sent\'{\i}s}\ \emph {et~al.}(2019)\citenamefont
  {Sent\'{\i}s}, \citenamefont {Monr\`as}, \citenamefont {Mu\~noz Tapia},
  \citenamefont {Calsamiglia},\ and\ \citenamefont
  {Bagan}}]{PhysRevX.9.041029}%
  \BibitemOpen
  \bibfield  {author} {\bibinfo {author} {\bibfnamefont {G.}~\bibnamefont
  {Sent\'{\i}s}}, \bibinfo {author} {\bibfnamefont {A.}~\bibnamefont
  {Monr\`as}}, \bibinfo {author} {\bibfnamefont {R.}~\bibnamefont {Mu\~noz
  Tapia}}, \bibinfo {author} {\bibfnamefont {J.}~\bibnamefont {Calsamiglia}}, \
  and\ \bibinfo {author} {\bibfnamefont {E.}~\bibnamefont {Bagan}},\ }\href
  {\doibase 10.1103/PhysRevX.9.041029} {\bibfield  {journal} {\bibinfo
  {journal} {Phys. Rev. X}\ }\textbf {\bibinfo {volume} {9}},\ \bibinfo {pages}
  {041029} (\bibinfo {year} {2019})}\BibitemShut {NoStop}%
\bibitem [{\citenamefont {Tancara}\ \emph {et~al.}(2023)\citenamefont
  {Tancara}, \citenamefont {Dinani}, \citenamefont {Norambuena}, \citenamefont
  {Fanchini},\ and\ \citenamefont {Coto}}]{PhysRevA.107.022402}%
  \BibitemOpen
  \bibfield  {author} {\bibinfo {author} {\bibfnamefont {D.}~\bibnamefont
  {Tancara}}, \bibinfo {author} {\bibfnamefont {H.~T.}\ \bibnamefont {Dinani}},
  \bibinfo {author} {\bibfnamefont {A.}~\bibnamefont {Norambuena}}, \bibinfo
  {author} {\bibfnamefont {F.~F.}\ \bibnamefont {Fanchini}}, \ and\ \bibinfo
  {author} {\bibfnamefont {R.}~\bibnamefont {Coto}},\ }\href {\doibase
  10.1103/PhysRevA.107.022402} {\bibfield  {journal} {\bibinfo  {journal}
  {Phys. Rev. A}\ }\textbf {\bibinfo {volume} {107}},\ \bibinfo {pages}
  {022402} (\bibinfo {year} {2023})}\BibitemShut {NoStop}%
\bibitem [{\citenamefont {Hu}\ \emph {et~al.}(2019)\citenamefont {Hu},
  \citenamefont {Wu}, \citenamefont {Cai}, \citenamefont {Ma}, \citenamefont
  {Mu}, \citenamefont {Xu}, \citenamefont {Wang}, \citenamefont {Song},
  \citenamefont {Deng}, \citenamefont {Zou},\ and\ \citenamefont
  {Sun}}]{doi:10.1126/sciadv.aav2761}%
  \BibitemOpen
  \bibfield  {author} {\bibinfo {author} {\bibfnamefont {L.}~\bibnamefont
  {Hu}}, \bibinfo {author} {\bibfnamefont {S.-H.}\ \bibnamefont {Wu}}, \bibinfo
  {author} {\bibfnamefont {W.}~\bibnamefont {Cai}}, \bibinfo {author}
  {\bibfnamefont {Y.}~\bibnamefont {Ma}}, \bibinfo {author} {\bibfnamefont
  {X.}~\bibnamefont {Mu}}, \bibinfo {author} {\bibfnamefont {Y.}~\bibnamefont
  {Xu}}, \bibinfo {author} {\bibfnamefont {H.}~\bibnamefont {Wang}}, \bibinfo
  {author} {\bibfnamefont {Y.}~\bibnamefont {Song}}, \bibinfo {author}
  {\bibfnamefont {D.-L.}\ \bibnamefont {Deng}}, \bibinfo {author}
  {\bibfnamefont {C.-L.}\ \bibnamefont {Zou}}, \ and\ \bibinfo {author}
  {\bibfnamefont {L.}~\bibnamefont {Sun}},\ }\href {\doibase
  10.1126/sciadv.aav2761} {\bibfield  {journal} {\bibinfo  {journal} {Science
  Advances}\ }\textbf {\bibinfo {volume} {5}},\ \bibinfo {pages} {eaav2761}
  (\bibinfo {year} {2019})}\BibitemShut {NoStop}%
\bibitem [{\citenamefont {Carolan}\ \emph {et~al.}(2020)\citenamefont
  {Carolan}, \citenamefont {Mohseni}, \citenamefont {Olson}, \citenamefont
  {Prabhu}, \citenamefont {Chen}, \citenamefont {Bunandar}, \citenamefont
  {Niu}, \citenamefont {Harris}, \citenamefont {Wong}, \citenamefont
  {Hochberg}, \citenamefont {Lloyd},\ and\ \citenamefont {Englund}}]{RN81}%
  \BibitemOpen
  \bibfield  {author} {\bibinfo {author} {\bibfnamefont {J.}~\bibnamefont
  {Carolan}}, \bibinfo {author} {\bibfnamefont {M.}~\bibnamefont {Mohseni}},
  \bibinfo {author} {\bibfnamefont {J.~P.}\ \bibnamefont {Olson}}, \bibinfo
  {author} {\bibfnamefont {M.}~\bibnamefont {Prabhu}}, \bibinfo {author}
  {\bibfnamefont {C.}~\bibnamefont {Chen}}, \bibinfo {author} {\bibfnamefont
  {D.}~\bibnamefont {Bunandar}}, \bibinfo {author} {\bibfnamefont {M.~Y.}\
  \bibnamefont {Niu}}, \bibinfo {author} {\bibfnamefont {N.~C.}\ \bibnamefont
  {Harris}}, \bibinfo {author} {\bibfnamefont {F.~N.~C.}\ \bibnamefont {Wong}},
  \bibinfo {author} {\bibfnamefont {M.}~\bibnamefont {Hochberg}}, \bibinfo
  {author} {\bibfnamefont {S.}~\bibnamefont {Lloyd}}, \ and\ \bibinfo {author}
  {\bibfnamefont {D.}~\bibnamefont {Englund}},\ }\href {\doibase
  10.1038/s41567-019-0747-6} {\bibfield  {journal} {\bibinfo  {journal} {Nature
  Physics}\ }\textbf {\bibinfo {volume} {16}},\ \bibinfo {pages} {322}
  (\bibinfo {year} {2020})}\BibitemShut {NoStop}%
\bibitem [{\citenamefont {Liang}\ \emph {et~al.}(2020)\citenamefont {Liang},
  \citenamefont {Shen}, \citenamefont {Li},\ and\ \citenamefont
  {Li}}]{PhysRevA.101.032323}%
  \BibitemOpen
  \bibfield  {author} {\bibinfo {author} {\bibfnamefont {J.-M.}\ \bibnamefont
  {Liang}}, \bibinfo {author} {\bibfnamefont {S.-Q.}\ \bibnamefont {Shen}},
  \bibinfo {author} {\bibfnamefont {M.}~\bibnamefont {Li}}, \ and\ \bibinfo
  {author} {\bibfnamefont {L.}~\bibnamefont {Li}},\ }\href {\doibase
  10.1103/PhysRevA.101.032323} {\bibfield  {journal} {\bibinfo  {journal}
  {Phys. Rev. A}\ }\textbf {\bibinfo {volume} {101}},\ \bibinfo {pages}
  {032323} (\bibinfo {year} {2020})}\BibitemShut {NoStop}%
\bibitem [{\citenamefont {Buhrman}\ \emph {et~al.}(2001)\citenamefont
  {Buhrman}, \citenamefont {Cleve}, \citenamefont {Watrous},\ and\
  \citenamefont {de~Wolf}}]{PhysRevLett.87.167902}%
  \BibitemOpen
  \bibfield  {author} {\bibinfo {author} {\bibfnamefont {H.}~\bibnamefont
  {Buhrman}}, \bibinfo {author} {\bibfnamefont {R.}~\bibnamefont {Cleve}},
  \bibinfo {author} {\bibfnamefont {J.}~\bibnamefont {Watrous}}, \ and\
  \bibinfo {author} {\bibfnamefont {R.}~\bibnamefont {de~Wolf}},\ }\href
  {\doibase 10.1103/PhysRevLett.87.167902} {\bibfield  {journal} {\bibinfo
  {journal} {Phys. Rev. Lett.}\ }\textbf {\bibinfo {volume} {87}},\ \bibinfo
  {pages} {167902} (\bibinfo {year} {2001})}\BibitemShut {NoStop}%
\bibitem [{\citenamefont {Ripper}\ \emph {et~al.}(2023)\citenamefont {Ripper},
  \citenamefont {Amaral},\ and\ \citenamefont {Temporão}}]{RN70}%
  \BibitemOpen
  \bibfield  {author} {\bibinfo {author} {\bibfnamefont {P.}~\bibnamefont
  {Ripper}}, \bibinfo {author} {\bibfnamefont {G.}~\bibnamefont {Amaral}}, \
  and\ \bibinfo {author} {\bibfnamefont {G.}~\bibnamefont {Temporão}},\ }\href
  {\doibase 10.1007/s11128-023-03961-y} {\bibfield  {journal} {\bibinfo
  {journal} {Quantum Information Processing}\ }\textbf {\bibinfo {volume}
  {22}},\ \bibinfo {pages} {220} (\bibinfo {year} {2023})}\BibitemShut
  {NoStop}%
\bibitem [{\citenamefont {Agliardi}\ \emph {et~al.}(2025)\citenamefont
  {Agliardi}, \citenamefont {O’Meara}, \citenamefont {Yogaraj}, \citenamefont
  {Ghosh}, \citenamefont {Sabino}, \citenamefont {Fernández-Campoamor},
  \citenamefont {Cortiana}, \citenamefont {Bernabé-Moreno}, \citenamefont
  {Tacchino}, \citenamefont {Mezzacapo},\ and\ \citenamefont
  {Shehab}}]{agliardi2023}%
  \BibitemOpen
  \bibfield  {author} {\bibinfo {author} {\bibfnamefont {G.}~\bibnamefont
  {Agliardi}}, \bibinfo {author} {\bibfnamefont {C.}~\bibnamefont {O’Meara}},
  \bibinfo {author} {\bibfnamefont {K.}~\bibnamefont {Yogaraj}}, \bibinfo
  {author} {\bibfnamefont {K.}~\bibnamefont {Ghosh}}, \bibinfo {author}
  {\bibfnamefont {P.}~\bibnamefont {Sabino}}, \bibinfo {author} {\bibfnamefont
  {M.}~\bibnamefont {Fernández-Campoamor}}, \bibinfo {author} {\bibfnamefont
  {G.}~\bibnamefont {Cortiana}}, \bibinfo {author} {\bibfnamefont
  {J.}~\bibnamefont {Bernabé-Moreno}}, \bibinfo {author} {\bibfnamefont
  {F.}~\bibnamefont {Tacchino}}, \bibinfo {author} {\bibfnamefont
  {A.}~\bibnamefont {Mezzacapo}}, \ and\ \bibinfo {author} {\bibfnamefont
  {O.}~\bibnamefont {Shehab}},\ }\href {\doibase 10.1088/2058-9565/ada08c}
  {\bibfield  {journal} {\bibinfo  {journal} {Quantum Science and Technology}\
  }\textbf {\bibinfo {volume} {10}},\ \bibinfo {pages} {025005} (\bibinfo
  {year} {2025})}\BibitemShut {NoStop}%
\bibitem [{\citenamefont {White}\ \emph {et~al.}(2024)\citenamefont {White},
  \citenamefont {Polino}, \citenamefont {Ghafari}, \citenamefont {Joch},
  \citenamefont {Villegas-Aguilar}, \citenamefont {Shalm}, \citenamefont
  {Verma}, \citenamefont {Huber},\ and\ \citenamefont
  {Tischler}}]{white2024robust}%
  \BibitemOpen
  \bibfield  {author} {\bibinfo {author} {\bibfnamefont {S.~J.~U.}\
  \bibnamefont {White}}, \bibinfo {author} {\bibfnamefont {E.}~\bibnamefont
  {Polino}}, \bibinfo {author} {\bibfnamefont {F.}~\bibnamefont {Ghafari}},
  \bibinfo {author} {\bibfnamefont {D.~J.}\ \bibnamefont {Joch}}, \bibinfo
  {author} {\bibfnamefont {L.}~\bibnamefont {Villegas-Aguilar}}, \bibinfo
  {author} {\bibfnamefont {L.~K.}\ \bibnamefont {Shalm}}, \bibinfo {author}
  {\bibfnamefont {V.~B.}\ \bibnamefont {Verma}}, \bibinfo {author}
  {\bibfnamefont {M.}~\bibnamefont {Huber}}, \ and\ \bibinfo {author}
  {\bibfnamefont {N.}~\bibnamefont {Tischler}},\ }\href {\doibase
  10.48550/arXiv.2404.16106} {\bibfield  {journal} {\bibinfo  {journal}
  {arXiv:2404.16106}\ } (\bibinfo {year} {2024}),\
  10.48550/arXiv.2404.16106}\BibitemShut {NoStop}%
\bibitem [{\citenamefont {Nguyen}\ \emph {et~al.}(2021)\citenamefont {Nguyen},
  \citenamefont {Tseng}, \citenamefont {Maslennikov}, \citenamefont {Gan},\
  and\ \citenamefont {Matsukevich}}]{nguyen2021experimental}%
  \BibitemOpen
  \bibfield  {author} {\bibinfo {author} {\bibfnamefont {C.-H.}\ \bibnamefont
  {Nguyen}}, \bibinfo {author} {\bibfnamefont {K.-W.}\ \bibnamefont {Tseng}},
  \bibinfo {author} {\bibfnamefont {G.}~\bibnamefont {Maslennikov}}, \bibinfo
  {author} {\bibfnamefont {H.~C.~J.}\ \bibnamefont {Gan}}, \ and\ \bibinfo
  {author} {\bibfnamefont {D.}~\bibnamefont {Matsukevich}},\ }\href {\doibase
  10.48550/arXiv.2103.10219} {\bibfield  {journal} {\bibinfo  {journal}
  {arXiv:2103.10219}\ } (\bibinfo {year} {2021}),\
  10.48550/arXiv.2103.10219}\BibitemShut {NoStop}%
\bibitem [{\citenamefont {Li}\ \emph {et~al.}(2022)\citenamefont {Li},
  \citenamefont {Barraza}, \citenamefont {Alvarado~Barrios}, \citenamefont
  {Solano},\ and\ \citenamefont
  {Albarr\'an-Arriagada}}]{PhysRevApplied.18.014047}%
  \BibitemOpen
  \bibfield  {author} {\bibinfo {author} {\bibfnamefont {Y.-D.}\ \bibnamefont
  {Li}}, \bibinfo {author} {\bibfnamefont {N.}~\bibnamefont {Barraza}},
  \bibinfo {author} {\bibfnamefont {G.}~\bibnamefont {Alvarado~Barrios}},
  \bibinfo {author} {\bibfnamefont {E.}~\bibnamefont {Solano}}, \ and\ \bibinfo
  {author} {\bibfnamefont {F.}~\bibnamefont {Albarr\'an-Arriagada}},\ }\href
  {\doibase 10.1103/PhysRevApplied.18.014047} {\bibfield  {journal} {\bibinfo
  {journal} {Phys. Rev. Appl.}\ }\textbf {\bibinfo {volume} {18}},\ \bibinfo
  {pages} {014047} (\bibinfo {year} {2022})}\BibitemShut {NoStop}%
\bibitem [{\citenamefont {Hong}\ \emph {et~al.}(1987)\citenamefont {Hong},
  \citenamefont {Ou},\ and\ \citenamefont {Mandel}}]{PhysRevLett.59.2044}%
  \BibitemOpen
  \bibfield  {author} {\bibinfo {author} {\bibfnamefont {C.~K.}\ \bibnamefont
  {Hong}}, \bibinfo {author} {\bibfnamefont {Z.~Y.}\ \bibnamefont {Ou}}, \ and\
  \bibinfo {author} {\bibfnamefont {L.}~\bibnamefont {Mandel}},\ }\href
  {\doibase 10.1103/PhysRevLett.59.2044} {\bibfield  {journal} {\bibinfo
  {journal} {Phys. Rev. Lett.}\ }\textbf {\bibinfo {volume} {59}},\ \bibinfo
  {pages} {2044} (\bibinfo {year} {1987})}\BibitemShut {NoStop}%
\bibitem [{\citenamefont {Garcia-Escartin}\ and\ \citenamefont
  {Chamorro-Posada}(2013)}]{PhysRevA.87.052330}%
  \BibitemOpen
  \bibfield  {author} {\bibinfo {author} {\bibfnamefont {J.~C.}\ \bibnamefont
  {Garcia-Escartin}}\ and\ \bibinfo {author} {\bibfnamefont {P.}~\bibnamefont
  {Chamorro-Posada}},\ }\href {\doibase 10.1103/PhysRevA.87.052330} {\bibfield
  {journal} {\bibinfo  {journal} {Phys. Rev. A}\ }\textbf {\bibinfo {volume}
  {87}},\ \bibinfo {pages} {052330} (\bibinfo {year} {2013})}\BibitemShut
  {NoStop}%
\bibitem [{\citenamefont {Zhang}\ \emph {et~al.}(2021)\citenamefont {Zhang},
  \citenamefont {Zhan}, \citenamefont {Liao}, \citenamefont {Zheng},
  \citenamefont {Jiang}, \citenamefont {Mi}, \citenamefont {Yao},\ and\
  \citenamefont {Zhang}}]{s41377-021-00608-4}%
  \BibitemOpen
  \bibfield  {author} {\bibinfo {author} {\bibfnamefont {A.}~\bibnamefont
  {Zhang}}, \bibinfo {author} {\bibfnamefont {H.}~\bibnamefont {Zhan}},
  \bibinfo {author} {\bibfnamefont {J.}~\bibnamefont {Liao}}, \bibinfo {author}
  {\bibfnamefont {K.}~\bibnamefont {Zheng}}, \bibinfo {author} {\bibfnamefont
  {T.}~\bibnamefont {Jiang}}, \bibinfo {author} {\bibfnamefont
  {M.}~\bibnamefont {Mi}}, \bibinfo {author} {\bibfnamefont {P.}~\bibnamefont
  {Yao}}, \ and\ \bibinfo {author} {\bibfnamefont {L.}~\bibnamefont {Zhang}},\
  }\href {\doibase 10.1038/s41377-021-00608-4} {\bibfield  {journal} {\bibinfo
  {journal} {Light: Sci. Appl.}\ }\textbf {\bibinfo {volume} {10}},\ \bibinfo
  {pages} {169} (\bibinfo {year} {2021})}\BibitemShut {NoStop}%
\bibitem [{\citenamefont {Cincio}\ \emph {et~al.}(2018)\citenamefont {Cincio},
  \citenamefont {Subaşı}, \citenamefont {Sornborger},\ and\ \citenamefont
  {Coles}}]{Cincio_2018}%
  \BibitemOpen
  \bibfield  {author} {\bibinfo {author} {\bibfnamefont {L.}~\bibnamefont
  {Cincio}}, \bibinfo {author} {\bibfnamefont {Y.}~\bibnamefont {Subaşı}},
  \bibinfo {author} {\bibfnamefont {A.~T.}\ \bibnamefont {Sornborger}}, \ and\
  \bibinfo {author} {\bibfnamefont {P.~J.}\ \bibnamefont {Coles}},\ }\href
  {\doibase 10.1088/1367-2630/aae94a} {\bibfield  {journal} {\bibinfo
  {journal} {New Journal of Physics}\ }\textbf {\bibinfo {volume} {20}},\
  \bibinfo {pages} {113022} (\bibinfo {year} {2018})}\BibitemShut {NoStop}%
\bibitem [{\citenamefont {Huggins}\ \emph {et~al.}(2020)\citenamefont
  {Huggins}, \citenamefont {Lee}, \citenamefont {Baek}, \citenamefont
  {O’Gorman},\ and\ \citenamefont {Whaley}}]{Huggins_2020}%
  \BibitemOpen
  \bibfield  {author} {\bibinfo {author} {\bibfnamefont {W.~J.}\ \bibnamefont
  {Huggins}}, \bibinfo {author} {\bibfnamefont {J.}~\bibnamefont {Lee}},
  \bibinfo {author} {\bibfnamefont {U.}~\bibnamefont {Baek}}, \bibinfo {author}
  {\bibfnamefont {B.}~\bibnamefont {O’Gorman}}, \ and\ \bibinfo {author}
  {\bibfnamefont {K.~B.}\ \bibnamefont {Whaley}},\ }\href {\doibase
  10.1088/1367-2630/ab867b} {\bibfield  {journal} {\bibinfo  {journal} {New
  Journal of Physics}\ }\textbf {\bibinfo {volume} {22}},\ \bibinfo {pages}
  {073009} (\bibinfo {year} {2020})}\BibitemShut {NoStop}%
\bibitem [{\citenamefont {Fanizza}\ \emph {et~al.}(2020)\citenamefont
  {Fanizza}, \citenamefont {Rosati}, \citenamefont {Skotiniotis}, \citenamefont
  {Calsamiglia},\ and\ \citenamefont {Giovannetti}}]{PhysRevLett.124.060503}%
  \BibitemOpen
  \bibfield  {author} {\bibinfo {author} {\bibfnamefont {M.}~\bibnamefont
  {Fanizza}}, \bibinfo {author} {\bibfnamefont {M.}~\bibnamefont {Rosati}},
  \bibinfo {author} {\bibfnamefont {M.}~\bibnamefont {Skotiniotis}}, \bibinfo
  {author} {\bibfnamefont {J.}~\bibnamefont {Calsamiglia}}, \ and\ \bibinfo
  {author} {\bibfnamefont {V.}~\bibnamefont {Giovannetti}},\ }\href {\doibase
  10.1103/PhysRevLett.124.060503} {\bibfield  {journal} {\bibinfo  {journal}
  {Phys. Rev. Lett.}\ }\textbf {\bibinfo {volume} {124}},\ \bibinfo {pages}
  {060503} (\bibinfo {year} {2020})}\BibitemShut {NoStop}%
\bibitem [{\citenamefont {Steinbrecher}\ \emph {et~al.}(2019)\citenamefont
  {Steinbrecher}, \citenamefont {Olson}, \citenamefont {Englund},\ and\
  \citenamefont {Carolan}}]{s41534-019-0174-7}%
  \BibitemOpen
  \bibfield  {author} {\bibinfo {author} {\bibfnamefont {G.~R.}\ \bibnamefont
  {Steinbrecher}}, \bibinfo {author} {\bibfnamefont {J.~P.}\ \bibnamefont
  {Olson}}, \bibinfo {author} {\bibfnamefont {D.}~\bibnamefont {Englund}}, \
  and\ \bibinfo {author} {\bibfnamefont {J.}~\bibnamefont {Carolan}},\ }\href
  {\doibase 10.1038/s41534-019-0174-7} {\bibfield  {journal} {\bibinfo
  {journal} {npj Quantum Information}\ }\textbf {\bibinfo {volume} {5}},\
  \bibinfo {pages} {60} (\bibinfo {year} {2019})}\BibitemShut {NoStop}%
\bibitem [{\citenamefont {Ewaniuk}\ \emph {et~al.}(2023)\citenamefont
  {Ewaniuk}, \citenamefont {Carolan}, \citenamefont {Shastri},\ and\
  \citenamefont {Rotenberg}}]{qute.202200125}%
  \BibitemOpen
  \bibfield  {author} {\bibinfo {author} {\bibfnamefont {J.}~\bibnamefont
  {Ewaniuk}}, \bibinfo {author} {\bibfnamefont {J.}~\bibnamefont {Carolan}},
  \bibinfo {author} {\bibfnamefont {B.~J.}\ \bibnamefont {Shastri}}, \ and\
  \bibinfo {author} {\bibfnamefont {N.}~\bibnamefont {Rotenberg}},\ }\href
  {\doibase https://doi.org/10.1002/qute.202200125} {\bibfield  {journal}
  {\bibinfo  {journal} {Advanced Quantum Technologies}\ }\textbf {\bibinfo
  {volume} {6}},\ \bibinfo {pages} {2200125} (\bibinfo {year}
  {2023})}\BibitemShut {NoStop}%
\bibitem [{\citenamefont {Wetzstein}\ \emph {et~al.}(2020)\citenamefont
  {Wetzstein}, \citenamefont {Ozcan}, \citenamefont {Gigan}, \citenamefont
  {Fan}, \citenamefont {Englund}, \citenamefont {Soljačić}, \citenamefont
  {Denz}, \citenamefont {Miller},\ and\ \citenamefont
  {Psaltis}}]{s41586-020-2973-6}%
  \BibitemOpen
  \bibfield  {author} {\bibinfo {author} {\bibfnamefont {G.}~\bibnamefont
  {Wetzstein}}, \bibinfo {author} {\bibfnamefont {A.}~\bibnamefont {Ozcan}},
  \bibinfo {author} {\bibfnamefont {S.}~\bibnamefont {Gigan}}, \bibinfo
  {author} {\bibfnamefont {S.}~\bibnamefont {Fan}}, \bibinfo {author}
  {\bibfnamefont {D.}~\bibnamefont {Englund}}, \bibinfo {author} {\bibfnamefont
  {M.}~\bibnamefont {Soljačić}}, \bibinfo {author} {\bibfnamefont
  {C.}~\bibnamefont {Denz}}, \bibinfo {author} {\bibfnamefont {D.~A.~B.}\
  \bibnamefont {Miller}}, \ and\ \bibinfo {author} {\bibfnamefont
  {D.}~\bibnamefont {Psaltis}},\ }\href {\doibase 10.1038/s41586-020-2973-6}
  {\bibfield  {journal} {\bibinfo  {journal} {Nature}\ }\textbf {\bibinfo
  {volume} {588}},\ \bibinfo {pages} {39} (\bibinfo {year} {2020})}\BibitemShut
  {NoStop}%
\bibitem [{\citenamefont {Zuo}\ \emph {et~al.}(2022)\citenamefont {Zuo},
  \citenamefont {Cao}, \citenamefont {Cao}, \citenamefont {Lai}, \citenamefont
  {Zeng},\ and\ \citenamefont {Du}}]{10.1117/1.AP.4.2.026004}%
  \BibitemOpen
  \bibfield  {author} {\bibinfo {author} {\bibfnamefont {Y.}~\bibnamefont
  {Zuo}}, \bibinfo {author} {\bibfnamefont {C.}~\bibnamefont {Cao}}, \bibinfo
  {author} {\bibfnamefont {N.}~\bibnamefont {Cao}}, \bibinfo {author}
  {\bibfnamefont {X.}~\bibnamefont {Lai}}, \bibinfo {author} {\bibfnamefont
  {B.}~\bibnamefont {Zeng}}, \ and\ \bibinfo {author} {\bibfnamefont
  {S.}~\bibnamefont {Du}},\ }\href {\doibase 10.1117/1.AP.4.2.026004}
  {\bibfield  {journal} {\bibinfo  {journal} {Advanced Photonics}\ }\textbf
  {\bibinfo {volume} {4}},\ \bibinfo {pages} {026004} (\bibinfo {year}
  {2022})}\BibitemShut {NoStop}%
\bibitem [{\citenamefont {Bernstein}\ \emph {et~al.}(2023)\citenamefont
  {Bernstein}, \citenamefont {Sludds}, \citenamefont {Panuski}, \citenamefont
  {Trajtenberg-Mills}, \citenamefont {Hamerly},\ and\ \citenamefont
  {Englund}}]{doi:10.1126/sciadv.adg7904}%
  \BibitemOpen
  \bibfield  {author} {\bibinfo {author} {\bibfnamefont {L.}~\bibnamefont
  {Bernstein}}, \bibinfo {author} {\bibfnamefont {A.}~\bibnamefont {Sludds}},
  \bibinfo {author} {\bibfnamefont {C.}~\bibnamefont {Panuski}}, \bibinfo
  {author} {\bibfnamefont {S.}~\bibnamefont {Trajtenberg-Mills}}, \bibinfo
  {author} {\bibfnamefont {R.}~\bibnamefont {Hamerly}}, \ and\ \bibinfo
  {author} {\bibfnamefont {D.}~\bibnamefont {Englund}},\ }\href {\doibase
  10.1126/sciadv.adg7904} {\bibfield  {journal} {\bibinfo  {journal} {Science
  Advances}\ }\textbf {\bibinfo {volume} {9}},\ \bibinfo {pages} {eadg7904}
  (\bibinfo {year} {2023})}\BibitemShut {NoStop}%
\bibitem [{\citenamefont {Chabaud}\ \emph {et~al.}(2021)\citenamefont
  {Chabaud}, \citenamefont {Markham},\ and\ \citenamefont
  {Sohbi}}]{Chabaud2021quantummachine}%
  \BibitemOpen
  \bibfield  {author} {\bibinfo {author} {\bibfnamefont {U.}~\bibnamefont
  {Chabaud}}, \bibinfo {author} {\bibfnamefont {D.}~\bibnamefont {Markham}}, \
  and\ \bibinfo {author} {\bibfnamefont {A.}~\bibnamefont {Sohbi}},\ }\href
  {\doibase 10.22331/q-2021-07-05-496} {\bibfield  {journal} {\bibinfo
  {journal} {{Quantum}}\ }\textbf {\bibinfo {volume} {5}},\ \bibinfo {pages}
  {496} (\bibinfo {year} {2021})}\BibitemShut {NoStop}%
\bibitem [{\citenamefont {Bacon}\ \emph {et~al.}(2006)\citenamefont {Bacon},
  \citenamefont {Chuang},\ and\ \citenamefont
  {Harrow}}]{PhysRevLett.97.170502}%
  \BibitemOpen
  \bibfield  {author} {\bibinfo {author} {\bibfnamefont {D.}~\bibnamefont
  {Bacon}}, \bibinfo {author} {\bibfnamefont {I.~L.}\ \bibnamefont {Chuang}}, \
  and\ \bibinfo {author} {\bibfnamefont {A.~W.}\ \bibnamefont {Harrow}},\
  }\href {\doibase 10.1103/PhysRevLett.97.170502} {\bibfield  {journal}
  {\bibinfo  {journal} {Phys. Rev. Lett.}\ }\textbf {\bibinfo {volume} {97}},\
  \bibinfo {pages} {170502} (\bibinfo {year} {2006})}\BibitemShut {NoStop}%
\bibitem [{\citenamefont {Durt}\ \emph {et~al.}(2010)\citenamefont {Durt},
  \citenamefont {Englert}, \citenamefont {Bengtsson},\ and\ \citenamefont
  {{\.Z}yczkowski}}]{doi:10.1142/S0219749910006502}%
  \BibitemOpen
  \bibfield  {author} {\bibinfo {author} {\bibfnamefont {T.}~\bibnamefont
  {Durt}}, \bibinfo {author} {\bibfnamefont {B.-G.}\ \bibnamefont {Englert}},
  \bibinfo {author} {\bibfnamefont {I.}~\bibnamefont {Bengtsson}}, \ and\
  \bibinfo {author} {\bibfnamefont {K.}~\bibnamefont {{\.Z}yczkowski}},\ }\href
  {\doibase 10.1142/S0219749910006502} {\bibfield  {journal} {\bibinfo
  {journal} {International Journal of Quantum Information}\ }\textbf {\bibinfo
  {volume} {08}},\ \bibinfo {pages} {535} (\bibinfo {year} {2010})}\BibitemShut
  {NoStop}%
\bibitem [{\citenamefont {Hou}\ \emph {et~al.}(2018)\citenamefont {Hou},
  \citenamefont {Tang}, \citenamefont {Shang}, \citenamefont {Zhu},
  \citenamefont {Li}, \citenamefont {Yuan}, \citenamefont {Wu}, \citenamefont
  {Xiang}, \citenamefont {Li},\ and\ \citenamefont {Guo}}]{RN71}%
  \BibitemOpen
  \bibfield  {author} {\bibinfo {author} {\bibfnamefont {Z.}~\bibnamefont
  {Hou}}, \bibinfo {author} {\bibfnamefont {J.-F.}\ \bibnamefont {Tang}},
  \bibinfo {author} {\bibfnamefont {J.}~\bibnamefont {Shang}}, \bibinfo
  {author} {\bibfnamefont {H.}~\bibnamefont {Zhu}}, \bibinfo {author}
  {\bibfnamefont {J.}~\bibnamefont {Li}}, \bibinfo {author} {\bibfnamefont
  {Y.}~\bibnamefont {Yuan}}, \bibinfo {author} {\bibfnamefont {K.-D.}\
  \bibnamefont {Wu}}, \bibinfo {author} {\bibfnamefont {G.-Y.}\ \bibnamefont
  {Xiang}}, \bibinfo {author} {\bibfnamefont {C.-F.}\ \bibnamefont {Li}}, \
  and\ \bibinfo {author} {\bibfnamefont {G.-C.}\ \bibnamefont {Guo}},\ }\href
  {\doibase 10.1038/s41467-018-03849-x} {\bibfield  {journal} {\bibinfo
  {journal} {Nature Communications}\ }\textbf {\bibinfo {volume} {9}},\
  \bibinfo {pages} {1414} (\bibinfo {year} {2018})}\BibitemShut {NoStop}%
\bibitem [{\citenamefont {Grice}\ and\ \citenamefont
  {Walmsley}(1997)}]{PhysRevA.56.1627}%
  \BibitemOpen
  \bibfield  {author} {\bibinfo {author} {\bibfnamefont {W.~P.}\ \bibnamefont
  {Grice}}\ and\ \bibinfo {author} {\bibfnamefont {I.~A.}\ \bibnamefont
  {Walmsley}},\ }\href {\doibase 10.1103/PhysRevA.56.1627} {\bibfield
  {journal} {\bibinfo  {journal} {Phys. Rev. A}\ }\textbf {\bibinfo {volume}
  {56}},\ \bibinfo {pages} {1627} (\bibinfo {year} {1997})}\BibitemShut
  {NoStop}%
\bibitem [{\citenamefont {Cramér}(1946)}]{cramer1999mathematical}%
  \BibitemOpen
  \bibfield  {author} {\bibinfo {author} {\bibfnamefont {H.}~\bibnamefont
  {Cramér}},\ }\href {\doibase doi:10.1515/9781400883868} {\emph {\bibinfo
  {title} {Mathematical Methods of Statistics (PMS-9)}}}\ (\bibinfo
  {publisher} {Princeton University Press},\ \bibinfo {address} {Princeton},\
  \bibinfo {year} {1946})\BibitemShut {NoStop}%
\bibitem [{\citenamefont {Braunstein}\ and\ \citenamefont
  {Caves}(1994)}]{PhysRevLett.72.3439}%
  \BibitemOpen
  \bibfield  {author} {\bibinfo {author} {\bibfnamefont {S.~L.}\ \bibnamefont
  {Braunstein}}\ and\ \bibinfo {author} {\bibfnamefont {C.~M.}\ \bibnamefont
  {Caves}},\ }\href {\doibase 10.1103/PhysRevLett.72.3439} {\bibfield
  {journal} {\bibinfo  {journal} {Phys. Rev. Lett.}\ }\textbf {\bibinfo
  {volume} {72}},\ \bibinfo {pages} {3439} (\bibinfo {year}
  {1994})}\BibitemShut {NoStop}%
\bibitem [{\citenamefont {Hayashi}\ \emph {et~al.}(2005)\citenamefont
  {Hayashi}, \citenamefont {Hashimoto},\ and\ \citenamefont
  {Horibe}}]{PhysRevA.72.032325}%
  \BibitemOpen
  \bibfield  {author} {\bibinfo {author} {\bibfnamefont {A.}~\bibnamefont
  {Hayashi}}, \bibinfo {author} {\bibfnamefont {T.}~\bibnamefont {Hashimoto}},
  \ and\ \bibinfo {author} {\bibfnamefont {M.}~\bibnamefont {Horibe}},\ }\href
  {\doibase 10.1103/PhysRevA.72.032325} {\bibfield  {journal} {\bibinfo
  {journal} {Phys. Rev. A}\ }\textbf {\bibinfo {volume} {72}},\ \bibinfo
  {pages} {032325} (\bibinfo {year} {2005})}\BibitemShut {NoStop}%
\bibitem [{\citenamefont {Kueng}\ \emph {et~al.}(2017)\citenamefont {Kueng},
  \citenamefont {Rauhut},\ and\ \citenamefont {Terstiege}}]{KUENG201788}%
  \BibitemOpen
  \bibfield  {author} {\bibinfo {author} {\bibfnamefont {R.}~\bibnamefont
  {Kueng}}, \bibinfo {author} {\bibfnamefont {H.}~\bibnamefont {Rauhut}}, \
  and\ \bibinfo {author} {\bibfnamefont {U.}~\bibnamefont {Terstiege}},\ }\href
  {\doibase https://doi.org/10.1016/j.acha.2015.07.007} {\bibfield  {journal}
  {\bibinfo  {journal} {Applied and Computational Harmonic Analysis}\ }\textbf
  {\bibinfo {volume} {42}},\ \bibinfo {pages} {88} (\bibinfo {year}
  {2017})}\BibitemShut {NoStop}%
\bibitem [{\citenamefont {Haah}\ \emph {et~al.}(2016)\citenamefont {Haah},
  \citenamefont {Harrow}, \citenamefont {Ji}, \citenamefont {Wu},\ and\
  \citenamefont {Yu}}]{10.1145/2897518.2897585}%
  \BibitemOpen
  \bibfield  {author} {\bibinfo {author} {\bibfnamefont {J.}~\bibnamefont
  {Haah}}, \bibinfo {author} {\bibfnamefont {A.~W.}\ \bibnamefont {Harrow}},
  \bibinfo {author} {\bibfnamefont {Z.}~\bibnamefont {Ji}}, \bibinfo {author}
  {\bibfnamefont {X.}~\bibnamefont {Wu}}, \ and\ \bibinfo {author}
  {\bibfnamefont {N.}~\bibnamefont {Yu}},\ }in\ \href {\doibase
  10.1145/2897518.2897585} {\emph {\bibinfo {booktitle} {Proceedings of the
  Forty-Eighth Annual ACM Symposium on Theory of Computing}}},\ \bibinfo
  {series and number} {STOC '16}\ (\bibinfo  {publisher} {Association for
  Computing Machinery},\ \bibinfo {address} {New York, NY, USA},\ \bibinfo
  {year} {2016})\ p.\ \bibinfo {pages} {913–925}\BibitemShut {NoStop}%
\bibitem [{\citenamefont {Flammia}\ \emph {et~al.}(2012)\citenamefont
  {Flammia}, \citenamefont {Gross}, \citenamefont {Liu},\ and\ \citenamefont
  {Eisert}}]{Flammia_2012}%
  \BibitemOpen
  \bibfield  {author} {\bibinfo {author} {\bibfnamefont {S.~T.}\ \bibnamefont
  {Flammia}}, \bibinfo {author} {\bibfnamefont {D.}~\bibnamefont {Gross}},
  \bibinfo {author} {\bibfnamefont {Y.-K.}\ \bibnamefont {Liu}}, \ and\
  \bibinfo {author} {\bibfnamefont {J.}~\bibnamefont {Eisert}},\ }\href
  {\doibase 10.1088/1367-2630/14/9/095022} {\bibfield  {journal} {\bibinfo
  {journal} {New Journal of Physics}\ }\textbf {\bibinfo {volume} {14}},\
  \bibinfo {pages} {095022} (\bibinfo {year} {2012})}\BibitemShut {NoStop}%
\bibitem [{\citenamefont {Bagan}\ \emph {et~al.}(2002)\citenamefont {Bagan},
  \citenamefont {Baig},\ and\ \citenamefont {Mu\~noz
  Tapia}}]{PhysRevLett.89.277904}%
  \BibitemOpen
  \bibfield  {author} {\bibinfo {author} {\bibfnamefont {E.}~\bibnamefont
  {Bagan}}, \bibinfo {author} {\bibfnamefont {M.}~\bibnamefont {Baig}}, \ and\
  \bibinfo {author} {\bibfnamefont {R.}~\bibnamefont {Mu\~noz Tapia}},\ }\href
  {\doibase 10.1103/PhysRevLett.89.277904} {\bibfield  {journal} {\bibinfo
  {journal} {Phys. Rev. Lett.}\ }\textbf {\bibinfo {volume} {89}},\ \bibinfo
  {pages} {277904} (\bibinfo {year} {2002})}\BibitemShut {NoStop}%
\bibitem [{\citenamefont {Mahler}\ \emph {et~al.}(2013)\citenamefont {Mahler},
  \citenamefont {Rozema}, \citenamefont {Darabi}, \citenamefont {Ferrie},
  \citenamefont {Blume-Kohout},\ and\ \citenamefont
  {Steinberg}}]{PhysRevLett.111.183601}%
  \BibitemOpen
  \bibfield  {author} {\bibinfo {author} {\bibfnamefont {D.~H.}\ \bibnamefont
  {Mahler}}, \bibinfo {author} {\bibfnamefont {L.~A.}\ \bibnamefont {Rozema}},
  \bibinfo {author} {\bibfnamefont {A.}~\bibnamefont {Darabi}}, \bibinfo
  {author} {\bibfnamefont {C.}~\bibnamefont {Ferrie}}, \bibinfo {author}
  {\bibfnamefont {R.}~\bibnamefont {Blume-Kohout}}, \ and\ \bibinfo {author}
  {\bibfnamefont {A.~M.}\ \bibnamefont {Steinberg}},\ }\href {\doibase
  10.1103/PhysRevLett.111.183601} {\bibfield  {journal} {\bibinfo  {journal}
  {Phys. Rev. Lett.}\ }\textbf {\bibinfo {volume} {111}},\ \bibinfo {pages}
  {183601} (\bibinfo {year} {2013})}\BibitemShut {NoStop}%
\bibitem [{\citenamefont {Qi}\ \emph {et~al.}(2017)\citenamefont {Qi},
  \citenamefont {Hou}, \citenamefont {Wang}, \citenamefont {Dong},
  \citenamefont {Zhong}, \citenamefont {Li}, \citenamefont {Xiang},
  \citenamefont {Wiseman}, \citenamefont {Li},\ and\ \citenamefont
  {Guo}}]{RN91}%
  \BibitemOpen
  \bibfield  {author} {\bibinfo {author} {\bibfnamefont {B.}~\bibnamefont
  {Qi}}, \bibinfo {author} {\bibfnamefont {Z.}~\bibnamefont {Hou}}, \bibinfo
  {author} {\bibfnamefont {Y.}~\bibnamefont {Wang}}, \bibinfo {author}
  {\bibfnamefont {D.}~\bibnamefont {Dong}}, \bibinfo {author} {\bibfnamefont
  {H.-S.}\ \bibnamefont {Zhong}}, \bibinfo {author} {\bibfnamefont
  {L.}~\bibnamefont {Li}}, \bibinfo {author} {\bibfnamefont {G.-Y.}\
  \bibnamefont {Xiang}}, \bibinfo {author} {\bibfnamefont {H.~M.}\ \bibnamefont
  {Wiseman}}, \bibinfo {author} {\bibfnamefont {C.-F.}\ \bibnamefont {Li}}, \
  and\ \bibinfo {author} {\bibfnamefont {G.-C.}\ \bibnamefont {Guo}},\ }\href
  {\doibase 10.1038/s41534-017-0016-4} {\bibfield  {journal} {\bibinfo
  {journal} {npj Quantum Information}\ }\textbf {\bibinfo {volume} {3}},\
  \bibinfo {pages} {19} (\bibinfo {year} {2017})}\BibitemShut {NoStop}%
\bibitem [{\citenamefont {Wang}\ \emph {et~al.}(2023)\citenamefont {Wang},
  \citenamefont {Zhan}, \citenamefont {Li}, \citenamefont {Xiao}, \citenamefont
  {Zhu}, \citenamefont {Qu}, \citenamefont {Lin}, \citenamefont {Yu},\ and\
  \citenamefont {Xue}}]{PhysRevLett.131.150803}%
  \BibitemOpen
  \bibfield  {author} {\bibinfo {author} {\bibfnamefont {X.}~\bibnamefont
  {Wang}}, \bibinfo {author} {\bibfnamefont {X.}~\bibnamefont {Zhan}}, \bibinfo
  {author} {\bibfnamefont {Y.}~\bibnamefont {Li}}, \bibinfo {author}
  {\bibfnamefont {L.}~\bibnamefont {Xiao}}, \bibinfo {author} {\bibfnamefont
  {G.}~\bibnamefont {Zhu}}, \bibinfo {author} {\bibfnamefont {D.}~\bibnamefont
  {Qu}}, \bibinfo {author} {\bibfnamefont {Q.}~\bibnamefont {Lin}}, \bibinfo
  {author} {\bibfnamefont {Y.}~\bibnamefont {Yu}}, \ and\ \bibinfo {author}
  {\bibfnamefont {P.}~\bibnamefont {Xue}},\ }\href {\doibase
  10.1103/PhysRevLett.131.150803} {\bibfield  {journal} {\bibinfo  {journal}
  {Phys. Rev. Lett.}\ }\textbf {\bibinfo {volume} {131}},\ \bibinfo {pages}
  {150803} (\bibinfo {year} {2023})}\BibitemShut {NoStop}%
\bibitem [{\citenamefont {Zhang}\ \emph {et~al.}(2022)\citenamefont {Zhang},
  \citenamefont {Li}, \citenamefont {Dou}, \citenamefont {Lu}, \citenamefont
  {Yang}, \citenamefont {Pang},\ and\ \citenamefont {Jin}}]{Zhang:22}%
  \BibitemOpen
  \bibfield  {author} {\bibinfo {author} {\bibfnamefont {C.-N.}\ \bibnamefont
  {Zhang}}, \bibinfo {author} {\bibfnamefont {H.}~\bibnamefont {Li}}, \bibinfo
  {author} {\bibfnamefont {J.-P.}\ \bibnamefont {Dou}}, \bibinfo {author}
  {\bibfnamefont {F.}~\bibnamefont {Lu}}, \bibinfo {author} {\bibfnamefont
  {H.-Z.}\ \bibnamefont {Yang}}, \bibinfo {author} {\bibfnamefont {X.-L.}\
  \bibnamefont {Pang}}, \ and\ \bibinfo {author} {\bibfnamefont {X.-M.}\
  \bibnamefont {Jin}},\ }\href {\doibase 10.1364/PRJ.463404} {\bibfield
  {journal} {\bibinfo  {journal} {Photon. Res.}\ }\textbf {\bibinfo {volume}
  {10}},\ \bibinfo {pages} {2388} (\bibinfo {year} {2022})}\BibitemShut
  {NoStop}%
\bibitem [{\citenamefont {Ichihara}\ \emph {et~al.}(2023)\citenamefont
  {Ichihara}, \citenamefont {Yoshida}, \citenamefont {Hong},\ and\
  \citenamefont {Horikiri}}]{PhysRevA.107.032608}%
  \BibitemOpen
  \bibfield  {author} {\bibinfo {author} {\bibfnamefont {M.}~\bibnamefont
  {Ichihara}}, \bibinfo {author} {\bibfnamefont {D.}~\bibnamefont {Yoshida}},
  \bibinfo {author} {\bibfnamefont {F.-L.}\ \bibnamefont {Hong}}, \ and\
  \bibinfo {author} {\bibfnamefont {T.}~\bibnamefont {Horikiri}},\ }\href
  {\doibase 10.1103/PhysRevA.107.032608} {\bibfield  {journal} {\bibinfo
  {journal} {Phys. Rev. A}\ }\textbf {\bibinfo {volume} {107}},\ \bibinfo
  {pages} {032608} (\bibinfo {year} {2023})}\BibitemShut {NoStop}%
\bibitem [{\citenamefont {Schuld}(2021)}]{schuld2021supervised}%
  \BibitemOpen
  \bibfield  {author} {\bibinfo {author} {\bibfnamefont {M.}~\bibnamefont
  {Schuld}},\ }\href {\doibase 10.48550/arXiv.2101.11020} {\bibfield  {journal}
  {\bibinfo  {journal} {arXiv:2101.11020}\ } (\bibinfo {year} {2021}),\
  10.48550/arXiv.2101.11020}\BibitemShut {NoStop}%
\bibitem [{\citenamefont {Bowie}\ \emph {et~al.}(2023)\citenamefont {Bowie},
  \citenamefont {Shrapnel},\ and\ \citenamefont {Kewming}}]{Bowie_2024}%
  \BibitemOpen
  \bibfield  {author} {\bibinfo {author} {\bibfnamefont {C.}~\bibnamefont
  {Bowie}}, \bibinfo {author} {\bibfnamefont {S.}~\bibnamefont {Shrapnel}}, \
  and\ \bibinfo {author} {\bibfnamefont {M.~J.}\ \bibnamefont {Kewming}},\
  }\href {\doibase 10.1088/2058-9565/acfba9} {\bibfield  {journal} {\bibinfo
  {journal} {Quantum Science and Technology}\ }\textbf {\bibinfo {volume}
  {9}},\ \bibinfo {pages} {015001} (\bibinfo {year} {2023})}\BibitemShut
  {NoStop}%
\bibitem [{\citenamefont {Paine}\ \emph {et~al.}(2023)\citenamefont {Paine},
  \citenamefont {Elfving},\ and\ \citenamefont
  {Kyriienko}}]{PhysRevA.107.032428}%
  \BibitemOpen
  \bibfield  {author} {\bibinfo {author} {\bibfnamefont {A.~E.}\ \bibnamefont
  {Paine}}, \bibinfo {author} {\bibfnamefont {V.~E.}\ \bibnamefont {Elfving}},
  \ and\ \bibinfo {author} {\bibfnamefont {O.}~\bibnamefont {Kyriienko}},\
  }\href {\doibase 10.1103/PhysRevA.107.032428} {\bibfield  {journal} {\bibinfo
   {journal} {Phys. Rev. A}\ }\textbf {\bibinfo {volume} {107}},\ \bibinfo
  {pages} {032428} (\bibinfo {year} {2023})}\BibitemShut {NoStop}%
\bibitem [{\citenamefont {Liu}\ and\ \citenamefont
  {Wang}(2018)}]{PhysRevA.98.062324}%
  \BibitemOpen
  \bibfield  {author} {\bibinfo {author} {\bibfnamefont {J.-G.}\ \bibnamefont
  {Liu}}\ and\ \bibinfo {author} {\bibfnamefont {L.}~\bibnamefont {Wang}},\
  }\href {\doibase 10.1103/PhysRevA.98.062324} {\bibfield  {journal} {\bibinfo
  {journal} {Phys. Rev. A}\ }\textbf {\bibinfo {volume} {98}},\ \bibinfo
  {pages} {062324} (\bibinfo {year} {2018})}\BibitemShut {NoStop}%
\bibitem [{\citenamefont {Sancho-Lorente}\ \emph {et~al.}(2022)\citenamefont
  {Sancho-Lorente}, \citenamefont {Rom\'an-Roche},\ and\ \citenamefont
  {Zueco}}]{PhysRevA.105.042432}%
  \BibitemOpen
  \bibfield  {author} {\bibinfo {author} {\bibfnamefont {T.}~\bibnamefont
  {Sancho-Lorente}}, \bibinfo {author} {\bibfnamefont {J.}~\bibnamefont
  {Rom\'an-Roche}}, \ and\ \bibinfo {author} {\bibfnamefont {D.}~\bibnamefont
  {Zueco}},\ }\href {\doibase 10.1103/PhysRevA.105.042432} {\bibfield
  {journal} {\bibinfo  {journal} {Phys. Rev. A}\ }\textbf {\bibinfo {volume}
  {105}},\ \bibinfo {pages} {042432} (\bibinfo {year} {2022})}\BibitemShut
  {NoStop}%
\bibitem [{\citenamefont {Roncallo}\ \emph {et~al.}(2024)\citenamefont
  {Roncallo}, \citenamefont {Morgillo}, \citenamefont {Macchiavello},
  \citenamefont {Maccone},\ and\ \citenamefont {Lloyd}}]{roncallo2024quantum}%
  \BibitemOpen
  \bibfield  {author} {\bibinfo {author} {\bibfnamefont {S.}~\bibnamefont
  {Roncallo}}, \bibinfo {author} {\bibfnamefont {A.~R.}\ \bibnamefont
  {Morgillo}}, \bibinfo {author} {\bibfnamefont {C.}~\bibnamefont
  {Macchiavello}}, \bibinfo {author} {\bibfnamefont {L.}~\bibnamefont
  {Maccone}}, \ and\ \bibinfo {author} {\bibfnamefont {S.}~\bibnamefont
  {Lloyd}},\ }\href {\doibase 10.48550/arXiv.2404.15266} {\bibfield  {journal}
  {\bibinfo  {journal} {arXiv:2404.15266}\ } (\bibinfo {year} {2024}),\
  10.48550/arXiv.2404.15266}\BibitemShut {NoStop}%
\bibitem [{\citenamefont {de~Guise}\ \emph {et~al.}(2018)\citenamefont
  {de~Guise}, \citenamefont {Di~Matteo},\ and\ \citenamefont
  {S\'anchez-Soto}}]{PhysRevA.97.022328}%
  \BibitemOpen
  \bibfield  {author} {\bibinfo {author} {\bibfnamefont {H.}~\bibnamefont
  {de~Guise}}, \bibinfo {author} {\bibfnamefont {O.}~\bibnamefont {Di~Matteo}},
  \ and\ \bibinfo {author} {\bibfnamefont {L.~L.}\ \bibnamefont
  {S\'anchez-Soto}},\ }\href {\doibase 10.1103/PhysRevA.97.022328} {\bibfield
  {journal} {\bibinfo  {journal} {Phys. Rev. A}\ }\textbf {\bibinfo {volume}
  {97}},\ \bibinfo {pages} {022328} (\bibinfo {year} {2018})}\BibitemShut
  {NoStop}%
\bibitem [{\citenamefont {Papoulis}\ and\ \citenamefont
  {Unnikrishna~Pillai}(2002)}]{papoulis2002probability}%
  \BibitemOpen
  \bibfield  {author} {\bibinfo {author} {\bibfnamefont {A.}~\bibnamefont
  {Papoulis}}\ and\ \bibinfo {author} {\bibfnamefont {S.}~\bibnamefont
  {Unnikrishna~Pillai}},\ }\href@noop {} {\emph {\bibinfo {title}
  {{Probability, Random Variables and Stochastic Processes}}}}\ (\bibinfo
  {publisher} {McGraw-Hill},\ \bibinfo {year} {2002})\BibitemShut {NoStop}%
\bibitem [{\citenamefont {Massar}\ and\ \citenamefont
  {Popescu}(1995)}]{PhysRevLett.74.1259}%
  \BibitemOpen
  \bibfield  {author} {\bibinfo {author} {\bibfnamefont {S.}~\bibnamefont
  {Massar}}\ and\ \bibinfo {author} {\bibfnamefont {S.}~\bibnamefont
  {Popescu}},\ }\href {\doibase 10.1103/PhysRevLett.74.1259} {\bibfield
  {journal} {\bibinfo  {journal} {Phys. Rev. Lett.}\ }\textbf {\bibinfo
  {volume} {74}},\ \bibinfo {pages} {1259} (\bibinfo {year}
  {1995})}\BibitemShut {NoStop}%
\bibitem [{\citenamefont {Mele}(2024)}]{Mele2024introductiontohaar}%
  \BibitemOpen
  \bibfield  {author} {\bibinfo {author} {\bibfnamefont {A.~A.}\ \bibnamefont
  {Mele}},\ }\href {\doibase 10.22331/q-2024-05-08-1340} {\bibfield  {journal}
  {\bibinfo  {journal} {{Quantum}}\ }\textbf {\bibinfo {volume} {8}},\ \bibinfo
  {pages} {1340} (\bibinfo {year} {2024})}\BibitemShut {NoStop}%
\bibitem [{\citenamefont {Bruß}\ and\ \citenamefont
  {Macchiavello}(1999)}]{BRU1999249}%
  \BibitemOpen
  \bibfield  {author} {\bibinfo {author} {\bibfnamefont {D.}~\bibnamefont
  {Bruß}}\ and\ \bibinfo {author} {\bibfnamefont {C.}~\bibnamefont
  {Macchiavello}},\ }\href {\doibase
  https://doi.org/10.1016/S0375-9601(99)00099-7} {\bibfield  {journal}
  {\bibinfo  {journal} {Physics Letters A}\ }\textbf {\bibinfo {volume}
  {253}},\ \bibinfo {pages} {249} (\bibinfo {year} {1999})}\BibitemShut
  {NoStop}%
\bibitem [{\citenamefont {Davis}\ and\ \citenamefont
  {Kahan}(1970)}]{doi:10.1137/0707001}%
  \BibitemOpen
  \bibfield  {author} {\bibinfo {author} {\bibfnamefont {C.}~\bibnamefont
  {Davis}}\ and\ \bibinfo {author} {\bibfnamefont {W.~M.}\ \bibnamefont
  {Kahan}},\ }\href {\doibase 10.1137/0707001} {\bibfield  {journal} {\bibinfo
  {journal} {SIAM Journal on Numerical Analysis}\ }\textbf {\bibinfo {volume}
  {7}},\ \bibinfo {pages} {1} (\bibinfo {year} {1970})}\BibitemShut {NoStop}%
\bibitem [{\citenamefont {O'Donnell}\ and\ \citenamefont
  {Wright}(2017)}]{10.1145/3055399.3055454}%
  \BibitemOpen
  \bibfield  {author} {\bibinfo {author} {\bibfnamefont {R.}~\bibnamefont
  {O'Donnell}}\ and\ \bibinfo {author} {\bibfnamefont {J.}~\bibnamefont
  {Wright}},\ }in\ \href {\doibase 10.1145/3055399.3055454} {\emph {\bibinfo
  {booktitle} {Proceedings of the 49th Annual ACM SIGACT Symposium on Theory of
  Computing}}},\ \bibinfo {series and number} {STOC 2017}\ (\bibinfo
  {publisher} {Association for Computing Machinery},\ \bibinfo {address} {New
  York, NY, USA},\ \bibinfo {year} {2017})\ p.\ \bibinfo {pages}
  {962–974}\BibitemShut {NoStop}%
\bibitem [{\citenamefont {Yu}(2020)}]{yu2020sample}%
  \BibitemOpen
  \bibfield  {author} {\bibinfo {author} {\bibfnamefont {N.}~\bibnamefont
  {Yu}},\ }\href {https://arxiv.org/abs/2009.04610} {\bibfield  {journal}
  {\bibinfo  {journal} {arXiv:2009.04610}\ } (\bibinfo {year}
  {2020})}\BibitemShut {NoStop}%
\end{thebibliography}%

\begin{widetext}
\newpage
\section*{Supplementary Information for ``Experimental benchmarking of quantum state overlap estimation strategies with photonic systems''}
\section{Overlap estimation strategy
performance with qubit pair sampling}
\label{Section-I}
In a general overlap estimation scenario, we consider $N$ copies of $\ket{\psi}$ and $M$ copies of $\ket{\phi}$, which can be represented as $\ket{\psi} = U\ket{0}$ and $\ket{\phi}= UW\ket{0}$, with $U, W$ in the special unitary group $SU(d)$. The overlap between two states is given by $c =|\braket{\psi|\phi}|^2 = |\braket{0|W|0}|^2$ and the overlap information is solely contained in $W$. An overlap estimation strategy involves measuring all states using a general measurement $\{E_k\}$ with outcome $k$ and outputting an estimation $\tilde{c}(k)$ of the overlap. We assess the performance of an overlap estimation strategy using the local approach mentioned in \cite{PhysRevLett.124.060503}, assuming that the overlap $c$ is fixed. For a fixed overlap $c$, the set of possible unitary $W$ is given by $w = \left\{W\in SU(d):|\bra{0}W\ket{0}|^2 = c\right\}$. We quantify the estimation precision by computing the average of the square error over all states with fixed overlap c and over all outcomes:
\begin{equation}
    v(c) = \int_{SU(d)}\int_{w}dUdW \sum_{k}[\tilde{c}(k)-c]^2 \text{Tr}\left[E_k\ket{\Phi}\bra{\Phi}\right],
    \label{general_eq_1}
\end{equation}
here $dU$ and $dW$ are Haar measure, and $\ket{\Phi}=\ket{\psi}^{\otimes N}\otimes \ket{\phi}^{\otimes M}=U^{\otimes(N+M)}\ket{0}^{\otimes N}\left(W\ket{0}\right)^{\otimes M}$. 
\par
In this work, we mainly consider a qubit case with $d=2$ and $M=N$. To estimate the value of Eq.~(\ref{general_eq_1}) experimentally, we randomly sample qubit pairs using the Haar measure. The Haar-distributed unitary matrix $U$ can be parameterized as \cite{PhysRevA.97.022328}:
\begin{equation}
    U=\left(\begin{array}{cc}
        e^{i(\beta-\omega)/2}\cos{\frac{\theta}2}  & -e^{-i(\beta+\omega)/2}\sin{\frac{\theta}2} \\
        e^{i(\beta+\omega)/2}\sin{\frac{\theta}2} & e^{-i(\beta-\omega)/2}\cos{\frac{\theta}2}
    \end{array}\right),
    \label{eq_U}
\end{equation}
where $\beta$ and $\omega$ are uniformly distributed on $[0,2\pi)$, and $\theta$ follows a probability density function (PDF) $p(\theta) = \frac{1}{2}\sin\theta$ on $[0,\pi]$. The Haar measure on $SU(2)$ is denoted by $dU = (1/8\pi^2)\sin\theta d\theta d\beta d\omega$. From the fixed overlap expression $c=|\braket{0|W|0}|^2$, we can derive the matrix form of $W$ as
\begin{equation}
    W=\left(\begin{array}{cc}
        \sqrt{c}e^{i(\varphi_g-\varphi)/2}     & -\sqrt{1-c}e^{-i(\varphi_g+\varphi)/2} \\
        \sqrt{1-c}e^{i(\varphi_g+\varphi)/2} & \sqrt{c}e^{-i(\varphi_g-\varphi)/2}
    \end{array}\right),
    \label{eq_W_matrix}
\end{equation}
where $\varphi_g$ and $\varphi$ follow the uniform distribution on $[0,2\pi)$. Consequently, the qubit pair takes the following form:
\begin{equation}
    \ket{\psi}=U\ket{0},\ \ket{\phi}=U(\sqrt{c}\ket{0}+e^{i\varphi}\sqrt{1-c}\ket{1}),
    \label{eq-qubit-pair}
\end{equation}
here we have omitted the global phase term $e^{i(\varphi_g-\varphi)/2}$ in $\ket{\phi}$. It is worth noting that sampling such a unitary $W$ is equivalent to sample a pure state with a fixed angle respect to $\ket{0}$ on the Bloch sphere. By sampling $U$ and $\varphi$, we obtain random qubit pairs with a fixed overlap, as illustrated in Fig. \ref{figure_qubit_sample}.
\begin{figure}[!ht]\centering
    \includegraphics[width=1\textwidth]{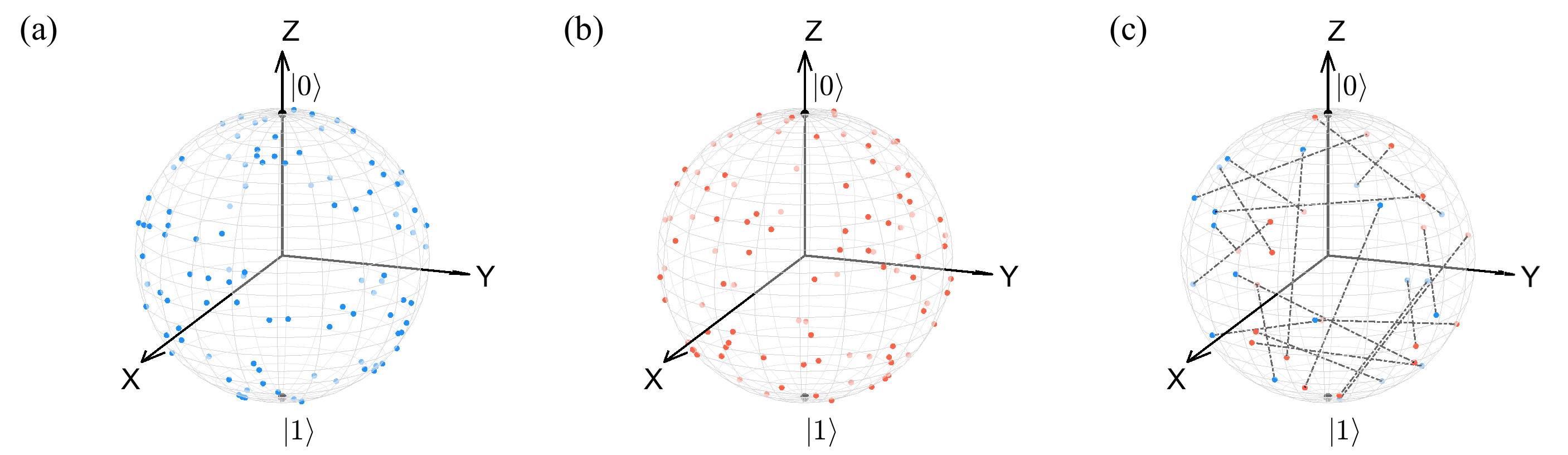}
    \caption{Bloch sphere representation of pure qubit pairs sampled at a fixed overlap $c = 0.5$. \textbf{a} 100 sample states of the qubit state $\ket{\psi}$.  \textbf{b} 100 sample states of the qubit state $\ket{\phi}$. \textbf{c} 20 out of 100 sample qubit pairs represented by dots on the sphere. The dashed lines inside the Bloch sphere have equal lengths, indicating the same overlap between different qubit pairs.}
        \label{figure_qubit_sample}
\end{figure}
\par
For an overlap estimation strategy denoted by $s$, which performs a measurement ${\hat{E}_k^{(s)}}$ on $N$ copies of the qubit pair in Eq.~(\ref{eq-qubit-pair}), and uses the estimator $\tilde{c}_{s}(k)$, the square error can be expressed as 
\begin{equation}
    v_{s}(c,N|U,\varphi) =  \sum_{k}[\tilde{c}_{s}(k)-c]^2 \text{Tr}\left[\hat{E}_k^{(s)}(\ket{\psi}\bra{\psi}\otimes\ket{\phi}\bra{\phi})^{\otimes N}\right], 
\end{equation}
which serves as a function of overlap $c$ and copy number $N$ for given $\ket{\psi}$ and $\ket{\phi}$. From Eq.~(\ref{general_eq_1}), the average square error of the strategy $s$ can be written as 
\begin{equation}
    v_{s}(c,N)=\frac{1}{2\pi}\int_U\int_{0}^{2\pi}v_{s}(c,N|U,\varphi)dUd{\varphi},
    \label{specific_var_eq}
\end{equation}
where $dU$ is the Haar measure of $SU(2)$. The estimators used in our work are unbiased or asymptotically unbiased, hence we also denote $v_{s}(c,N)$ in Eq.~(\ref{specific_var_eq}) as average variance of the strategy s. Our analysis shows that as the number of copies $N$ increases, the average variance $v_{s}(c,N)$ decreases with the scale of $O(1/N)$. Therefore, in the limit of $N\to \infty$, the average variance multiplied by the copy number $N$, i.e., $Nv_{s}(c) = \lim_{N\to \infty}N\cdot v_{s}(c,N)$, only depends on the overlap $c$. We adopt the scaled average variance $Nv_{s}(c)$ as a performance assessment for strategy $s$.
\section{Tomography-Tomography and Tomography-Projection}
\subsection{Average infidelity of pure state tomography based on MUB}
\label{sec_2a}
In this section, we derive an asymptotic solution for the average infidelity in pure qubit state tomography (QST) based on mutually unbiased bases (MUB) and maximum likelihood estimation (MLE). The main text mentioned two local overlap estimation strategies: tomography-tomography (TT) and tomography-projection (TP), both of which require QST based on MUB, i.e., $\{\ket{0},\ket{1}\}$, $\{\ket{+},\ket{-}\}$, $\{\ket{L},\ket{R}\}$ with $\ket{\pm}=(\ket{0}\pm \ket{1})/\sqrt{2}$ and $\ket{L},\ket{R}=(\ket{0}\pm i\ket{1})/\sqrt{2}$. To illustrate the tomography process, we focus on  the state $\ket{\psi}$ as an example. Given a total of $N$ copies of $\ket{\psi}$, we measure three Pauli operators $(\hat{\sigma}_x,\hat{\sigma}_y,\hat{\sigma}_z)$ on $N^{\prime} = N/3$ copies of $\ket{\psi}$ and record the statistics of obtaining the outcome $+1$ for each operator, denoted as $(n_x,n_y,n_z)$. The probabilities of obtaining the outcome $+1$ for the three Pauli operators are as follows
\begin{equation}
    p_x = |\braket{\psi|+}|^2 = \frac{1}{2}(1+\sin{\theta}\cos{\omega}),\
    p_y = |\langle\psi\ket{L}|^2 = \frac{1}{2}(1+\sin{\theta}\sin{\omega}),\
    p_z = |\braket{\psi|0}|^2 = \cos^{2}\frac{\theta}{2},
    \label{eq_prob_xyx}
\end{equation}
where $\ket{\psi} = U\ket{0}$ and $U$ is defined in Eq.~(\ref{eq_U}). Given that $\ket{\psi}$ is pure, we obtain the reconstructed state $\ket{\tilde{\psi}}$ by finding  a straightforward yet approximate solution to the maximum likelihood equations in the following form:
\begin{equation}
    \ket{\tilde{\psi}} = \sqrt{\frac{n_z}{N^{\prime}}}\ket{0}+\sqrt{1-\frac{n_z}{N^{\prime}}}\frac{(\frac{2n_x}{N^{\prime}}-1)+i(\frac{2n_y}{N^{\prime}}-1)}{\sqrt{(\frac{2n_x}{N^{\prime}}-1)^2+(\frac{2n_y}{N^{\prime}}-1)^2}}\ket{1}.
    \label{eq_psie_xyx}
\end{equation}
The tomography result is determined by the outcome $(n_x,n_y,n_z)$ and follows three independent binomial distributions with a PDF as follows:
\begin{equation}
    \begin{split}
        P_{tomo}\left(\tilde{\psi}(n_x,n_y,n_z)|N,U\right)= \text{Bin}(n_x,N^{\prime},p_x)\text{Bin}(n_y,N^{\prime},p_y)\text{Bin}(n_z,N^{\prime},p_z),
    \end{split}
    \label{eq_Ptomo}
\end{equation}
where $\text{Bin}(k,N,p)=\binom{N}{k}p^{k}(1-p)^{N-k}$ is the PDF of the binomial distribution, and  $p_x$, $p_y$,  $p_z$ are defined in Eq.~(\ref{eq_prob_xyx}). The fidelity between the true state and the reconstructed state is defined as $F = |\braket{\psi|\tilde{\psi}}|^2$ for $\ket{\psi}$ and $\ket{\tilde{\psi}}$. For a given $U$, the fidelity of the above tomography procedure is given by
\begin{small}
    \begin{equation}
    \begin{split}
       F_{U}= &  \sum_{n_x,n_y,n_z} |\braket{\psi|\tilde{\psi}}|^2 P_{tomo}\left(\tilde{\psi}(n_x,n_y,n_z)|N,U\right) \\
       = & \frac{1}{2}\left \langle  1+\left(\frac{2n_z}{N^{\prime}}-1\right)\cos{\theta}+2\sin{\theta}\sqrt{\frac{n_z}{N^{\prime}}\left(1-\frac{n_z}{N^{\prime}}\right)}
       \frac{\left(\frac{2n_x}{N^{\prime}}-1\right)\cos{\omega}+ \left(\frac{2n_y}{N^{\prime}}-1\right)\sin{\omega}}{\sqrt{\left(\frac{2n_x}{N^{\prime}}-1\right)^2+\left(\frac{2n_y}{N^{\prime}}-1\right)^2}}\right \rangle_{n_x,n_y,n_z} \\ 
       = & \frac{1}{2} \left[1+\left\langle\frac{2n_z}{N^{\prime}}-1\right\rangle_{n_z}\cos{\theta}+2\sin{\theta}
       \left\langle\sqrt{\frac{n_z}{N^{\prime}}\left(1-\frac{n_z}{N^{\prime}}\right)}\right\rangle_{n_z}
       \left\langle\frac{\left(\frac{2n_x}{N^{\prime}}-1\right)\cos{\omega}+ \left(\frac{2n_y}{N^{\prime}}-1\right)\sin{\omega}}{\sqrt{\left(\frac{2n_x}{N^{\prime}}-1\right)^2+\left(\frac{2n_y}{N^{\prime}}-1\right)^2}} \right\rangle_{n_x,n_y}
       \right],
    \end{split}
    \label{eq_FU}
\end{equation}
\end{small}
where the notation $\langle\cdot\rangle_{j}$ represents the expectation with respect to the random variable $j$. We introduce the following notations: 
\begin{equation}
    X = 2n_x/N^{\prime} - 1,\ Y = 2n_y/N^{\prime} - 1,\ Z = n_z/N^{\prime}.
    \label{eq_XYZ_def}
\end{equation}
 \begin{figure}[!ht]
    \centering
    \includegraphics[width=0.35\textwidth]{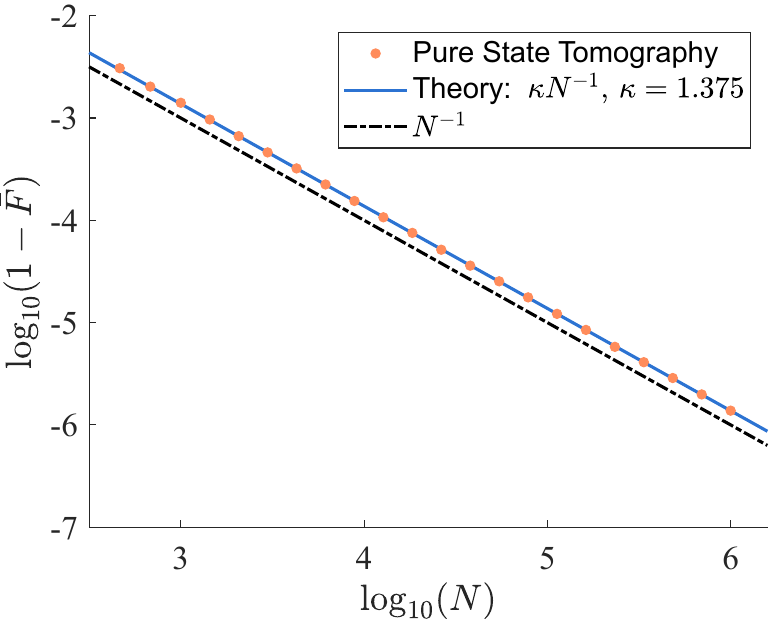}
    \caption{Average infidelity $1-\overline{F}$ v.s. tomography copy number $N$ for Monte Carlo simulations of the pure state tomography. For each $N$, we uniformly sample 1000 different qubit states and perform 20 repeated tomographic simulations for each qubit. The simulated results (orange dots) show excellent agreement with the theoretical result (blue line).}
    \label{fig:tomo scale}
\end{figure}
Using these notations, we have the following expressions:
\begin{equation}
    \begin{split}
        & \langle X \rangle_{n_x} = \sin{\theta}\cos{\omega},\ \sigma^2_{X} = \langle (X-\langle X \rangle_{n_x})^2 \rangle_{n_x} = \frac{1-\sin^2{\theta}\cos^2{\omega}}{N^{\prime}},\\
        & \langle Y \rangle_{n_y} = \sin{\theta}\sin{\omega},\ \sigma^2_{Y} = \langle (Y-\langle Y \rangle_{n_y})^2 \rangle_{n_y} = \frac{1-\sin^2{\theta}\sin^2{\omega}}{N^{\prime}},\\
        & \langle Z \rangle_{n_z} = \cos^2{\frac{\theta}{2}}, \ \sigma^2_{Z} = \langle (Z-\langle Z \rangle_{n_z})^2 \rangle_{n_z} = \frac{\sin^2{\theta}}{4N^{\prime}},
    \end{split}
    \label{XYZ-mu-var}
\end{equation}
where we can observe that the higher central moments of $X$, $Y$ and $Z$ are $O(1/{N^{\prime}}^2)$ orders. Additionally, we define two functions as follows:
 \begin{equation}
     f(Z) = \sqrt{Z(1-Z)},\ g(X,Y) = \frac{X \cos{\omega} + Y \sin{\omega} }{ \sqrt{X^2+Y^2} }.
     \label{eq_fg}
 \end{equation}
 From the probability theory~\cite{papoulis2002probability}, the expectations of Eq.~(\ref{eq_fg}) can be expanded as
 \begin{equation}
 \begin{split}
      \langle f(Z) \rangle_{n_z} = & f( \langle Z \rangle_{n_z} ) + \frac{d^2f}{dZ^2}\bigg|_{\langle Z \rangle_{n_z}} \frac{\sigma^2_{Z}}{2}+O(\frac{1}{{N^{\prime}}^2}) \approx \frac{\sin{\theta}}{2}-\frac{1}{4N^{\prime}\sin{\theta}} \\
      \langle g(X,Y) \rangle_{n_x,n_y}= & g(\langle X \rangle_{n_x},\langle Y \rangle_{n_y}) + \frac{\partial^2g}{\partial X^2}\bigg|_{\langle X \rangle_{n_x},\langle Y \rangle_{n_y}}\frac{\sigma^2_{X}}{2}+\frac{\partial^2g}{\partial Y^2}\bigg|_{\langle X \rangle_{n_x},\langle Y \rangle_{n_y}}\frac{\sigma^2_{Y}}{2}+O(\frac{1}{{N^{\prime}}^2})\\
      \approx & 1 - \frac{\sin^2{\omega}(1-\sin^2{\theta}\cos^2{\omega})}{2N^{\prime}\sin^2{\theta}}- \frac{\cos^2{\omega}(1-\sin^2{\theta}\sin^2{\omega})}{2N^{\prime}\sin^2{\theta}} \\
      = & 1- \frac{1-2\sin^2{\theta}\sin^2{\omega}\cos^2{\omega}}{2N^{\prime}\sin^2{\theta}}.
 \end{split}
 \label{eq_fzgxy_mean}
 \end{equation}
 $F_U$ in Eq.~(\ref{eq_FU}) is then given by
 \begin{equation}
     \begin{split}
         F_U = & \frac{1}{2}\left[1+ (2\langle Z \rangle_{n_z}-1) \cos{\theta} +2\sin{\theta}\langle f(Z) \rangle_{n_z} \langle g(X,Y) \rangle_{n_x,n_y} \right] \\
         = & 1- \frac{1-\sin^2{\theta}\sin^2{\omega}\cos^2{\omega}}{2N^{\prime}} + O(\frac{1}{{N^{\prime}}^2}).
     \end{split}
 \end{equation}
After averaging over $U$, the average fidelity of pure qubit tomography based on MUB is then given by
\begin{equation}
    \begin{split}
        \overline{F} = \int_{U}F_U dU = & \frac{1}{4\pi}\int_{0}^{\pi}\int_{0}^{2\pi}(1- \frac{1-\sin^2{\theta}\sin^2{\omega}\cos^2{\omega}}{2N^{\prime}})\sin{\theta} d\theta d\omega+ O(\frac{1}{{N^{\prime}}^2})\\
        = & 1 - \frac{11}{24N^{\prime}} + O(\frac{1}{{N^{\prime}}^2}) \approx 1- \frac{11}{8N},
    \end{split}
    \label{eq_Fbar}
\end{equation}
 where $N = 3N^{\prime}$. Neglecting the high-order terms, the average infidelity $1-\overline{F} = 11/8N$ scales as $O(1/N)$ and the scaled average infidelity is given by 
 \begin{equation}
     \kappa = N(1-\overline{F}) = \frac{11}{8}.
 \end{equation}
 \par
 Furthermore, we validate this theoretical result through Monte Carlo simulations of pure state tomography. Figure~\ref{fig:tomo scale} presents the plot of simulated average infidelity against the copy number $N$ in tomography. We fit the simulated data to the form $1-\overline{F} = \kappa N^{-1}$ and find $\kappa = 1.377 \pm 0.006$ which agree with the theoretical result. Furthermore, We note that with an optimal guess of the reconstructed state using the Bayesian approach in \cite{PhysRevLett.89.277904}, the coefficient $\kappa$ can be improved to $13/12$ with static measurements. By using the adaptive approach for pure qubit tomography, $\kappa$ can be further improved to 1, saturating the collective measurement bound given by \cite{PhysRevLett.74.1259}.
\subsection{Tomography error analysis}
\label{sec:tomo-error}
In this section, we derive the asymptotic results for the average error distribution of the reconstructed states given by QST. Starting from Eq.~(\ref{eq-qubit-pair}), we can express the orthogonal states of $\ket{\psi}$ and $\ket{\phi}$ as
\begin{equation}
    \ket{\psi_\perp} = U\ket{1},\ \ket{\phi_\perp} = U(\sqrt{1-c}\ket{0}-e^{i\varphi}\sqrt{c}\ket{1}),
\end{equation}
here, we have $\braket{\psi|\phi_\perp}=\sqrt{1-c}$, $\braket{\psi_\perp|\phi_\perp}=-e^{i\varphi}\sqrt{c}$ and $\braket{\psi_\perp|\phi}=e^{i\varphi}\sqrt{1-c}$. The reconstructed states of $\ket{\phi}$ and $\ket{\psi}$ by QST can be represented as 
\begin{equation}
\begin{split}
        \ket{\tilde{\psi}} = \cos{\frac{\chi_1}{2}}\ket{\psi} + \sin{\frac{\chi_1}{2}}e^{i\zeta_1}\ket{\psi_{\perp}},\ \ket{\tilde{\phi}} =\cos{\frac{\chi_2}{2}}\ket{\phi} + \sin{\frac{\chi_2}{2}}e^{i\zeta_2}\ket{\phi_{\perp}},
\end{split}
\label{estimate_states}
\end{equation}
where $\chi_j\in [0,\pi]$ (for $j=1,2$) denote the deviation of the estimate states from the true state, and $\zeta_j\in [0,2\pi)$ (for $j=1,2$) are random phases introduced by QST. The joint PDFs of $\chi_j$ and $\zeta_j$ are denoted by $p(\chi_1,\zeta_1|N,U)$ and $p(\chi_2,\zeta_2|c,N,U,\varphi)$, which depend on the specific values of $U$ and $\varphi$. In this supplementary materials, we use the notation $\braket{\cdot}$ to represent the expectation respect to these distributions under a given pair of $U$ and $\varphi$. For instance, $\braket{\chi_1} = \int_{\chi_1}\int_{\zeta_1}\chi_1p(\chi_1,\zeta_1|N,U)d\chi_1d\zeta_1$. On the other hand, the notation $\overline{\braket{\cdot}}$ denotes the average expectation over all possible values of $U$ and $\varphi$. For example
\begin{equation}
    \overline{\braket{\chi_2}} = \frac{1}{2\pi}\int_U\int_{0}^{2\pi}\braket{\chi_2}dUd\varphi =\frac{1}{2\pi}\int_U\int_{0}^{2\pi}\left[\int_{\chi_2}\int_{\zeta_2}\chi_2p(\chi_2,\zeta_2|c,N,U,\varphi)d\chi_2d\zeta_2\right]dUd\varphi.
\end{equation}
\par
Taking $\ket{\tilde{\psi}}$ as an example to analyze the error distribution for the pure state tomography, we temporarily omit the lower indices of $\chi_j$ and $\zeta_j$ in this section. Considering a Bloch sphere with that $z$ axis represents the true state $\ket{\psi}$, then $\chi$ and $\zeta$ in $\ket{\tilde{\psi}}$ correspond the polar and azimuthal angles, respectively, as illustrated in Fig.~\ref{fig-tomo-error}. To simplify the analysis, we introduce a coordinate transformation:
\begin{equation}
    t^c = \sin{\chi} \cos\zeta,\ t^s = \cos{\chi} \sin\zeta,
    \label{eq-tc-ts}
\end{equation}
the two error variables $(t^c,t^s)$ follows a joint PDF denoted as $p(t^c,t^s|N,U)$, which can be derived from $p(\chi,\zeta|N,U)$ with the transformation. By building a relationship between $(t^c,t^s)$ and three independent variables $(X,Y,Z)$ defined in Eq.~(\ref{eq_psie_xyx}), namely, the recording statistics $(n_x,n_y,n_z)$ in the tomography, we can derive the means and variances of $(t^c,t^s)$ for a given $U$ and integrate them over $SU(2)$ to get the average moments.
\begin{figure}[!hbt]
    \centering
\includegraphics[width=\textwidth]{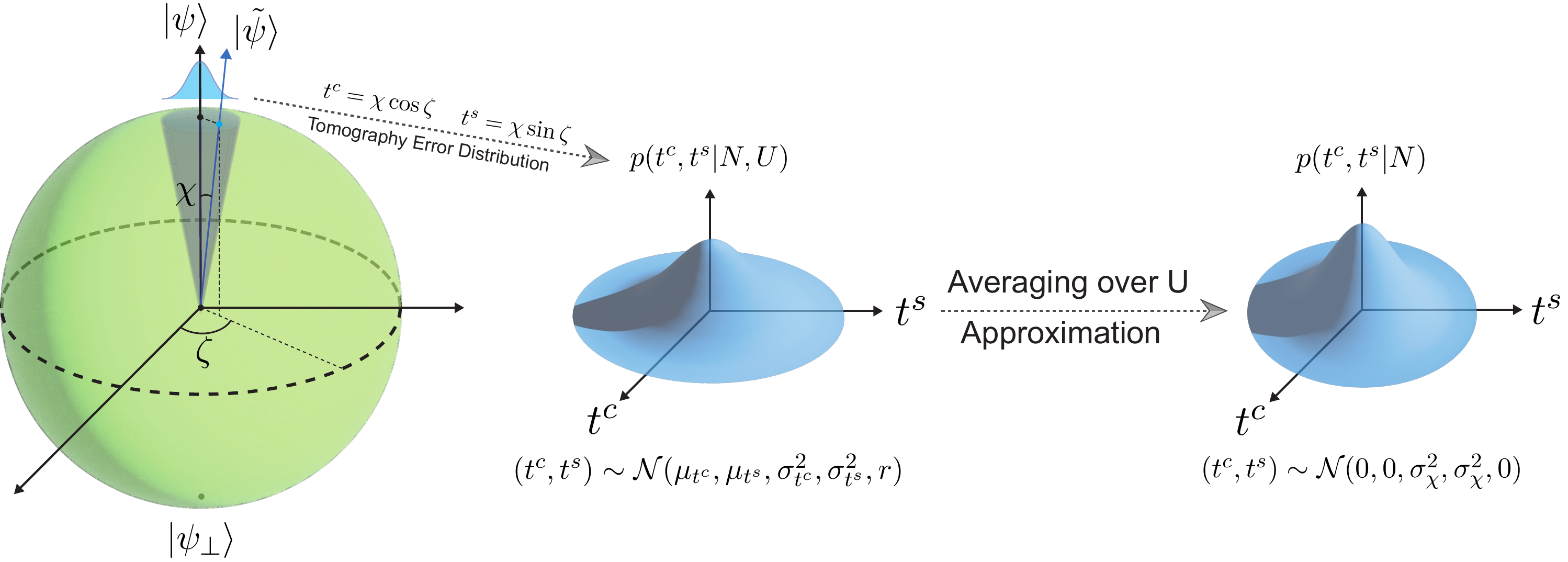}
    \caption{Schematic of the error distribution in pure qubit MUB tomography. $t^c$ and $t^s$ are obtained through coordinate transformation from $\zeta$ and $\chi$. The error probability distribution $p(t^c,t^s|N,U)$, under a specific unitary operator $U$, can be approximated as a bivariate Gaussian distribution of $t^c$ and $t^s$. When averaging over all possible $U$, the average distribution $p(t^c,t^s|N)$ can be approximated as a bivariate Gaussian distribution with the same variance for the marginal distributions. This approximation implies that $t^c$ and $t^s$ are statistically independent in the average case.}
    \label{fig-tomo-error}
\end{figure}
\par
Firstly, we rewrite the $\ket{\tilde{\psi}}$ in Eq.~(\ref{estimate_states}) on the computational basis $\{\ket{0},\ket{1}\}$ as follows:
\begin{equation}
\begin{split}
\ket{\tilde{\psi}}& =  \sqrt{A}\ket{0} + e^{iB}\sqrt{1-A}\ket{1},\ A =  \frac12\left(1 + \cos{\chi}\cos{\theta}-\sin{\chi}\cos{(\beta-\zeta)}\sin{\theta}\right),\\
       \cos{B}& =  \frac{G_X}{2\sqrt{A(1-A)}},\ G_X = [\cos{\chi}\sin{\theta} + \sin{\chi}\cos{(\beta-\zeta)}\cos{\theta}]\cos{\omega} + \sin{\chi}\sin{(\beta-\zeta)}\sin{\omega},\\
       \sin{B}& =  \frac{G_Y}{2\sqrt{A(1-A)}},\ G_Y = [\cos{\chi}\sin{\theta} + \sin{\chi}\cos{(\beta-\zeta)}\cos{\theta}]\sin{\omega} - \sin{\chi}\sin{(\beta-\zeta)}\cos{\omega},
\end{split}
\label{eq_gxgy}
\end{equation}
where $\theta$, $\beta$ and $\omega$ are from the parameterized $U$ defined in Eq.~(\ref{eq_U}) and we have discarded a global phase. Recalling $X$, $Y$ and $Z$ defined in Eq.~(\ref{eq_psie_xyx}), we have another expression of $\ket{\tilde{\psi}}$ as 
\begin{equation}
    \ket{\tilde{\psi}} = \sqrt{Z}\ket{0}+\sqrt{1-Z}\frac{X+iY}{\sqrt{X^2+Y^2}}.
    \label{eq_psi_e2}
\end{equation}
Combining two expressions of $\ket{\tilde{\psi}}$ in Eq.~(\ref{eq_gxgy}) and Eq.~(\ref{eq_psi_e2}), we can get the following equations:
\begin{equation}
    \begin{split}
        G_X = \frac{2X\sqrt{Z(1-Z)}}{\sqrt{X^2+Y^2}},\ 
        G_Y = \frac{2Y\sqrt{Z(1-Z)}}{\sqrt{X^2+Y^2}},\
        A = Z.
    \end{split}
    \label{eq_xyz_origin}
\end{equation}
We can rewrite $G_X$ and $G_Y$ with $(t^c,t^s)$ as follows: 
\begin{equation}
    \begin{split}
        G_X = & \left[\frac{\cos{\beta}\cos{\omega}}{\cos{\theta}} + \sin{\beta}\sin{\omega} \right]t^c + \left[\frac{\sin{\beta}\cos{\omega}}{\cos{\theta}} - \cos{\beta}\sin{\omega} \right]t^s + (2Z-1)\cos{\omega}\tan{\theta}, \\
        G_Y = & \left[\frac{\cos{\beta}\sin{\omega}}{\cos{\theta}} - \sin{\beta}\cos{\omega} \right]t^c + \left[\frac{\sin{\beta}\sin{\omega}}{\cos{\theta}} + \cos{\beta}\cos{\omega} \right]t^s + (2Z-1)\sin{\omega}\tan{\theta}.
    \end{split}
\end{equation}
Here we have build a relationship between $(X, Y, Z)$  and $(t^c,t^s)$ through $G_X$ and $G_Y$. Then we can express $t^c$ and $t^s$ as follows:
\begin{equation}
    \begin{split}
        t^c & = (\cos{\theta}\cos{\beta}\cos{\omega} + \sin{\beta}\sin{\omega} ) G_X + (\cos{\theta}\cos{\beta}\sin{\omega} - \sin{\beta}\cos{\omega})G_Y - (2Z-1)\sin{\theta}\cos{\beta}, \\
        t^s & = (\cos{\theta}\sin{\beta}\cos{\omega} - \cos{\beta}\sin{\omega} ) G_X + (\cos{\theta}\sin{\beta}\sin{\omega} + \cos{\beta}\cos{\omega})G_Y - (2Z-1)\sin{\theta}\sin{\beta}. \\
    \end{split}
    \label{eq_tcts}
\end{equation}
In the case of large $N$, 
with the same approach in Eq.~(\ref{eq_fzgxy_mean}), the means of $G_X$ and $G_Y$ can be derived through the moments of $(X,Y,Z)$ in Eq.~(\ref{XYZ-mu-var}) as
\begin{equation}
\begin{split}
        \langle G_X \rangle & = \sin{\theta}\cos{\omega} -\frac{\cos{\omega}}{\sin{\theta}N^{\prime}}+\frac{(1+3\cos{2\omega})\sin^2{\omega}\cos{\omega}\sin{\theta}}{2N^{\prime}} + O(\frac{1}{{N^{\prime}}^2}),\\
        \langle G_Y \rangle & = \sin{\theta}\sin{\omega} -\frac{\sin{\omega}}{\sin{\theta}N^{\prime}}+\frac{(1-3\cos{2\omega})\sin{\omega}\cos^2{\omega}\sin{\theta}}{2N^{\prime}} + O(\frac{1}{{N^{\prime}}^2}),\\
\end{split}
\label{eq_gxgy_mean}
\end{equation}
with $N^{\prime} = N/3$. Combining Eq.~(\ref{eq_tcts}) with Eq.~(\ref{eq_gxgy_mean}), the means of $t^c$ and $t^s$ are then given by
\begin{equation}
\begin{split}
        \mu_{t^c} & =\langle t^c \rangle  = \frac{\sin{\beta}}{4N^{\prime}} \sin{\theta}\sin{4\omega}+\frac{\cos{\beta}}{4N^{\prime}}\left(\sin{\theta}\cos{\theta}\sin^2{2\omega}-4\cot{\theta}\right)+O(\frac{1}{{N^{\prime}}^2}), \\
        \mu_{t^s} & =\langle t^s \rangle  = 
        -\frac{\cos{\beta}}{4N^{\prime}}\sin{\theta}\sin{4\omega}+\frac{\sin{\beta}}{4N^{\prime}}\left(\sin{\theta}\cos{\theta}\sin^2{2\omega}-4\cot{\theta}\right)+O(\frac{1}{{N^{\prime}}^2}).
\end{split}
\end{equation}
Here, we show that $\mu_{t^c}$ and $\mu_{t^s}$ scale as $O(1/N)$, resulting from the asymptotic unbiasedness of maximum likelihood estimation. To obtain the variance of $(t^c,t^s)$, we expand $t^c$ and $t^s$ in Eq.~(\ref{eq_tcts}) as functions of $(X,Y,Z)$ up to the first order terms around the mean point $(\braket{X},\braket{Y},\braket{Z})$, which are given by:
\begin{equation}
    \begin{split}
        t^c & =  \sin{\beta}\sin{\omega}X-\sin{\beta}\cos{\omega}Y - \frac{\cos{\beta}}{\sin{\theta}}(2Z-1-\cos{\theta}) + \text{high-order terms},\\
        t^s & =  -\cos{\beta}\sin{\omega}X+\cos{\beta}\cos{\omega}Y - \frac{\sin{\beta}}{\sin{\theta}}(2Z-1-\cos{\theta}) + \text{high-order terms}. \\
    \end{split}
\end{equation}  
The leading order of variance is only determined by the first-order terms in $t_c$ and $t_s$. Consequently, we can derive the variance ($\sigma_{t^c}^2$, $\sigma_{t^s}^2$) and correlation coefficient $r$ as
\begin{equation}
    \begin{split}
        \sigma_{t^c}^2 & =  \frac{1-2\sin^2{\theta}\sin^2{\omega}\cos^2{\omega}\sin^2{\beta}}{N^{\prime}}+O(\frac{1}{{N^{\prime}}^2}),\ 
        \sigma_{t^s}^2 =  \frac{1-2\sin^2{\theta}\sin^2{\omega}\cos^2{\omega}\cos^2{\beta}}{N^{\prime}}+O(\frac{1}{{N^{\prime}}^2}), \\
        r & =  \left[\frac{\sin{2\beta}\left( 1-\sin^2{\theta}\sin^2{\omega}\cos^2{\omega} \right)}{N^{\prime}}+O(\frac{1}{{N^{\prime}}^2})\right]/\sigma_{t^c}\sigma_{t^s}.
    \end{split}
\end{equation}
\par
Through integrating the results over $U$, we can get the means, variances and second-order raw moments in the average case as follows:
\begin{equation}
\begin{split}
\overline{\mu_{t^c}} = \overline{\braket{t^c}}= O(\frac{1}{N^2}),\ \ \ \overline{\sigma^2_{t^c}} = \frac{11}{4N} + O(\frac{1}{N^2}),\ \ \ \overline{\braket{(t^c)^2}} & = \overline{\sigma^2_{t^c}} + \overline{\braket{t^c}^2} = \frac{11}{4N} + O(\frac{1}{N^2}),
\\ \overline{\mu_{t^s}} = \overline{\braket{t^s}} = O(\frac{1}{N^2}),\ \ \ \overline{\sigma^2_{t^s}}  = \frac{11}{4N} + O(\frac{1}{N^2}),\ \ \ \overline{\braket{(t^s)^2}} & = \overline{\sigma^2_{t^s}} + \overline{\braket{t^s}^2} = \frac{11}{4N} + O(\frac{1}{N^2}),
\\ \overline{\braket{t^ct^s}} &  = \overline{r\sigma_{t^c}\sigma_{t^s}} = O(\frac{1}{N^2}),
\end{split}
\end{equation}
where $\overline{\braket{\cdot}}$ denotes
\begin{equation}
    \overline{\braket{\cdot}}= \int_U \braket{\cdot} dU = \frac{1}{8\pi^2}\int_{0}^{\pi}\int_{0}^{2\pi}\int_{0}^{2\pi}(\cdot)\sin{\theta} d\theta d\beta d\omega.
\end{equation}
\par
In the asymptotic limit, the average fidelity shown in Eq.~(\ref{eq_Fbar}) can be derived through the distribution of $\chi$ as 
\begin{equation}
    \overline{F} = \overline{\left\langle \cos^2{\frac{\chi}2} \right\rangle} \approx \overline{\left\langle 1-\frac{\chi^2}{4} \right\rangle} = 1 - \frac{\overline{\left\langle \chi^2 \right\rangle}}4,
\label{eq_infidelity}
\end{equation}
where we use the approximation $\cos^2{\chi/2}\approx 1-\chi^2/4$. Together with Eq.~(\ref{eq_Fbar}), we can build the relationship between $\overline{\left\langle \chi^2 \right\rangle}$ and the scaled average infidelity $\kappa$ as
\begin{equation}
    \overline{\left\langle \chi^2 \right\rangle}= \frac{4\kappa}{N},\ \kappa = \frac{11}{8}.
    \label{eq-chi2-mean}
\end{equation}
The result can also be derived from the average moments of $(t^c,t^s)$ as
\begin{equation}
    \overline{\braket{\chi^2}} \approx \overline{\braket{\sin^2{\chi}}} = \overline{\braket{(t^c)^2+(t^s)^2}} = \overline{\braket{(t^c)^2}}+\overline{\braket{(t^s)^2}}= \frac{11}{2N} + O(\frac{1}{N^2})\approx \frac{4\kappa}{N},
\end{equation}
where we use the approximation $\sin{\chi}\approx\chi$ . We define $\sigma_{\chi}^2 = 2\kappa/N $, and find that
\begin{equation}
    \overline{\sigma_{t_c}^2}\approx \overline{\braket{(t^c)^2}} \approx
    \overline{\sigma_{t_s}^2}\approx
     \overline{\braket{(t^s)^2}}\approx
 \frac{\overline{\left\langle \chi^2 \right\rangle}}2\approx \sigma_{\chi}^2 = \frac{2\kappa}{N}.
 \label{eq-sigma-chi}
\end{equation}
\par
To gain an intuitive understanding of the results, we can approximate the joint PDF $p(t^c,t^s|N,U)$ as a bivariate Gaussian distribution $\mathcal{N}(\mu_{t^c},\mu_{t^s},\sigma^2_{t^c},\sigma^2_{t^s},r)$ with means $(\mu_{t^c},\mu_{t^s})$, variances $(\sigma^2_{t^c},\sigma^2_{t^s})$, and correlation coefficient $r$. Taking the average over $U$ yields an average distribution $p(t^c,t^s|N) = \overline{p(t^c,t^s|N,U)}$, which can be approximated as $\mathcal{N}(0,0,\sigma^2_{\chi},\sigma^2_{\chi},0)$ by neglecting terms that scale as $O(1/N^2)$. This implies that averaging over $SU(2)$ naturally generates a symmetric distribution of tomography errors, as depicted in Fig.~\ref{fig-tomo-error}. Then the average distributions of $\chi$ and $\zeta$ can be approximated by Rayleigh distribution and uniform distribution with PDFs as $\overline{p(\chi,\zeta|N,U)} \sim p(\chi|N)p(\zeta),\ p(\chi|N) = \chi/\sigma^2_{\chi}\exp{(-\chi^2/2\sigma^2_{\chi})},\ p(\zeta)=1/2\pi$.
\par
The above we have discussed is about $\chi_1$ and $\zeta_1$ in $\ket{\tilde{\psi}}$, and the same goes for $\chi_2$ and $\zeta_2$ in $\ket{\tilde{\phi}}$. With the invariant of Haar measure, we can show that 
\begin{equation}
\begin{split}
        \overline{p(\chi_2,\zeta_2|c,N,U,\varphi)} & = \frac{1}{2\pi}\int_U\int_0^{2\pi}p(\chi_2,\zeta_2|c,N,U,\varphi)dUd\varphi \\
        & = \frac{1}{2\pi}\int_{UW^{\dagger}}\int_0^{2\pi}p(\chi_2,\zeta_2|c,N,UW^{\dagger},\varphi)d(UW^{\dagger})d\varphi \\
        & = \int_U p(\chi_1,\zeta_1|N,U)dU = \overline{p(\chi_1,\zeta_1|N,U)},
\end{split}
\end{equation}
where $p(\chi_2,\zeta_2|c,N,U,\varphi) = p(\chi_1,\zeta_1|N,UW)$ and $W$ is defined in Eq.~(\ref{eq_W_matrix}). The average distributions of the tomography error for both $\ket{\psi}$ and $\ket{\phi}$ is identical, thanks to the reference-frame average.

\subsection{Tomography-Tomography strategy precision}
In this section, we derive the theoretical results for the average overlap estimation precision in the TT strategy. The overlap between two reconstructed states $\ket{\tilde{\psi}}$ and $\ket{\tilde{\phi}}$ in Eq.~(\ref{estimate_states}) is given by
\begin{equation}
\begin{split}
     \tilde{c}_{tt}=|\braket{\tilde{\phi}|\tilde{\psi}}|^2 =\ & \frac12
     \bigg[1+(2c-1)\cos{\chi_1}\cos{\chi_2} + 2\sqrt{c(1-c)}\big[\cos{(\zeta_1-\varphi)}\sin{\chi_1}\cos{\chi_2}+\cos{\zeta_2}\cos{\chi_1}\sin{\chi_2}\big] \\
     & - (2c-1)\cos{(\zeta_1-\varphi)}\cos{\zeta_2}\sin{\chi_1}\sin{\chi_2} - \sin{(\zeta_1-\varphi)}\sin{\zeta_2}\sin{\chi_1}\sin{\chi_2}\bigg].
\end{split}
\label{estimate_c}
\end{equation}
\par
\textbf{Average mean.} Firstly, we derive the average mean of the estimated overlap in TT strategy. With the error variables defined in Eq.~(\ref{eq-tc-ts}), we rewrite the expression of $\tilde{c}_{tt}$ by ignoring terms higher than the second order as
\begin{equation}
    \begin{split}
     \tilde{c}_{tt} \approx~& c +\sqrt{c(1-c)}\left(t_1^c\cos{\varphi}+t_1^s\sin{\varphi}+t_2^c\right) -\frac{2c-1}{4}(\chi_1^2 +\chi_2^2) \\
     & -\frac{2c-1}{2}(t_1^c\cos{\varphi}+t_1^s\sin{\varphi})t_2^c-\frac{1}{2}(t_1^c\sin{\varphi}-t_1^s\cos{\varphi})t_2^s,
\end{split}
\label{estimate-ctt-2}
\end{equation}
then the mean is given by
\begin{equation}
\begin{split}
        \overline{\langle \tilde{c}_{tt} \rangle} & =  \frac{1}{2\pi}\int_{U}\int_{\varphi}\int_{\zeta_1,\zeta_2,\chi_1,\chi_2} \tilde{c}_{tt} p(\chi_1,\zeta_1|N,U)p(\chi_2,\zeta_2|c,N,U,\varphi)dUd\varphi d\zeta_1 d\zeta_2 d\chi_1 d\chi_2\\
        & = c-\frac{2c-1}{4}\left(\overline{\braket{\chi_1^2}}+\overline{\braket{\chi_2^2}}\right)+O(\frac{1}{N^2}) \\
        & = c-\frac{2\kappa}{N}(2c-1)+O(\frac{1}{N^2}) ,
\end{split}
\label{eq_tt_mean}
\end{equation}
where we use $\overline{\braket{t^c_1}},\overline{\braket{t^s_1}},\overline{\braket{t^c_2}},\overline{\braket{t^s_2}},\braket{t_1^ct_2^c},\braket{t_1^st_2^c},\braket{t_1^ct_2^s},\braket{t_1^st_2^s}= O(1/N^2)$ and Eq.~(\ref{eq-chi2-mean}). Here, we show that the overlap estimator is asymptotically unbiased in TT strategy. 
\par
\textbf{Average variance.} With ignoring terms higher than the second order, the squared error can be expressed as
\begin{equation}
\begin{split}
     (\tilde{c}_{tt}-c)^2 & \approx c(1-c)\left(t_1^c\cos{\varphi}+t_1^s\sin{\varphi}+t_2^c\right)^2 
     \\ & = c(1-c)\left[
     (t_1^c)^2\cos^2{\varphi}+(t_1^s)^2\sin^2{\varphi}+(t_2^c)^2+t_1^ct_1^s\sin{2\varphi}+ 2t_1^ct_2^c\cos{\varphi}+2t_1^st_2^c\sin{\varphi} 
     \right].
\end{split}
\end{equation}
The average variance of overlap for TT strategy is then given by
\begin{equation}
\begin{split}
     v_{tt} = \overline{\left \langle(\tilde{c}_{tt}-c)^2 \right \rangle} &
     = c(1-c)\left( \frac{1}{2}\overline{\braket{(t^c_1)^2}} +\frac{1}{2}\overline{\braket{(t^s_1)^2}} + \overline{\braket{(t^c_2)^2}}  \right) +O(\frac{1}{N^2}) \\
     & =\frac{c(1-c)}{2}\left(\overline{\braket{\chi_1^2}}+\overline{\braket{\chi_2^2}}   \right) +O(\frac{1}{N^2}) \\
     & = \frac{4\kappa c(1-c)}{N} +O(\frac{1}{N^2}).
\end{split}
\label{vtt_1}
\end{equation}
with Eq.~(\ref{eq-sigma-chi}). We have shown a direct relationship between the tomography infidelity and overlap estimation precision in the TT strategy. The overlap estimation error in the TT strategy originates from tomography error, and both errors scale as $O(1/N)$. 

\textbf{Fisher information.} We treat $\tilde{c}$ as a random variable and aim to determine its PDF $p(\tilde{c}|c,N)$ which depends on the true overlap $c$. After averaging over $U$, we approximate the average tomography error distribution $p(t^c,t^s|N)$ as the bivariate Gaussian distribution $\mathcal{N}(0,0,\sigma^2_{\chi},\sigma^2_{\chi},0)$. Then $\chi_j$ will follow the Rayleigh distribution, while $\zeta_1-\varphi$ and $\zeta_2$ follow the uniform distribution. $t_1^{\prime} = \chi_1\cos{(\zeta_1-\varphi)}$ and $t_2^{\prime} = \chi_2\cos{\zeta_2}$ are independent and both follow the same Gaussian distribution with the PDF $p(t_j^{\prime}) = 1/\sqrt{2\pi}\sigma_{\chi}\exp(-(t_j^{\prime})^2/2\sigma_{\chi}^2),\ j=1,2$ . By retaining only the first-order terms, we can rewrite Eq.~(\ref{estimate_c}) as follows:
\begin{equation}
    \frac{\tilde{c}_{tt}-c}{\sqrt{c(1-c)}} \approx \chi_1\cos{(\zeta_1-\varphi)}+\chi_2\cos{\zeta_2} = t_1^{\prime}+t_2^{\prime} = T \sim \mathcal{N}(0,2\sigma_{\chi}^2),
\end{equation}
where $\mathcal{N}(\mu,\sigma^2)$ denotes the Gaussian distribution with mean $\mu$ and variance $\sigma^2$. We can observe that the estimated overlap $\tilde{c}_{tt}$ will asymptotically follow a Gaussian distribution $\mathcal{N}(c,2c(1-c)\sigma_{\chi}^2)$. Furthermore, we can also rederive the variance $v_{tt}$ as mentioned in Eq.~(\ref{vtt_1}). The Fisher information (FI) of overlap in TT strategy is then given by
\begin{equation}
\begin{split}
        I_{tt}(c)  =~& \left \langle \left(\frac{\partial \log{p(\tilde{c}|c,N)}}{\partial c}\right)^2 \right \rangle_{\tilde{c}} \\ 
         =~& \frac{(c-2(1-c)(2c-1)\sigma_{\chi}^2)^2}{16c^2(1-c)^4\sigma_{\chi}^4} + \frac{c-2(1-c)(2c-1)\sigma_{\chi}^2}{4c(1-c)^4\sigma_{\chi}^4} \left \langle \tilde{c} \right \rangle_{\tilde{c}}\\
         & + \frac{c(2c^2+2c-1)-2(1-c)(2c-1)^2\sigma^2_{\chi}}{8c^3(1-c)^4\sigma_{\chi}^4} \left \langle \tilde{c}^2 \right \rangle_{\tilde{c}}
        - \frac{(2c-1)}{4c^2(1-c)^4\sigma_{\chi}^4} \left \langle \tilde{c}^3 \right \rangle_{\tilde{c}} + \frac{(2c-1)^2}{16c^4(1-c)^4\sigma_{\chi}^4} \left \langle \tilde{c}^4 \right \rangle_{\tilde{c}} \\
         =~& \frac{1-4c(1-c)}{2c^2(1-c)^2} + \frac{1}{2c(1-c)\sigma_{\chi}^2}
         = \frac{1-4c(1-c)}{2c^2(1-c)^2} + \frac{N}{4\kappa c(1-c)} \approx \frac{N}{4\kappa c(1-c)}
\end{split}
\end{equation}
where we use the expressions of high order moments of Gaussian distribution in the third equal sign.
\subsection{Tomography-Projection strategy precision}
In this section, we derive the theoretical results for the average overlap estimation precision in the TP strategy. After the tomography procedure, we get the reconstructed state $\ket{\tilde{\phi}}$ as in Eq.~(\ref{estimate_states}). Then in the projection procedure, the probability of successfully projecting the state $\ket{\psi}$ onto $\ket{\tilde{\phi}}$ is given by
\begin{equation}
    p_{tp} = |\braket{\psi|\tilde{\phi}}|^2=\frac12\big[ 1 + (2c-1)\cos{\chi_2}+ 2\sqrt{c(1-c)}\cos{\zeta_2}\sin{\chi_2} \big].
\end{equation}
The expression of $p_{tp}$ can be expand to the second order as follows:
\begin{equation}
    p_{tp} \approx c + \sqrt{c(1-c)}t_2^c - \frac{2c-1}{4}\chi_2^2.
    \label{eq_ptp}
\end{equation}
 After conducting $N$ trials of projection, the number of successful projections follows a binomial distribution, denoted as $k \sim \text{Bin}(k,N, p_{tp})$. The overlap estimator in TP strategy is given by $\tilde{c}_{tp}=k/N$.
\par
\textbf{Average mean.} The average mean of $\tilde{c}_{tp}$ is given by
\begin{equation}
\begin{split}
    \overline{\braket{\tilde{c}_{tp}}} & 
    = \int_U\int_{\zeta_2,\chi_2}\left\langle \frac{k}{N}\right\rangle_{k}p(\chi_2,\zeta_2|N,U)dU d\zeta_2 d\chi_2 = \overline{\left\langle p_{tp} \right\rangle} \\
    & = c- \frac{2c-1}{4}\overline{\braket{\chi_2^2}} = c-\frac{\kappa}{N}(2c-1) + O(\frac{1}{N^2}),
\end{split}
\end{equation}
where $\langle\cdot\rangle_k$ denotes the expectation with respect to $k$ according to the binomial distribution and $\langle\cdot\rangle$ represents the expectation with respect to variables in $p_{tp}$. We neglect the integration respect to $\varphi$ due to the reference-frame average. Here, we show that the overlap estimator is asymptotically unbiased in TP strategy.
\par
\textbf{Average variance.} The average variance of TP strategy is given by
\begin{equation}
\begin{split}
v_{tp} = \int_U\int_{\zeta_2,\chi_2}\left\langle (\frac{k}{N}-c)^2\right\rangle_{k}p(\chi_2,\zeta_2|N,U)dUd\zeta_2 d\chi_2. \\
\end{split}
    \label{tp_error}
\end{equation}
From the binomial distribution, the expectation of squared error with respect to $k$ can be derived as
\begin{equation}
\begin{split}
        \left\langle (\frac{k}{N}-c)^2\right\rangle_{k} = & \frac{1}{N^2}\left\langle k^2\right\rangle_{k}-\frac{2c}{N}\left\langle k\right\rangle_{k}+c^2 
        = \frac{1}{N^2}\cdot \left[Np_{tp}(1-p_{tp}+N^2p^2_{tp})\right]-\frac{2c}{N}\cdot Np_{tp}+c^2 \\
        = & \frac{p_{tp}(1-p_{tp})}{N}+(p_{tp}-c)^2.
\end{split}
\label{Ek_expand}
\end{equation}
The expectations of $p_{tp}(1-p_{tp})$ and $(p_{tp}-c)^2$ in the average case are given by
\begin{equation}
\begin{split}
\overline{\braket{p_{tp}(1-p_{tp})}} & = c(1-c) + O(\frac{1}{N}), \\
 \overline{\left\langle(p_{tp}-c)^2\right\rangle} & = c(1-c)\overline{\braket{(t_2^c)^2}} + O(\frac{1}{N^2}) = \frac{2\kappa c(1-c)}{N}+ O(\frac{1}{N^2}).
\end{split}
\end{equation}
Then $v_{tp}$ in Eq.~(\ref{tp_error}) can be derived as
\begin{equation}
    \begin{split}
        v_{tp} = \overline{\left\langle \frac{p_{tp}(1-p_{tp})}{N}\right\rangle}+\overline{\left\langle(p_{tp}-c)^2\right\rangle}
        = \frac{(1+2\kappa)c(1-c)}{N}+O(\frac{1}{N^2}).
    \end{split}
    \label{eq_vtp}
\end{equation} 
The variance of TP strategy can be decomposed into two components. The first component, $\overline{\braket{p_{tp}(1-p_{tp})}}/N$, denotes the error introduced by the finite number of projections. The second component, $\overline{\braket{(p_{tp}-c)^2}}$, represents the error introduced by the deviation of $\ket{\tilde{\phi}}$ from $\ket{\phi}$ during the tomography procedure.
\par
\textbf{Fisher information.} In TP strategy, the probability distribution of $k$ after averaging respect to $\chi_2$ and $\zeta_2$ can be expressed as:
\begin{equation}
    p(k|c,N) = \overline{\left\langle \binom{N}{k}p_{tp}^k(1-p_{tp})^{N-k}  \right\rangle},
\end{equation}
The FI of overlap in TP strategy with $N$ copies of qubit pairs is then given by
\begin{equation}
    I_{tp}(c) = \left\langle \left(\frac{\partial \log{p(k|c,N)}}{\partial c}\right)^2  \right\rangle_{k}.
    \label{FI_tp}
\end{equation}
It is challenging to calculate this expression analytically. Instead, we numerically evaluate Eq.~(\ref{FI_tp}) with $N=900$, which corresponds to our experimental setting. The numerical results are shown in Fig.~\ref{fig:FI_TP_num}, indicating that the estimator in the TP strategy saturates the Cramér-Rao bound.
\begin{figure}[!ht]
    \centering
    \includegraphics[width=0.35\textwidth]{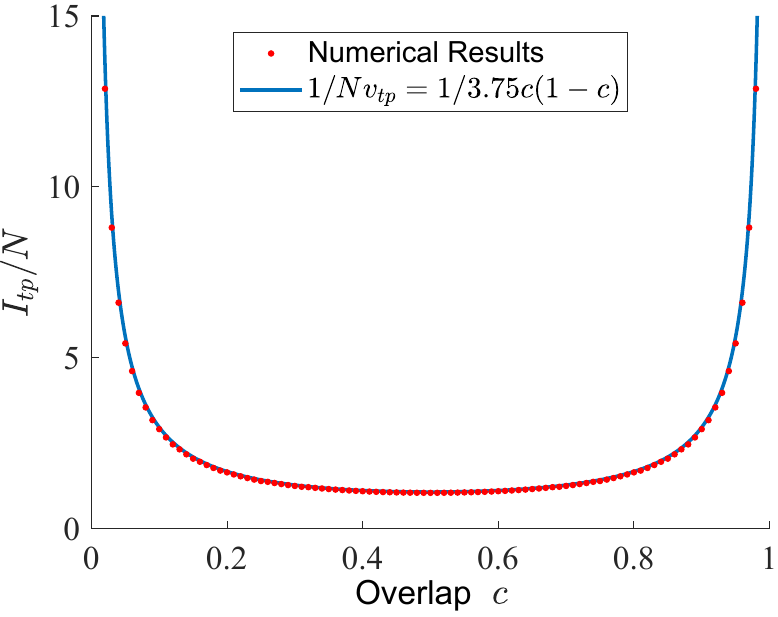}
    \caption{Fisher information of overlap in Tomography-Projection strategy. Numerical results (red dots) are given by calculating the unit FI $I_{tp}/N$ with $N = 900$. The inverse of scaled average variance (blue line) in TP strategy is derived form Eq.~(\ref{eq_vtp}).}
    \label{fig:FI_TP_num}
\end{figure}
\subsection{Numerical results of average variance for TT and TP strategies}
In this section, we show the numerical verification of the theoretical results of average overlap estimation variance in TT and TP strategies. In TP strategy, for a given reconstructed state $\ket{\tilde{\phi}}$, the probability of recording $k$ successful projection out of $N$ measurements is given by
\begin{equation}
    P_{proj}(k|\tilde{\phi},N)=\binom{N}{k}(|\braket{\tilde{\phi}|\psi}|^2)^k(1-|\braket{\tilde{\phi}|\psi}|^2)^{N-k}.
\end{equation}
Under a specific choice of $U$ and $\varphi$, the expressions of overlap estimation variance for TT and TP strategy are given by
\begin{footnotesize}
    \begin{align}
    v_{tt}(c,N|U,\varphi) = & \sum_{n_x^1,n_y^1,n_z^1=0}^{N/3}\sum_{n_x^2,n_y^2,n_z^2=0}^{N/3}\left(|\braket{\tilde{\phi}|\tilde{\psi}}|^2-c \right)^2 \times P_{tomo}(\tilde{\phi}(n_x^1,n_y^1,n_z^1)|c,N,U,\varphi)\times P_{tomo}(\tilde{\psi}(n_x^2,n_y^2,n_z^2)|N,U), \label{eq.vtt}
    \\
    v_{tp}(c,N|U,\varphi) = & \sum_{n_x,n_y,n_z=0}^{N/3}\left[ \sum_{k=0}^{N}(\frac{k}{N}-c)^2P_{proj}(k|\tilde{\phi},N)\right]\times P_{tomo}(\tilde{\phi}(n_x,n_y,n_z)|c,N,U,\varphi), \label{eq.vtp}         
\end{align}
\end{footnotesize}
where $P_{tomo}$ is defined in Eq.~(\ref{eq_Ptomo}). By averaging the variance with respect to $U$ and $\varphi$ and multiplying by the copy number $N$, the scaled average variance can be expressed as
\begin{equation}
    Nv(c)=\frac{N}{2\pi}\int_U\int_{0}^{2\pi}v(c,N|U,\varphi)dUd{\varphi},
    \label{eq:average}
\end{equation}
where $v(c,N|\varphi,U)$ is given by Eq.~\eqref{eq.vtt} for TT strategy and Eq.~\eqref{eq.vtp} for TP strategy. We employ Monte Carlo integration technique to numerically compute the integration in Eq.~\eqref{eq:average} for both the TT and TP strategies. For a given overlap value and an estimation strategy, we randomly sample 1000 pairs of $U$ and $\varphi$ using the Haar measure, and then calculate the average variance as the integration value using these samples. As illustrated in Fig.~\:\ref{fig:all_num}, we select 41 overlap values for both TT and TP to calculate the scaled average variance with $N = 900$ copies. We obtain the numerical expressions of $Nv(c)$ by using $Nv(c) = \alpha c(1-c)+\beta$ as the fitting formula. With a large $N$, the undetermined coefficient $\beta = O(1/N)$ term can be ignored.
\begin{figure}[!ht]
    \centering
    \includegraphics[width=0.65\textwidth]{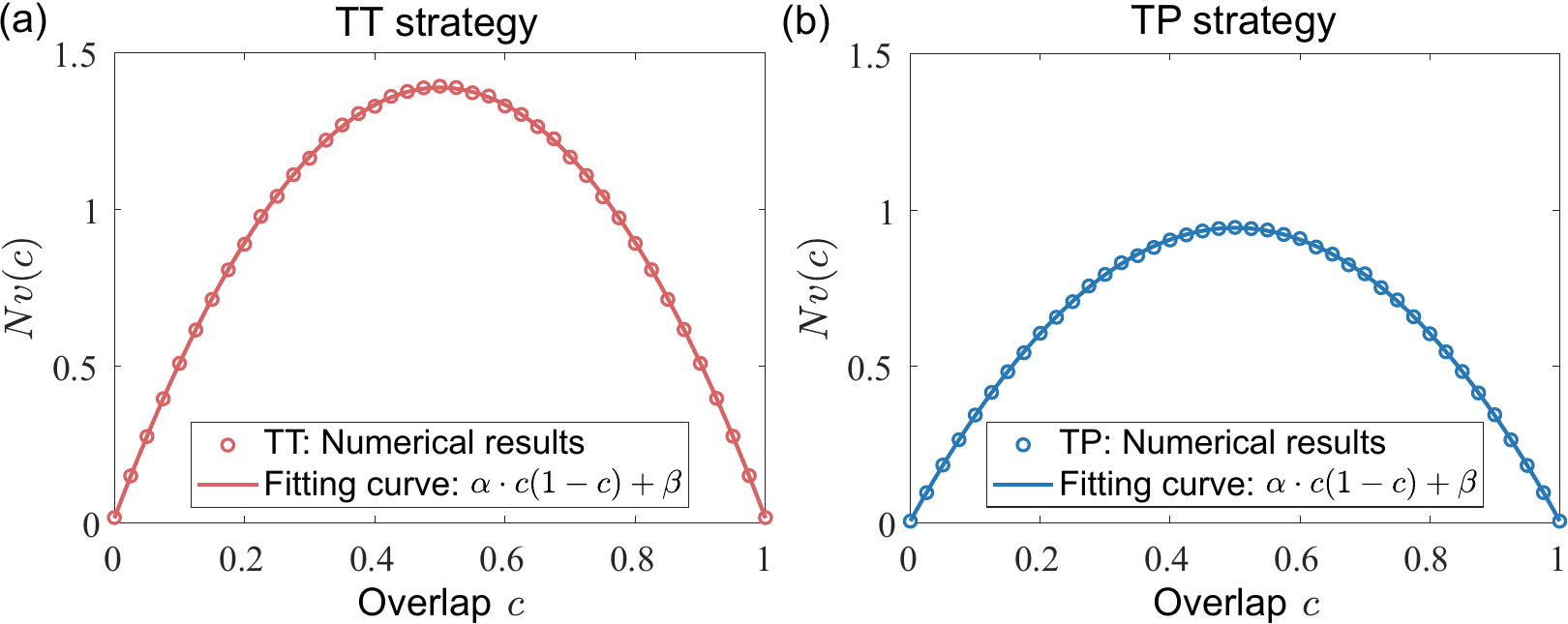}
    \caption{Numerical calculation results and the fitting curves of the scaled average variance $Nv(c)$ in two local strategies. \textbf{a} Tomography-Tomography (TT) strategy. The fitting expression is $Nv(c) = 5.496c(1-c)+0.016$ with the $95\%$ confidence bound of $\alpha$ as $(5.484,5.507)$. \textbf{b} Tomography-Projection (TP) strategy. The fitting expression is $Nv(c) = 3.746c(1-c)+0.006$ with the $95\%$ confidence bound of $\alpha$ as $(3.737,3.754)$. The coefficients of determination $R^2$ for both fittings are 1.000.}
    \label{fig:all_num}
\end{figure}
\par
\subsection{TT and TP strategies for high-dimensional quantum states}
The analysis of the precision of TT and TP strategies for estimating the overlap of two unknown qubits can be generalized to high-dimensional quantum states. As discussed in Section~\ref{Section-I},
without loss of generality, we can express two $d$-dimensional quantum states (qudits) $\ket{\psi},\ \ket{\phi}\in \mathbb{C}^d$ as:
\begin{equation}
\begin{split}
        \ket{\psi} & =U\ket{0} =UW\left(\sqrt{c}\ket{0}+\sqrt{1-c}\left|\phi^{r,0}_\perp\right\rangle \right) =\sqrt{c}\ket{\phi}+\sqrt{1-c}\ket{\phi^r_\perp}, \\\quad\ket{\phi} & = UW\ket{0}= U\left(\sqrt{c}\ket{0}+\sqrt{1-c}\left|\psi^{r,0}_\perp\right\rangle \right) = \sqrt{c}\ket{\psi}+\sqrt{1-c}\ket{\psi^r_\perp},
\end{split}
\label{expand_qudits}
\end{equation}
where we assume a real-valued unsquared overlap $\langle\psi|\phi\rangle= \braket{0|W|0} =\sqrt{c}$. Here, $|\psi_\perp^r\rangle$ and $|\phi_\perp^r\rangle$ are orthogonal to $|\psi\rangle$ and $|\phi\rangle$, respectively, and satisfy $\langle\phi^r_\perp|\psi^r_\perp\rangle = -\sqrt{c}$. Both $|\psi^{r,0}_\perp\rangle$ and $|\phi^{r,0}_\perp\rangle$ are orthogonal to $\ket{0}$.
\par
We consider a sufficient-copy scenario for tomography ($N \gg d$), ensuring the tomography fidelity approaching 1. In this regime, a general tomography approach reconstructs a qudit with an average fidelity of $\overline{F} = 1 - \kappa/N + O(1/N^2)$. Our analysis focuses on the leading-order term, $\kappa/N$, which dominates the average infidelity for large $N$. Similar to Eq.~\eqref{estimate_states}, the reconstructed states for $\ket{\psi}$ and $\ket{\phi}$, along with their corresponding probability distributions, can be expressed as
\begin{equation}
\begin{split}
        \ket{\tilde{\psi}} & = \cos{\frac{\chi_1}{2}}\ket{\psi} + \sin{\frac{\chi_1}{2}}\ket{\psi^{t}_{\perp}} = U\left(\cos{\frac{\chi_1}{2}}\ket{0} + \sin{\frac{\chi_1}{2}}\ket{\psi^{t,0}_{\perp}}  \right),\quad p(\chi_1,\ket{\psi^t_\perp}|N,U),\\ 
        \ket{\tilde{\phi}} & = \cos{\frac{\chi_2}{2}}\ket{\phi} + \sin{\frac{\chi_2}{2}}\ket{\phi^{t}_{\perp}}=UW\left(\cos{\frac{\chi_2}{2}}\ket{0} + \sin{\frac{\chi_2}{2}}\ket{\phi^{{t,0}}_{\perp}}\right),\quad p(\chi_2,\ket{\phi^{t}_{\perp}}|N,UW),
\end{split}
\label{estimate_qudits}
\end{equation}
where $\ket{\psi^{t}_{\perp}}$ and $\ket{\phi^{t}_{\perp}}$ are orthogonal to $\ket{\psi}$ and $\ket{\phi}$, respectively, residing within two $(d-1)$-dimensional subspaces. The average fidelity of tomography implies $\overline{\left\langle \chi_1^2 \right\rangle}=\overline{\left\langle \chi_2^2 \right\rangle} = 4\kappa/N+O(1/N^2)$. Note that $\ket{\psi^{r,0}_\perp}$ from Eq.~\eqref{expand_qudits} and $|\psi^{t,0}_\perp\rangle$ from Eq.~\eqref{estimate_qudits} both reside in the $(d-1)$-dimensional subspace orthogonal to $\ket{0}$, spanned by $\{\ket{1},\ket{2},..,\ket{d-1}\}$. We consider an approximation of the joint PDF of $\chi_1$ and $\ket{\psi_\perp^{t,0}}$ after averaging over $U$. In the reconstructed state $\ket{\tilde{\psi}}$, the parameter $\chi_1$ denotes the magnitude of tomography error, and $\ket{\psi_\perp^t}$ represents the error direction in the subspace orthogonal to $\ket{\psi}$. Averaging over the Haar-distributed of $\ket{\psi}$ makes the error direction isotropic and independent of the error magnitude. Therefore, for a sufficiently large number of copies $N$ used for tomography, we approximate the distribution of $\ket{\psi_\perp^{t,0}}=U^{\dagger}\ket{\psi_\perp^{t}}$ as Haar distributed in the $(d-1)$-dimensional subspace orthogonal to $\ket{0}$, which means $\int_Up(\chi_1,\ket{\psi^{t}_\perp}|N,U) dU \approx p_{\chi}(\chi_1|N)$. Then, suppose there is a function with the form $f(\chi_1,\braket{\psi_\perp^{r}|\psi_\perp^{t}})=f_1(\chi_1)f_2(\braket{\psi_\perp^{r}|\psi_\perp^{t}})$, the average of $f(\chi_1,\braket{\psi_\perp^{r}|\psi_\perp^{t}})$ can be shown as 
\begin{equation}
    \overline{\left\langle f_1(\chi_1)f_2(\braket{\psi_\perp^{r}|\psi_\perp^{t}})\right\rangle} \approx \overline{\left\langle f_1(\chi_1)\right\rangle}\cdot\overline{\left\langle f_2(\braket{\psi_\perp^{r,0}|\psi_\perp^{t,0}})\right\rangle}.
\end{equation}
For example, the following holds:
\begin{equation}
\begin{split}
        \overline{\left\langle{\chi^2_1}\left|\braket{\psi_\perp^{r}|\psi_\perp^{t}}\right|^2\right\rangle} & =  \int_U\int_{\chi_1}\int_{\psi^{t}_\perp}\chi^2_1\left|\braket{\psi_\perp^{r}|\psi_\perp^{t}}\right|^2 p\left(\chi_1,\ket{\psi^t_\perp}|N,U\right) dUd\chi_1d\psi^{t}_\perp,\\
        & = \int_{\chi_1}\int_{\psi^{t,0}_\perp\in\mathbb{C}^{d-1}}\chi^2_1\left|\braket{\psi_\perp^{r,0}|\psi_\perp^{t,0}}\right|^2 \left(\int_Up(\chi_1,U\ket{\psi^{t,0}_\perp}|N,U) dU\right) d\chi_1d\psi^{t,0}_\perp,\\
        & \approx  \int_{\chi_1}\chi^2_1p_{\chi}(\chi_1|N)d\chi_1\int_{\psi^{t,0}_\perp\in\mathbb{C}^{d-1}}\left|\braket{\psi_\perp^{r,0}|\psi_\perp^{t,0}}\right|^2 d\psi^{t,0}_\perp \\
        & = \overline{\braket{\chi_1^2}}\overline{\left|\braket{\psi_\perp^{r,0}|\psi_\perp^{t,0}}\right|^2},
\end{split}
\end{equation}
where $d\psi^{t}_\perp$ and $d\psi^{t,0}_\perp$ denote the Haar measure. We use $\braket{\psi_\perp^{r}|\psi_\perp^{t}} = \braket{\psi_\perp^{r,0}|\psi_\perp^{t,0}}$ and approximate the joint PDF of $\chi_1$ and $\ket{\psi_\perp^{t,0}}$, noting that $\braket{\psi_\perp^{r,0}|\psi_\perp^{t,0}}$ is independent of $U$. We further consider the average inner product between $\ket{\psi^{r,0}_\perp}$ and $\ket{\psi^{t,0}_\perp}$ as
\begin{equation}
\begin{split}
        \overline{\braket{\psi^{r,0}_\perp|\psi^{t,0}_\perp}} 
        & = \int_{\psi^{t,0}_\perp\in \mathbb{C}^{d-1}}\left(\int_{\omega}\braket{\psi^{r,0}_\perp|\psi^{t,0}_\perp}dW\right)\left(\int_U\int_{\chi_1}p(\chi_1,U\ket{\psi^{t,0}_\perp}|N,U)dUd\chi_1\right)d\psi^{t,0}_\perp, \\
        & \approx  \int_{\psi^{t,0}_\perp\in \mathbb{C}^{d-1}}\left(\int_{\omega}\braket{\psi^{r,0}_\perp|\psi^{t,0}_\perp}dW\right)d\psi^{t,0}_\perp \\
        & = \int_{\psi^{r,0}_\perp\in \mathbb{C}^{d-1}}\int_{\psi^{t,0}_\perp\in \mathbb{C}^{d-1}}\braket{\psi^{r,0}_\perp|\psi^{t,0}_\perp}d\psi^{r,0}_\perp d\psi^{t,0}_\perp \\
        & = 0, 
\end{split}
\label{eq_psirt}
\end{equation}
where $d\psi^{r,0}_\perp$ and $d\psi^{t,0}_\perp$ denote the Haar measure.  The second equality in Eq.~\eqref{eq_psirt} arises because $\ket{\psi^{r,0}_\perp}$ depends only on $W$ subject to the constraints $\braket{0|W|0}=\sqrt{c}$ and $\braket{0|\psi^{r,0}_\perp}=0$. Therefore, integrating over $W$ (i.e., $\int_{\omega}\braket{\psi^{r,0}_\perp|\psi^{t,0}_\perp}dW$) is equivalent to integrating over the Haar-random $\ket{\psi^{r,0}_\perp}$ in the $(d-1)$-dimensional subspace. This result reflects the fact that the average inner product of two Haar-random, $(d-1)$-dimensional states is zero, i.e., $\overline{\braket{\psi^{r,0}_\perp|\psi^{t,0}_\perp}}=0$. For the (squared) overlap between $\ket{\psi^{r,0}_\perp}$ and $\ket{\psi^{t,0}_\perp}$, we have 
\begin{equation}
    \begin{split}
        \overline{\left|\braket{\psi^{r,0}_\perp|\psi^{t,0}_\perp}\right|^2} 
        \approx \int_{\psi^{t,0}_\perp\in \mathbb{C}^{d-1}}\int_{\psi^{r,0}_\perp\in\mathbb{C}^{d-1}}\left|\braket{\psi^{r,0}_\perp|\psi^{t,0}_\perp}\right|^2d\psi^{r,0}_\perp d\psi^{t,0}_\perp = \frac{1}{d-1},
\end{split}
\end{equation}
which represents the average overlap between two Haar-random states in the $(d-1)$-dimensional subspace orthogonal to $\ket{0}$~\cite{Mele2024introductiontohaar}. Due to the Haar measure, we can express $\braket{\psi^{r,0}_\perp|\psi^{t,0}_\perp}$ as $|\braket{\psi^{r,0}_\perp|\psi^{t,0}_\perp}|e^{i\zeta}$ where $\zeta$ is uniformly distributed. Consequently, $\overline{\left(\braket{\psi^{r,0}_\perp|\psi^{t,0}_\perp}\right)^2}=\overline{\left(\braket{\psi^{t,0}_\perp|\psi^{r,0}_\perp}\right)^2}=0$. Therefore, $\overline{(\text{Re}[\braket{\psi^{r,0}_\perp|\psi^{t,0}_\perp}])^2}=1/2(d-1)$. Analogous results for $\chi_2$, $\braket{\phi_\perp^{r}|\phi_\perp^{t}}$ and $\braket{\phi_\perp^{r,0}|\phi_\perp^{t,0}}$ can be obtained using the same approach. In summary, the following relationships hold:
\begin{equation}
    \begin{split}
        \overline{\braket{\psi_\perp^{r,0}|\psi_\perp^{t,0}}} = \overline{\braket{\phi_\perp^{r,0}|\phi_\perp^{t,0}}} = \overline{\text{Re}\left[\braket{\psi_\perp^{r,0}|\psi_\perp^{t,0}}\right]} =\overline{\text{Re}\left[\braket{\phi_\perp^{r,0}|\phi_\perp^{t,0}}\right]} & = 0,
        \\\overline{\left|\braket{\psi_\perp^{r,0}|\psi_\perp^{t,0}}\right|^2} = \overline{\left|\braket{\phi_\perp^{r,0}|\phi_\perp^{t,0}}\right|^2}=2\overline{\left(\text{Re}\left[\braket{\psi_\perp^{r,0}|\psi_\perp^{t,0}}\right]\right)^2} = 2\overline{\left(\text{Re}\left[\braket{\phi_\perp^{r,0}|\phi_\perp^{t,0}}\right]\right)^2} & = \frac{1}{d-1}.
\end{split}
\label{eq_rt_relation}
\end{equation}
\par
For the TT strategy, the overlap estimator is given by
\begin{equation}
\begin{split}
     \tilde{c}_{tt} & =|\braket{\tilde{\psi}|\tilde{\phi}}|^2\\
     & =\left|\cos{\frac{\chi_1}{2}}\cos{\frac{\chi_2}{2}}\braket{\psi|\phi}+\cos{\frac{\chi_1}{2}\sin{\frac{\chi_2}{2}}\braket{\psi|\phi_\perp^t} +\sin{\frac{\chi_1}{2}}\cos{\frac{\chi_2}{2}}\braket{\psi_\perp^t|\phi}+\sin{\frac{\chi_1}{2}}\sin{\frac{\chi_2}{2}}}\braket{\psi_\perp^t|\phi_\perp^t}\right|^2\\
     & =\left|\cos{\frac{\chi_1}{2}}\cos{\frac{\chi_2}{2}}\sqrt{c}+\cos{\frac{\chi_1}{2}\sin{\frac{\chi_2}{2}}\sqrt{1-c}\braket{\phi_\perp^r|\phi_\perp^t} +\sin{\frac{\chi_1}{2}}\cos{\frac{\chi_2}{2}}\sqrt{1-c}\braket{\psi_\perp^t|\psi_\perp^r}+\sin{\frac{\chi_1}{2}}\sin{\frac{\chi_2}{2}}}\braket{\psi_\perp^t|\phi_\perp^t}\right|^2\\
     & \approx \left|\left(1-\frac{\chi_1^2+\chi_2^2}{8}\right)\sqrt{c}+\frac{\sqrt{1-c}}{2}\left(\chi_2\braket{\phi_\perp^r|\phi_\perp^t}+\chi_1\braket{\psi_\perp^t|\psi_\perp^r}\right)+\frac{\chi_1\chi_2}{4}\braket{\psi_\perp^t|\phi_\perp^t}\right|^2 \\
     & \approx \left(1-\frac{\chi_1^2+\chi_2^2}{4}\right)c + \left(\chi_1\text{Re}[\braket{\psi_\perp^r|\psi_\perp^t}]+\chi_2\text{Re}[\braket{\phi_\perp^r|\phi_\perp^t}]\right)\sqrt{c(1-c)}  + \left(\chi_1^2|\braket{\psi_\perp^r|\psi_\perp^t}|^2 + \chi_2^2|\braket{\phi_\perp^r|\phi_\perp^t}|^2\right)\frac{1-c}{4}
      \\
     &\ \ +\chi_1\chi_2\text{Re}\left[\braket{\psi_\perp^t|\psi_\perp^r}\braket{\phi_\perp^r|\phi_\perp^t}  \right]\frac{1-c}{2}+\frac{\chi_1\chi_2}{2}\text{Re}\left[\braket{\psi_\perp^t|\phi_\perp^t}\right]\sqrt{c},
\end{split}
\label{estimate-ctt-qudit}
\end{equation}
where terms higher than second order in $\chi_i$ and $\chi_j$ have been neglected. We further assume that $\ket{\psi_\perp^{t,0}}$ and $\ket{\phi_\perp^{t,0}}$ as independent and identically distributed, meaning $\int_Up(\chi_1,\psi^t_\perp|N,U)p(\chi_2,\phi^{t}_{\perp}|N,UW)dU\approx p_{\chi}(\chi_1|N)p_{\chi}(\chi_2|N)$. We expect the approximation applied to the probability distribution to introduce only higher-order errors, scaling as $O(1/N^2)$. Consequently, we can derive that $\overline{\braket{\psi_\perp^t|\psi_\perp^r}\braket{\phi_\perp^r|\phi_\perp^t}}=0$ and $ \overline{\braket{\psi_\perp^t|\phi_\perp^t}}=0$. The average mean of the TT strategy is then given by:
\begin{equation}
    \overline{\braket{\tilde{c}_{tt}}} = c-\frac{dc-1}{4(d-1)}\left(\overline{\braket{\chi_1^2}}+\overline{\braket{\chi_2^2}}\right)+O(\frac{1}{N^2})=c-\frac{2\kappa(dc-1)}{N(d-1)}+O(\frac{1}{N^2}),
\end{equation}
which is the general form of Eq.~\eqref{eq_tt_mean}. Considering only the first-order terms of $\chi_1$ and $\chi_2$ in Eq.~\eqref{estimate-ctt-qudit}, together with Eq.~\eqref{eq_rt_relation}, the average variance of the TT strategy is:
\begin{equation}
\begin{split}
        v_{tt}=\overline{\braket{(\tilde{c}_{tt}-c)^2}} & = c(1-c)\overline{\left(\chi_1\text{Re}[\braket{\psi_\perp^r|\psi_\perp^t}]+\chi_2\text{Re}[\braket{\phi_\perp^r|\phi_\perp^t}]\right)^2} +O(\frac{1}{N^2}) \\
        & =  c(1-c)\left(\overline{\braket{\chi_1^2}}\overline{\left(\text{Re}\left[\braket{\psi_\perp^{r,0}|\psi_\perp^{t,0}}\right]\right)^2}+\overline{\braket{\chi_2^2}}\overline{\left(\text{Re}\left[\braket{\phi_\perp^{r,0}|\phi_\perp^{t,0}}\right]\right)^2} \right) +O(\frac{1}{N^2}) \\
        &=\frac{4\kappa c(1-c)}{(d-1)N} +O(\frac{1}{N^2}).
\end{split}
\end{equation}
For $d=2$, this reduces to the qubit case in Eq.~\eqref{vtt_1}.
\par
For the TP strategy, the success projection probability is given by
\begin{equation}
\begin{split}
        p_{tp} & = |\braket{\psi|\tilde{\phi}}|^2= \left| \cos{\frac{\chi_2}{2}}\sqrt{c}+\sin{\frac{\chi_2}{2}}\sqrt{1-c}\braket{\phi_\perp^r|\phi_\perp^t}\right|^2 \\
        & = \cos^2{\frac{\chi_2}{2}}\cdot c + \sin{\chi_2}\text{Re}\left[\braket{\phi_\perp^r|\phi_\perp^t}\right]\sqrt{c(1-c)}+\sin^2{\frac{\chi_2}{2}}(1-c)\left|\braket{\phi_\perp^r|\phi_\perp^t}\right|^2 \\
        & \approx c + \chi_2\text{Re}\left[\braket{\phi_\perp^r|\phi_\perp^t}\right]\sqrt{c(1-c)}-\frac{\chi_2^2}{4}\left[\left(\left|\braket{\phi_\perp^r|\phi_\perp^t}\right|^2+1\right)c-\left|\braket{\phi_\perp^r|\phi_\perp^t}\right|^2\right],
\end{split}
\end{equation}
which is the general form of Eq.~\eqref{eq_ptp}. The average mean of the TP strategy is
\begin{equation}
    \begin{split}
    \overline{\braket{\tilde{c}_{tp}}} & = \overline{\left\langle p_{tp} \right\rangle}
    = c - \frac{\overline{\braket{\chi_2^2}}}{4}\frac{dc-1}{d-1}+ O(\frac{1}{N^2}) = c - \frac{\kappa (dc-1)}{N(d-1)}+ O(\frac{1}{N^2}).
\end{split}
\end{equation}
Similar to the qubit case, the average variance of the TP strategy can be decomposed into two parts:
\begin{equation}
    \begin{split}
\overline{\braket{p_{tp}(1-p_{tp})}} & = c(1-c) + O(\frac{1}{N}), \\
 \overline{\left\langle(p_{tp}-c)^2\right\rangle} & = c(1-c)\overline{\braket{\chi_2^2}}\overline{\left(\text{Re}\left[\braket{\phi_\perp^{r,0}|\phi_\perp^{t,0}}\right]\right)^2} + O(\frac{1}{N^2}) = \frac{2\kappa c(1-c)}{(d-1)N}+ O(\frac{1}{N^2}).
\end{split}
\end{equation}
The average variance for TP strategy is then
\begin{equation}
\begin{split}
        v_{tp} = \overline{\left\langle \frac{p_{tp}(1-p_{tp})}{N}\right\rangle}+\overline{\left\langle(p_{tp}-c)^2\right\rangle}=\left(\frac{2\kappa}{d-1}+1\right)\frac{c(1-c)}{N}+O(\frac{1}{N^2}).
\end{split}
\end{equation}
For $d=2$, this reduces to the qubit case in Eq.~\eqref{eq_vtp}.
\par
The overlap estimation errors for the TT and TP strategies can be also decomposed into two components: a tomography error ($v_{tomo}$) and a projection error ($v_{proj}$):
\begin{equation}
    v_{tomo} = \frac{2\kappa c(1-c)}{(d-1)N},\quad v_{proj}=\frac{c(1-c)}{N}.
\end{equation}
Note that $v_{tomo}$ includes a factor $1/(d-1)$, while $v_{proj}$ remains the same as in the single-qubit case. Similarly, the average variances can then be expressed as $v_{tt}=2v_{tomo}$ and $v_{tp}=v_{tomo}+v_{proj}$.
\subsection{TT and TP strategies performance with high-dimensional state tomography}
Here, we analyze the scaled average infidelity $\kappa$ of high-dimensional quantum states using various measurement approaches in the sufficient-copy scenario. Consider the tomography of a $d$-dimensional pure quantum state $\ket{\psi}$ using $N$ copies. Therefore, the total quantum state is $\ket{\psi}^{\otimes N}$. We examine three categories of tomography measurements:
\par
\textbf{Joint measurements across all copies.} The first category involves joint measurements on $\ket{\psi}^{\otimes N}$. The optimal measurement approach~\cite{PhysRevA.72.032325} achieves an average fidelity of $\overline{F}=(N+1)/(N+d)$~\cite{BRU1999249}. Therefore, the optimal scaled average infidelity is $\kappa_{opt}=d-1$. Applying this optimal tomography to the TT and TP strategies for overlap estimation effectively yields the \textit{estimate-and-estimate} and \textit{estimate-and-project} strategies described in~\cite{PhysRevLett.124.060503}, with $v_{tt} = 4c(1-c)/N$ and $v_{tp}=3c(1-c)/N$, respectively.
\par
\textbf{Independent measurements on each copy. }
The second category allows arbitrary measurements within each copy of $\ket{\psi}$, but restricts measurements to be independent across copies. For a quantum state $\rho$ of rank at most $r$, using only independent, non-adaptive measurements on each copy, the optimal tomography approach achieves a sample complexity of $N=O(dr^2/\epsilon^2)$ to estimate $\rho$ within trace distance $\epsilon$~\cite{KUENG201788,10.1145/2897518.2897585}, where trace distance is defined as $T(\rho,\tilde{\rho})=\text{tr}|\rho-\tilde{\rho}|/2$. For pure states ($r=1$, $\rho = \ket{\psi}\bra{\psi}$), the estimation $\tilde{\rho}$ from the approach in~\cite{KUENG201788} is generally not exact rank-1. A rank-1 estimate $\ket{\tilde{\psi}}\bra{\tilde{\psi}}$ can be constructed from the eigenvector $\ket{\tilde{\psi}}$ corresponding to the largest eigenvalue of $\tilde{\tilde{\rho}}$. 
Given $T(\rho,\tilde{\rho}) \leq \epsilon$, let the sorted eigenvalues of $\rho$ and $\tilde{\rho}$ be $(1, 0, \dots, 0)$ and $(\lambda_1, \lambda_2, \dots, \lambda_d)$ with $\lambda_1 \geq \lambda_2 \geq \dots \geq \lambda_d$, respectively. Weyl's inequality implies $|1-\lambda_1|\le\|\tilde{\rho}-\rho\|_2\leq \text{tr}|\rho-\tilde{\rho}|\leq 2\epsilon$ and $\max_{j\ne 1}\{|\lambda_j|\}\leq 2\epsilon$. Applying the Davis-Kahan ($\sin\theta$) theorem~\cite{doi:10.1137/0707001} with a spectral gap $\delta \geq 1-2\epsilon$, the infidelity between $|\psi\rangle$ and $|\tilde{\psi}\rangle$ is bounded by:
\begin{equation}
    \sqrt{1-|\braket{\psi|\tilde{\psi}}|^2}\leq \frac{\|\tilde{\rho}-\rho\|_2}{\delta}\leq \frac{2\epsilon}{1-2\epsilon}\approx 2\epsilon = O\left(\sqrt{\frac{d}{N}}\right).
\end{equation}
Therefore, post-processing $\tilde{\rho}$ to obtain $\ket{\tilde{\psi}}$ introduces at most a constant factor of 2 in the error. Consequently, $1-|\braket{\psi|\tilde{\psi}}|^2\leq 4\epsilon^2 = O(d/N)\sim\kappa_{ind}/N$, where $\kappa_{ind}=O(d)$ denotes the optimal scaled infidelity for tomography using independent measurements. Combining this into the TT and TP strategies, yields the following average variances: 
\begin{equation}
    v_{tt}=\frac{O(d)\cdot4c(1-c)}{(d-1)N}\sim O\left(\frac{c(1-c)}{N}\right),\quad v_{tp}=\left(2\frac{O(d)}{d-1}+1\right)\frac{c(1-c)}{N}\sim O\left(\frac{c(1-c)}{N}\right).
    \label{eq_v_ind}
\end{equation}
These average variances become independent for $d\gg 1$, similar to the optimal joint measurement approach in the first category. This stems from the same $N\sim O(d/\epsilon^2)$ scaling of the sample complexity for tomography of rank-1 states, regardless of whether joint or independent measurements are used~\cite{10.1145/2897518.2897585,10.1145/3055399.3055454}.
\par
\textbf{Local measurements on each qubit of multi-qubit states.} Now we consider the situation that the $d$-dimensional quantum state is composed of $n$ qubits ($d=2^n$).  If one can perform a joint measurement on all qubits of each copy, it reduces to the second approach discussed above. Here we consider a more practical tomography approach involving only local measurements on each qubit of each copy of the $n$-qubit state, such as Pauli measurements~\cite{Flammia_2012,yu2020sample}. In~\cite{Flammia_2012}, the authors show that using $N=O(r^2d^2\log d/\epsilon^2)=O(r^2 4^n n/\epsilon^2)$ copies allows reconstruction of a rank-$r$ $\rho$ with trace distance less than $\epsilon$. Thus, for pure state tomography using local measurements, the scaled infidelity is given by $\kappa_{loc}=O(4^n n)$. The average variances for the TT and TP strategies under this restriction are:
\begin{equation}
\begin{split}
    v_{tt} & =\frac{O(4^n n)\cdot4c(1-c)}{(2^n-1)N}\sim O\left(\frac{2^n nc(1-c)}{N}\right),\\
    v_{tp} & =\left(2\frac{O(4^n n)}{2^n-1}+1\right)\frac{c(1-c)}{N}\sim O\left(\frac{2^n nc(1-c)}{N}\right),
\end{split}
\end{equation}
where the approximations hold when $d\gg1$. However, the projective measurement on $\ket{\tilde{\psi}}$ in the TP strategy is generally non-local. Therefore, restricting to local operations effectively precludes the TP strategy.
\par
Table~\ref{tab:tt_tp_hd} summarizes the results of average variances. We emphasize that these results are derived asymptotically under the sufficient-copy assumption ($N\gg d$). This assumption arises from the requirement of a sufficiently small $\epsilon$ in our average variance analysis, combined with the tomography sample complexity scaling of $N\sim O(d/\epsilon^2)$ or $O(4^n n/\epsilon^2)$. In practice, implementing joint measurements across all copies becomes significantly challenging as $N$ increases. Independent measurements on each copy mitigate this issue and also yield dimension-independent average variances for TT and TP, as shown in Eq.~\eqref{eq_v_ind}, although implementation challenges persist for large $d$. Restricting measurements to local operations explicitly introduces dimension-dependent average variances and generally renders the TP strategy impractical.
\begin{table*}[htbp]
    \centering
    \renewcommand\arraystretch{1.5}
\setlength{\tabcolsep}{4mm}{
    \begin{tabular}{lccc}
    \toprule
    \  & Joint measurements & Independent measurements & Local measurements \\
    \hline
    $\kappa$ & $d-1$ & $O(d)$ & $O(4^n n)$\\
    \hline
    $v_{tt}$ & $4c(1-c)/N$ & $O(c(1-c)/N)$ & $O(2^n nc(1-c)/N)$ \\
    \hline
    $v_{tp}$ & $3c(1-c)/N $ & $O(c(1-c)/N)$ & $O(2^n nc(1-c)/N)$ \\
    \hline
    \end{tabular}}
    \caption{The scaled average infidelity $\kappa$, and average variances $v_{tt}$ and $v_{tp}$ for the TT and TP strategies with different tomography approaches for high-dimensional quantum states in the sufficient-copy scenario ($N\gg d$). For local measurements, we consider each copy is an $n$-qubit state, therefore $d=2^n$.}
    \label{tab:tt_tp_hd}
\end{table*}
\par
In the following, we consider the limited-copy scenario, where $N=\alpha d$ with $\alpha\sim O(1)$. We focus on the regime where $d\gg 1$, making this scenario common. In this case, qudit tomography suffers from both information incompleteness and significant statistical errors, resulting in poor and biased estimations. Therefore, the overlap estimators for the TT and TP strategies are generally biased. Following the supplemental material of~\cite{PhysRevLett.124.060503}, which derives the mean square error (MSE) for \textit{estimate-and-project} (Eq.~(106) in~\cite{PhysRevLett.124.060503}) and \textit{estimate-and-estimate} (Eq.~(78) in~\cite{PhysRevLett.124.060503}) strategies,  we can calculate MSE for the TT strategy with the optimal joint measurement tomography as follows:
\begin{equation}
    \begin{split}
        v_{tt}(c,N,d) & = \frac{(2N+d)[ (c(cd-2)(d+1)+2)(2N+d+1)+2c(1-c)N^2 ]}{(N+d)^2(N+d+1)^2} \\
        & = \frac{(2\alpha+1)d[(c(cd-2)(d+1)+2)(2\alpha d+d+1)+2c(1-c)\alpha^2 d^2]}{(\alpha+1)^2d^2(\alpha d+d+1)^2}\\
        & = \frac{(2\alpha+1)[c^2(2\alpha+1)d+2c(1-c)\alpha^2]}{(\alpha+1)^4d}+O\left(\frac{1}{\alpha^2d^2}\right) \\
        & = \frac{(2\alpha+1)^2}{(\alpha+1)^4} c^2 + \frac{2(2\alpha+1)\alpha^2}{(\alpha+1)^4d}c(1-c)+O\left(\frac{1}{\alpha^2d^2}\right) \\
        & = \frac{(2\alpha+1)^2}{(\alpha+1)^4} c^2 + \frac{2(2\alpha+1)\alpha^3}{(\alpha+1)^4 N}c(1-c)+O\left(\frac{1}{N^2}\right).
    \end{split}
\end{equation}
Since $\alpha\sim O(1)$, the term $(2\alpha+1)^2c^2/(\alpha+1)^4$ introduces a constant-order error in the overlap estimation, leading to an MSE of $v_{tt} = O(1)$ for the TT strategy and demonstrating its inefficiency in the limited-copy scenario. To recover the sufficient-copy scenario, where the first term is negligible compared to the $O(1/N)$ terms, it requires $1/\alpha^2\sim o(1/N)$, or equivalently, $N\sim\omega(d^2)$. In this case, the constant term vanishes, and the second term dominates, yielding $v_{tt} = 4c(1-c)/N$, consistent with our previous discussion. Similarly, the MSE for the TP strategy with the optimal joint measurement tomography is:
\begin{equation}
    \begin{split}
        v_{tp}(c,N,d) & = \frac{c^2(d^2N+dN-3N^2+N)+c(3N^2-dN-5N)+3N+d-1}{N(N+d)(N+d+1)} \\
        & = \frac{c^2(\alpha d^2+\alpha d-3\alpha^2 d+\alpha)d+c(3\alpha^2d^2-\alpha d^2-5\alpha d)+3\alpha d+d-1}{\alpha(\alpha+1)(\alpha d+d+1)d^2} \\
        & = \frac{c^2 d+c(3\alpha-1 )}{(\alpha+1)^2d} +\frac{c^2(1-3\alpha)}{(\alpha+1)^2d}+O\left(\frac{1}{\alpha^2d^2}\right) \\
        & = \frac{c^2}{(\alpha+1)^2} + \frac{3\alpha-1 }{(\alpha+1)^2d}c(1-c)+O\left(\frac{1}{\alpha^2d^2}\right) \\
        & = \frac{c^2}{(\alpha+1)^2} + \frac{\alpha(3\alpha-1) }{(\alpha+1)^2N}c(1-c)+O\left(\frac{1}{N^2}\right).
    \end{split}
\end{equation}
Again, the first term represents an $O(1)$ error when $\alpha\sim O(1)$. When $\alpha\gg 1$, it becomes negligible, and the TP strategy becomes asymptotically unbiased, with an MSE consistent with the average variance $v_{tp}=3c(1-c)/N$. Therefore, due to the limitations of tomography in the limited-copy scenario, both TT and TP strategies provide biased overlap estimations with mean square errors scaling as $O(1)$.
\par
\section{Optical swap test with experimental imperfections}
\subsection{Ideal optical swap test} 
The optical swap test (OST) is a modified version of the swap test that can be implemented practically via a multi-mode Hong-Ou-Mandel interference (HOMI), which uses a non-polarizing beam-splitter (NPBS) to perform the interference between two photons encoded by $\ket{\psi}$ and $\ket{\phi}$, respectively. For the overlap estimation task, we can utilize the OST to estimate the overlap by recording the test results as either ``pass" or ``fail" over $N$ trials. As illustrated in Fig.~\ref{fig:HOM_multimode_schematic}, we consider an ideal case that two perfect indistinguishable photons have been prepared as a two-qudit joint state, which can be expressed as
\begin{equation}
    \ket{\psi}_1\otimes\ket{\phi}_2 = (\sum_{i=1}^{d}\alpha_i \hat{a}^{\dagger}_{i,1}\ket{0}_1)\otimes(\sum_{j=1}^{d}\beta_j \hat{a}^{\dagger}_{j,2}\ket{0}_2),
\end{equation}
here $\hat{a}^{\dagger}_{i,1}$ and $\hat{a}^{\dagger}_{j,2}$ denote photon creation operator in $i$ mode at input port 1 of NPBS and $j$ mode at input port 2 respectively, $\ket{0}_1$ and $\ket{0}_2$ represent the vacuum state for two input sides. The balanced NPBS has following transformation on creation operators of input modes: $\hat{a}^{\dagger}_{i,1}\to(\hat{a}^{\dagger}_{i,3}+\hat{a}^{\dagger}_{i,4})/\sqrt2$, $\hat{a}^{\dagger}_{j,2}\to(\hat{a}^{\dagger}_{j,3}-\hat{a}^{\dagger}_{j,4})/\sqrt2$, where 3 and 4 denote two output ports. The output field can be written as
\begin{small}
    \begin{equation}
    \begin{split}
        \ket{\Psi^{\text{out}}} =&\ \frac12\sum_{i=1}^{d}\sum_{j=1}^{d}\alpha_i\beta_j (\hat{a}^{\dagger}_{i,3}+\hat{a}^{\dagger}_{i,4})(\hat{a}^{\dagger}_{j,3}-\hat{a}^{\dagger}_{j,4})\ket{0}_3\ket{0}_4 \\
        =&\ \sum_{i=1}^{d}\frac{\alpha_i\beta_i}{\sqrt 2} (\ket{2_i}_3\ket{0}_4 - \ket{0}_3\ket{2_i}_4)+\sum_{1\leq i<j\leq d}\frac{\alpha_i\beta_j+\alpha_j\beta_i}2 (\ket{1_i1_j}_3\ket{0}_4-\ket{0}_3\ket{1_i1_j}_4)\\ 
        &\ +\sum_{i \neq j}^{d}\frac{\alpha_i\beta_j-\alpha_j\beta_i}2 \ket{1_i}_3\ket{1_j}_4,
    \end{split}
    \label{eq-ost-ideal}
\end{equation}
\end{small}
where $\ket{1_i1_j}_3$ or $\ket{1_i1_j}_4$ describe two photons occupying the same output port 3 or 4, but in different mode $i$ and $j$. After the interference, photon detectors are used to detect the photon distribution of the output filed. In the OST, ``pass" outcomes correspond to cases where both photons are detected in the same output port of the interference NPBS, such as $\ket{1_i1_j}_3\ket{0}_4$ and $\ket{2_i}_3\ket{0}_4$. Conversely, cases where the photons are detected in different output ports like $\ket{1_i}_3\ket{1_j}_4$, denote the ``fail" outcomes. From Eq.~\eqref{eq-ost-ideal}, the probability of ``fail" outcomes for OST is given by
\begin{equation}
    \begin{split}
        p(f) & = \sum_{i\neq j}^{d}|\frac{\alpha_i\beta_j-\alpha_j\beta_i}{2}|^2 = \sum_{i=1,j=1}^{d}|\frac{\alpha_i\beta_j-\alpha_j\beta_i}{2}|^2 \\
        & = \frac{1}{2}\sum_{i=1,j=1}^{d}|\alpha_i|^2|\beta_j|^2-\frac{1}{2}(\sum_{i=1}^{d}\alpha_i\beta_i^{*})(\sum_{j=1}^{d}\alpha_j^{*}\beta_j) \\
        & = \frac{1}{2}(1-|\braket{\psi|\phi}|^2),
    \end{split}
    \label{eq-pf-iost}
\end{equation}
with the (unsquared) overlap $\braket{\psi|\phi}=\sum_{j}\alpha_j^{*}\beta_j$ and the normalization conditions $\sum_{i}|\alpha_i|^2=\sum_{i}|\beta_i|^2=1$. Hence, the overlap between two states determines the probability of ``fail" outcomes in the OST. After $N$ trials of OST, if we faithfully record the counts of ``pass" and ``fail" outcomes, we can use the ``fail'' counts $k_f$ to estimate the overlap as $\tilde{c}_{ost}=1-2k_f/N$. The overlap estimation variance through the ideal OST strategy is then given by $v(c,N)=(1-c^2)/N$, where $c = |\braket{\psi|\phi}|^2$ is the true overlap.
\par
\begin{figure}[!t]
    \centering
    \includegraphics[width=0.35\textwidth]{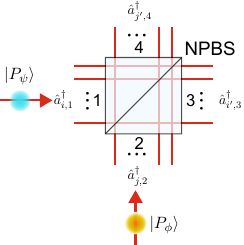}
    \caption{Schematic of multimode Hong-Ou-Mandel interference. The non-polarizing beam splitter (NPBS) has two input ports (1 and 2) and two output ports (3 and 4). $\ket{P_{\psi}}$ and $\ket{P_{\phi}}$ represent the quantum states of the photon encoded by the qudits $\ket{\psi}$ and $\ket{\phi}$, respectively. The photons are distributed among $d$ discrete modes, such as path modes and polarization modes. The creation operators for the input ports 1 and 2 are denoted as $\hat{a}^{\dagger}_{i,1}$ and $\hat{a}^{\dagger}_{j,2}$, while for the output ports 3 and 4, they are represented as $\hat{a}^{\dagger}_{i^{\prime},3}$ and $\hat{a}^{\dagger}_{j^{\prime},4}$.}
    \label{fig:HOM_multimode_schematic}
\end{figure}
\subsection{OST with partially distinguishable photons and non-balanced beam-splitter} 
\label{OST-spectral}
In a HOMI experiment, photon pairs generated by spontaneous parametric down conversion are usually not perfectly indistinguishable due to slightly different spectral mode or spatial mode. In this section, we consider the case that two photons with discrete modes being encoded as $\ket{\psi}$ and $\ket{\phi}$, and have different properties on their non-encoded modes where we regard them as internal modes. The spectral mode in the internal modes is mainly considered. To begin with, we describe these photons as follows:
\begin{equation}
    \ket{P_{\psi}} =\int d\omega S_1(\omega)\sum_{i=1}^d \alpha_i\hat{a}^{\dagger}_{i}(\omega)\ket{0}, \ \ket{P_{\phi}} =\int d\omega S_2(\omega)\sum_{j=1}^d \beta_j\hat{a}^{\dagger}_{j}(\omega)\ket{0},
\end{equation}
here $\ket{P_{\psi}}$ and $\ket{P_{\phi}}$ denote the quantum states of the photons whose discrete modes are encoded by $\ket{\psi}$ and $\ket{\phi}$, respectively. $S_1(\omega)$ and $S_2(\omega)$ represent the spectral amplitudes of two photons, which may be not identical. The input state on ports 1 and 2 of the NPBS is then given by
\begin{equation}
    \begin{split}
        \ket{\Phi^{\text{in}}} = \ket{P_{\psi}}_1\ket{P_{\phi}}_2 =  \int d\omega_1 \int d\omega_2 S_1(\omega_1) S_2(\omega_2)\sum_{i,j=1}^d \alpha_i\beta_j\hat{a}^{\dagger}_{i,1}(\omega_1)\hat{a}^{\dagger}_{j,2}(\omega_2) \ket{0}_1\ket{0}_2.
    \end{split}
\end{equation}
We consider that the NPBS in experiment is not perfectly balanced, with reflectivity $\eta$. The non-balanced NPBS transforms the creation operators as follows: $\hat{a}^{\dagger}_{i,1}(\omega)\to \sqrt{1-\eta}\hat{a}^{\dagger}_{i,3}(\omega) + \sqrt{\eta}\hat{a}^{\dagger}_{i,4}(\omega)$, $\hat{a}^{\dagger}_{i,2}(\omega)\to \sqrt{\eta}\hat{a}^{\dagger}_{i,3}(\omega) - \sqrt{1-\eta}\hat{a}^{\dagger}_{i,4}(\omega)$. The output state on ports 3 and 4 is then given by
\begin{footnotesize}
\begin{equation}
    \begin{split}
        \ket{\Phi^{\text{out}}} =  &  \sum_{i,j=1}^d\alpha_i\beta_j\int d\omega_1 \int d\omega_2 S_1(\omega_1) S_2(\omega_2) \bigg[\sqrt{1-\eta}\hat{a}^{\dagger}_{i,3}(\omega_1) + \sqrt{\eta}\hat{a}^{\dagger}_{i,4}(\omega_1)\bigg]\bigg[\sqrt{\eta}\hat{a}^{\dagger}_{j,3}(\omega_2) - \sqrt{1-\eta}\hat{a}^{\dagger}_{j,4}(\omega_2)\bigg]\ket{0}_3\ket{0}_4 \\
        = &  \sum_{i,j=1}^d\alpha_i\beta_j\int d\omega_1 \int d\omega_2 S_1(\omega_1) S_2(\omega_2) \times \bigg[ \eta\hat{a}^{\dagger}_{i,4}(\omega_1)\hat{a}^{\dagger}_{j,3}(\omega_2)-(1-\eta)\hat{a}^{\dagger}_{i,3}(\omega_1)\hat{a}^{\dagger}_{j,4}(\omega_2) \\ & +\sqrt{\eta(1-\eta)}\hat{a}^{\dagger}_{i,3}(\omega_1)\hat{a}^{\dagger}_{j,3}(\omega_2) - \sqrt{\eta(1-\eta)}\hat{a}^{\dagger}_{i,4}(\omega_1)\hat{a}^{\dagger}_{j,4}(\omega_2) \bigg] \ket{0}_3\ket{0}_4, \\
    \end{split}
\end{equation}
\end{footnotesize}
where we assume that two photons arrived at NPBS simultaneously and time delay between two photons is zero. The projector of detecting one photon in mode $m$ of port 3, and another one in mode $n$ of port 4, which is corresponding to one possible ``fail'' outcome of the OST, is shown as
\begin{equation}
    \begin{split}
        \hat{P}_{m,3}\otimes\hat{P}_{n,4}= & \int d\omega_3\hat{a}^{\dagger}_{m,3}(\omega_3)\ket{0}_3\bra{0}_3\hat{a}_{m,3}(\omega_3)\otimes \int d\omega_4\hat{a}^{\dagger}_{n,4}(\omega_4)\ket{0}_4\bra{0}_4\hat{a}_{n,4}(\omega_4).
    \end{split}
    \label{24}
\end{equation}
We use $p(f,m,n)$ to denote the probability of detecting this kind of ``fail'' outcome, and it can be derived as follows:
\begin{footnotesize}
    \begin{equation}
    \begin{split}
        p(f,m,n) = & \bra{\Psi^{\text{out}}}\hat{P}_{m,3}\otimes\hat{P}_{n,4}\ket{\Psi^{\text{out}}} \\
        = & \sum_{i,j,k,l=1}^{d}\alpha_i^{*}\beta_j^{*}\alpha_k\beta_l\int d\omega_1\int d\omega_2 \int d\omega_1^{'} \int d\omega_2^{'}\int d\omega_3\int d\omega_4 S^*_1(\omega_1)S^*_2(\omega_2)S_1(\omega_1^{'})S_2(\omega_2^{'}) \\
        & \times \bra{0}_3 \bra{0}_4  \left[ \eta\hat{a}_{i,4}(\omega_1)\hat{a}_{j,3}(\omega_2)\hat{a}^{\dagger}_{m,3}(\omega_3)\hat{a}^{\dagger}_{n,4}(\omega_4)-(1-\eta)\hat{a}_{i,3}(\omega_1)\hat{a}_{j,4}(\omega_2)\hat{a}^{\dagger}_{m,3}(\omega_3)\hat{a}^{\dagger}_{n,4}(\omega_4)\right]\ket{0}_3\ket{0}_4\\
        & \times \bra{0}_3 \bra{0}_4 \left[ \eta\hat{a}_{m,3}(\omega_3)\hat{a}_{n,4}(\omega_4)\hat{a}^{\dagger}_{k,4}(\omega_1^{'})\hat{a}^{\dagger}_{l,3}(\omega_2^{'})-(1-\eta)\hat{a}_{m,3}(\omega_3)\hat{a}_{n,4}(\omega_4)\hat{a}^{\dagger}_{k,3}(\omega_1^{'})\hat{a}^{\dagger}_{l,4}(\omega_2^{'}) \right]\ket{0}_3\ket{0}_4\\
        = & \sum_{i,j,k,l=1}^{d}\alpha_i^{*}\beta_j^{*}\alpha_k\beta_l\int d\omega_1\int d\omega_2 \int d\omega_1^{'} \int d\omega_2^{'}\int d\omega_3\int d\omega_4 S^*_1(\omega_1)S^*_2(\omega_2)S_1(\omega_1^{'})S_2(\omega_2^{'}) \\
        & \times (\eta \delta(\omega_1-\omega_4)\delta_{in}\delta(\omega_2-\omega_3)\delta_{jm}-(1-\eta)\delta(\omega_1-\omega_3)\delta_{im}\delta(\omega_2-\omega_4)\delta_{jn})\\
        & \times (\eta \delta(\omega_1^{'}-\omega_4)\delta_{kn}\delta(\omega_2^{'}-\omega_3)\delta_{lm}-(1-\eta)\delta(\omega_1^{'}-\omega_3)\delta_{km}\delta(\omega_2^{'}-\omega_4)\delta_{ln})\\
        = & |\alpha_n|^2|\beta_m|^2 \eta^2 \int d\omega_3 |S_2(\omega_3)|^2 \int d\omega_4 |S_1(\omega_4)|^2 + |\alpha_m|^2|\beta_n|^2(1-\eta)^2 \int d\omega_3 |S_1(\omega_3)|^2 \int d\omega_4 |S_2(\omega_4)|^2 \\
        & - \alpha_n^{*}\beta_m^{*}\alpha_m\beta_n\eta(1-\eta) \left |\int d\omega S_2^{*}(\omega) S_1(\omega)\right|^2-\alpha_m^{*}\beta_n^{*}\alpha_n\beta_m\eta(1-\eta) \left|\int d\omega S_2^{*}(\omega) S_1(\omega)\right|^2 \\
        = & |\alpha_n|^2|\beta_m|^2 \eta^2 + |\alpha_m|^2|\beta_n|^2(1-\eta)^2 - (\alpha_n^{*}\beta_m^{*}\alpha_m\beta_n+\alpha_m^{*}\beta_n^{*}\alpha_n\beta_m)\eta(1-\eta)\Gamma,
    \end{split}
    \label{25}
\end{equation}
\end{footnotesize}
here $\delta(\omega)$ and $\delta_{ij}$ denote Dirac delta function and Kronecker delta respectively, and $\Gamma=\left|\int d\omega S_2^{*}(\omega) S_1(\omega)\right|^2$ represents indistinguishability of spectral modes of two photons, with the normalized conditions $\int d\omega \left |S_1(\omega) \right|^2 =\int d\omega \left |S_2(\omega) \right|^2 = 1$. The spectral indistinguishability $\Gamma$ should be distinguished from the overlap $c$ between the discrete modes. For the second equation in Eq.~\eqref{25}, terms with odd number of operators in one port, such as $\bra{0}_3\bra{0}_4\hat{a}_3\hat{a}_3\hat{a}_3^{\dagger}\hat{a}_4^{\dagger}\ket{0}_3\ket{0}_4$ and $\bra{0}_3\bra{0}_4\hat{a}_3\hat{a}_4\hat{a}_4^{\dagger}\hat{a}_4^{\dagger}\ket{0}_3\ket{0}_4$, become zero and have been discarded. Summing Eq.~\eqref{25} over $m$ and $n$, we can get the probability of the ``fail" outcomes
\begin{equation}
    \begin{split}
        p(f) = \sum_{m,n}p(f,m,n)
        & =  1-2\eta+2\eta^2-2\eta(1-\eta)\Gamma(\sum_{m=1}^{d}\alpha_m\beta_m^{*})(\sum_{n=1}^{d}\alpha_n^{*}\beta_n) \\
        & =  1-2\eta+2\eta^2-2\eta(1-\eta)\Gamma c,
    \end{split}
    \label{26}
\end{equation}
where $c = |\braket{\psi|\phi}|^2 = \left|\sum_{m=1}^{d}\alpha_m\beta_m^{*}\right|^2$ is the overlap between discrete modes of two photons. When $\eta = 0.5$ and $\Gamma = 1$, Eq.~\eqref{26} becomes the usual form $p(f) = (1-c)/2$. In order to estimate the overlap $c$ without bias, from Eq.~\eqref{26}, the estimator of $c$ should be corrected as
\begin{equation}
    \tilde{c}_{ost}=\frac{1-2\eta+2\eta^2}{2\eta(1-\eta)\Gamma}-\frac{k_f}{2\eta(1-\eta)N\Gamma},
    \label{eq-rev-ost}
\end{equation}
where $k_f$ is the number of ``fail'' outcomes out of $N$ rounds of OST. 
\par
\subsection{OST with pseudo photon-number-resolving detectors}
In photonic experiments, deterministic photon-number-resolving detectors are not always available. In cases where we rely on threshold single photon detectors, such as avalanche photodiodes, the accurate measurement of multi-photon bunching term probabilities becomes challenging. This limitation affects the overlap estimation performance of the OST strategy when only pseudo photon-number-resolving detectors (PPNRD) are used. In the worst-case scenario, we consider an OST setup where the PPNRDs response ``pass" outcomes, such as $\ket{1_i1_j}_3\ket{0}_4$ and $\ket{2_i}_3\ket{0}_4$ in Eq.~\eqref{eq-ost-ideal}, with only a probability of  $1/2$. The probability of detecting a ``pass" outcome is the same as the probability of losing it, which is $(1+c)/4$. Therefore, the probability of detecting an outcome, which includes both ``pass" and ``fail" outcomes, can be expressed as
\begin{equation}
    p(D) = p(f,D) + p(p,D) = \frac{1-c}{2} + \frac{1+c}{4}= \frac{3-c}{4},
\end{equation}
where $f$ and $p$ denote ``fail" and ``pass" outcomes of the OST, respectively, and $D$ represents the outcome captured by PPNRDs. The conditional PDF for ``fail'' and ``pass'' outcomes is given by
\begin{equation}
    p(f|D) = \frac{p(f,D)}{p(D)} = \frac{2-2c}{3-c},\ p(p|D) = \frac{p(p,D)}{p(D)} = \frac{1+c}{3-c},
    \label{prob_fdpd}
\end{equation}
where the overlap $c\in\left[0,1\right]$. In this PPNRD scenario, the number of copies of quantum states used in the OST strategy is also non-deterministic due to the counting procedure based on detection. To accurately count the number of consumed copies of quantum states, a detected ``pass" outcome will correspond to two rounds of the OST. In our analysis, we assume that we have detected $k_p$ ``pass" and $k_f$ ``fail" outcomes, and $N$ is even. For $N$ pairs of states $\ket{\psi}\ket{\phi}$, $k_f$ and $k_p$ must satisfy the restriction: $k_f + 2k_p = N$ if $k_f$ is even, $k_f + 2k_p = N+1$ if $k_f$ is odd and the last outcome is ``pass". The conditional PDF of $k_f$ is given by
\begin{equation}
    p(k_f|D)=
    \left\{
    \begin{array}{lc}
        \binom{\frac{N+k_f}2}{k_f}(\frac{2-2c}{3-c})^{k_f}(\frac{1+c}{3-c})^{\frac{N-k_f}{2}},
        k_f\text{ is even} \\
        \binom{\frac{N+k_f-1}{2}}{k_f}(\frac{2-2c}{3-c})^{k_f}(\frac{1+c}{3-c})^{\frac{N-k_f+1}{2}},
        k_f\text{ is odd}  \\
    \end{array}\right.
    ,\ k_f \in \{0, 1 ,...,N\}.
    \label{new-2}
\end{equation}
With the PDF of $k_f$, we can get the normalizing condition, the first and the second order raw moments of $k_f$ as follows:
\begin{equation}
\begin{split}
        \left\langle 1 \right\rangle_{k_f} = &  \frac{1+(\frac{1+c}{3-c})^{N+1}}{1+(\frac{1+c}{3-c})}-\frac{(\frac{1+c}{3-c})^{N+1}-(\frac{1+c}{3-c})  }{1+(\frac{1+c}{3-c})} = 1, \\
        \left\langle k_f \right\rangle_{k_f} = &  \frac{1}{8}\left[(1-c)(4N+1+c) -\right(\frac{1+c}{3-c} \left)^{N}(1-c^2) \right] = \frac{1}{8}(1-c)(4N+1+c) + O(\frac{1}{N}),\\
        \left\langle k_f^2 \right\rangle_{k_f} = &  \frac{N}{4}(1-c)[2+c(1-c)+N(1-c)] + O(1),
\end{split}
\end{equation}
where the expression $(\frac{1+c}{3-c})^{N}(1-c^2)$ scales as $O(1/N)$. The mean of estimated overlap in the OST strategy is given by
\begin{equation}
    \left\langle 1-\frac{2k_f}{N} \right\rangle_{k_f} = 1-\frac{2}{N}\left\langle k_f \right\rangle_{k_f} = c + O(\frac{1}{N}),
\end{equation}
here we show that the overlap estimator in OST strategy is asymptotically unbiased. Then the overlap estimation variance for OST strategy in this PPNRD scenario is given by
\begin{equation}
    \begin{split}
        v_{ost}(c,N) = & \left\langle (1-\frac{2k_f}{N}-c)^2 \right\rangle_{k_f} 
         =  (1-c)^2 -\frac{4}{N}(1-c)\left\langle k_f \right\rangle_{k_f}+\frac{4}{N^2} \left\langle k_f^2 \right\rangle_{k_f}\\
         = & \frac{(3-c)(1-c^2)}{2N} + O(\frac{1}{N^2})
         \approx \frac{(3-c)(1-c^2)}{2N}.
    \end{split}
    \label{vost}
\end{equation}
Compared with the variance $v(c,N)=(1-c^2)/N$ through the ideal OST, the additional factor $(3-c)/2$ in Eq.~\eqref{vost} reflects the precision reduction introduced by the PPNRDs. 
\subsection{OST strategy precision with our experimental setup}
The preceding discussion highlights the feasibility of using the OST strategy to estimate the overlap $c$ without bias, even in the presence of imperfect HOMI equipment and partially indistinguishable photons. In this section, we present the theoretical results on the precision of the OST strategy using the experimental setup mensioned in the maint text, as depicted in Fig.~\ref{fig:HOM_setup}. The detector is employed to detect ``pass" outcomes from the output port 3 (4) of NPBS-1 with a detection probability of $1/2$, similar to the PPNRD scenario discussed earlier. We note that the corrected estimator in Eq.~(\ref{eq-rev-ost}) is more sensitive to variations in $\Gamma$ compared to $\eta$, especially when $\Gamma=1$ and $\eta=0.5$. For our experimental setup, the reflectivity of NPBS-1 is approximately 0.53, and the imperfections of NPBS can be neglected. Therefore, we utilize $ \tilde{c}_{ost} = (1 - 2k_f/N)/\Gamma$ to estimate the overlap. In this case, we modify the overlap in Eq.~\eqref{vost} as $c^{\prime} = c\cdot\Gamma$. Here, $c$ represents the overlap between two quantum states that are encoded on polarization of two photons, and is the parameter we aim to estimate. Using the basic property of variance, we derive the overlap estimation variance for the OST strategy under our experimental setup as
\begin{equation}
    v_{ost}(c,N) = \frac{(3-\Gamma c)(1-\Gamma^2 c^2)}{2N\Gamma^2}.
    \label{27}
\end{equation}
The value of the spectral mode indistinguishability $\Gamma$ can be calibrated by performing the OST between two photons with identical encoded states on polarization. In our setup, $\Gamma$ is measured to be $0.965\pm0.008$, obtained as the average maximum HOMI visibility when different polarized photon pairs are used as inputs. We can express the variance of the OST strategy more insightfully as follows:
\begin{equation}
    v_{ost}(c,N) = \frac{(3-\Gamma c)}2\times \left(\frac{1-\Gamma^2}{N\Gamma^2}+\frac{1-c^2}{N}\right).
\end{equation}
This expression reveals three distinct contributions to the variance. The overlap-dependent factor $(3-\Gamma c)/2$, arising from the PPNRDs, reaches its maximum at $c=0$ and minimum at $c=1$. This behavior indicates that the detrimental effects of PPNRDs are most pronounced when the overlap is small. The overlap-independent term $(1-\Gamma^2)/N\Gamma^2$ captures the impact of partial photon indistinguishability, leading to a constant reduction in precision across the entire range of overlaps. The final term $(1-c^2)/N$ matches the variance of both the ideal swap test and the SCM strategy. This observation confirms that, the OST strategy and the SCM strategy would exhibit identical performance without experimental imperfections.
\begin{figure}[!ht]
    \centering
    \includegraphics[width=0.4\textwidth]{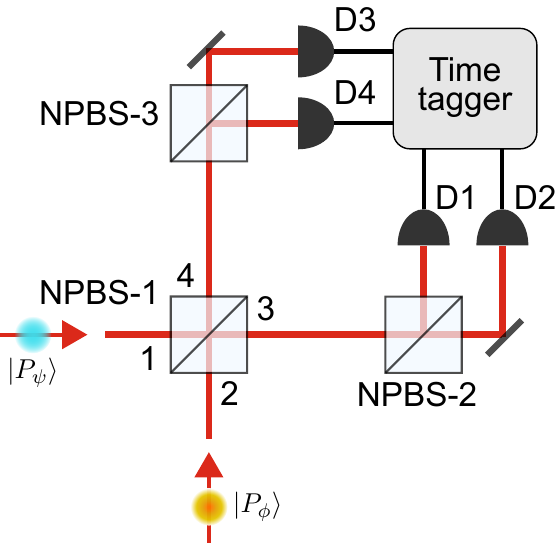}
    \caption{Experimental setup for Hong-Ou-Mandel interference and pseudo photon-number-resolving detection. Two photons $\ket{P_{\psi}}$ and $\ket{P_{\phi}}$, with their polarization modes encoded as $\ket{\psi}$ and $\ket{\phi}$ respectively, undergo interference at NPBS-1. The pseudo photon-number-resolving detector (PPNRD) is achieved using NPBS-2 (NPBS-3) and two single photon counting modules (SPCMs). The electrical signals form SPCMs (D1 to D4) are processed by time tagger (Time Tagger Ultra, Swabian) to produce coincidence counts. NPBS: non-polarizing beam splitter.}
    \label{fig:HOM_setup}
\end{figure}
\par
\textbf{Fisher information in OST strategy.} Considering the spectral mode indistinguishability $\Gamma<1$, we rewrite the probabilities in Eq.~(\ref{prob_fdpd}) as follows:
\begin{equation}
    p(F) = p(f|D) = \frac{2-2\Gamma c}{3-\Gamma c},\ p(P) = p(p|D)  = \frac{1+\Gamma c}{3-\Gamma c},
    \label{prob_fdpd_2}
\end{equation}
where $p(F)$ and $p(P)$ represent the original detection probabilities for ``fail'' and ``pass'' outcomes. From the Bernoulli distribution, the Fisher information of overlap in a detected OST event is given by
\begin{equation}
\begin{split}
       I_{ost} = & \frac{1}{p(F)}(\frac{\partial p(F) }{\partial c})^2 + \frac{1}{p(P)}(\frac{\partial p(P) }{\partial c})^2
    =  \frac{8\Gamma^2}{(3-\Gamma c)^3(1-\Gamma c)}+\frac{16\Gamma^2}{(3-\Gamma c)^3(1+\Gamma c)} \\
 = & \frac{8\Gamma^2}{(3-\Gamma c)^2(1-\Gamma^2 c^2)}.
\end{split}
\end{equation}
Compared with other overlap estimation strategies, the OST strategy should take into account the photon loss in PPNRD. For $N$ copies of state pairs, the total Fisher information of overlap is given by $N^{\prime}I_{ost} = 2N\Gamma^2/(3-\Gamma c)(1-\Gamma^2 c^2)$, where $N^{\prime}=(3-\Gamma c)N/4$ is the mean number of detected events. We define the effective Fisher information of overlap per state pair as
\begin{equation}
    I_{ost}^{e} = \frac{N^{\prime}I^{ost}}{N} = \frac{2\Gamma^2}{(3-\Gamma c)(1-\Gamma^2 c^2)} \approx \frac{1}{Nv_{ost}(c,N)},
\end{equation}
where $v_{ost}(c,N)$ is defined in Eq.~(\ref{27}). Therefore, we show that the estimator used in OST under our PPNRD scenario saturates the corresponding Cramér-Rao bound asymptotically.
\section{Experimental details}
\textbf{Photon source.} Light pulses with 150 fs duration, centered at 830 nm, from a ultrafast Ti-Sapphire Laser (Coherent Mira-HP; 76 MHz repetition rate) are firstly frequency doubled in a $\beta$-type barium borate ($\beta$-BBO) crystal to generate a second harmonic beam with 415 nm wavelength. Then the upconversion beam is then utilized to pump another $\beta$-BBO with phase-matched cut angle for type-\uppercase\expandafter{\romannumeral2} beam-like degenerate spontaneous down conversion (SPDC) which produces pairs of photons, denoted as signal and idler. The signal and idler photons possess distinct emergence angles and spatially separate from each other. After passing through two clean-up filters with a 3 nm bandwidth, the photons are coupled into separate single-mode fibers. The idler mode is detected by a single photon counting module (SPCM, Excelitas Technologies) with a detection efficiency of approximately 55\%, serving as a trigger. This configuration enables the photon source module to function as a herald single-photon source (HSPS). The signal mode is directed to the Tomography, Projection, and SCM modules, as mentioned in the main text, for further experimental operations. Additionally, both the signal and idler photons are directed to the OST module, where they undergo Hong-Ou-Mandel interference.


\textbf{State preparation.} In TP, TT and OST strategy experiments, qubits are encoded in the polarization degree of freedom of photons, i.e., $\{\ket{0}=\ket{H},\ket{1}=\ket{V}\}$, where $\ket{H}$ and $\ket{V}$ represent horizontal and vertical polarization, respectively. With a electronically controlled half wave-plate (E-HWP, Newport PR50PP Motorized Rotation Stage) followed by a liquid crystal phase retarder (LCPR, Thorlabs, LCC1113-B), the single photon state is prepared as
\begin{equation}
    \ket{\psi}\text{or}\ket{\phi}=\cos{2\theta}\ket{H}+e^{i\alpha}\sin{2\theta}\ket{V}\text{,}
    \label{7}
\end{equation}
where $\theta$ is the E-HWP angle and $\alpha$ is the relative phase between two polarization modes added by LCPR. In experiments, we configure HWPs and LCPRs sequentially for state preparation, ensuring each setting is complete before recording the measurement results. The precision of state preparation has been characterized by quantum state tomography, with an average fidelity up to 0.9989, as shown in Fig.~\ref{fig:tomo_fidelity}.
\begin{figure}[!t]
    \centering
    \includegraphics[width=0.6\textwidth]{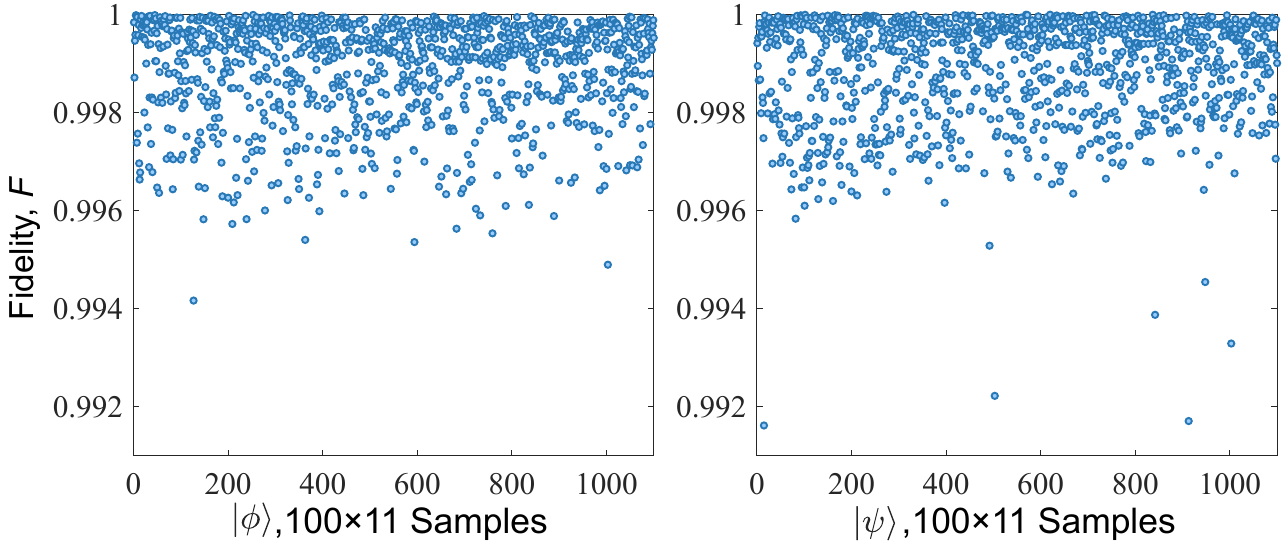}
    \caption{Fidelity of state preparation in TT strategy. The fidelity is defined as $F = |\braket{\phi|\tilde{\phi}}|^2$, where $\ket{\phi}$ is the target state and $\ket{\tilde{\phi}}$ is the state reconstructed by quantum state tomography using 180,000 measurements.}
    \label{fig:tomo_fidelity}
\end{figure}
In the SCM strategy experiment, the single photon will be prepared as a two-qubit joint state $\ket{\psi}\ket{\phi}$. After encoding the first qubit $\ket{\psi}$ in the same form as Eq.~\eqref{7}, a birefringent calcite beam splitter (BD) splits the single photon into two path modes, resulting in a path-polarised entangled state
\begin{equation}
    \cos2\theta_1\ket{s_0}\ket{H}+e^{i\alpha_1}\sin{2\theta_1}\ket{s_1}\ket{V}\text{,}
\end{equation}
where $s_0$ and $s_1$ denote the lower and upper path modes, respectively. In the path mode $s_1$, a HWP with angle $45^{\circ}$ rotates $\ket{V}$ to $\ket{H}$. Another HWP with angle $0^{\circ}$ is placed in path mode $s_0$ for optical path compensation. Then the state becomes
\begin{equation}
    \cos2\theta_1\ket{s_0}\ket{H}+e^{i\alpha_1}\sin{2\theta_1}\ket{s_1}\ket{H}\text{.}
\end{equation}
The second qubit $\ket{\phi}$ is encoded on the single photon using an E-HWP and a phase retarder implemented with an E-HWP sandwiched by two quarter-wave plates (QWP-HWP-QWP configuration, QHQ) to circumvent the non-uniform adding phase on different position of the LCPR. The QHQ phase retarder implements a unitary transformation in each path mode using two QWPs set at $45^{\circ}$ in combination with an E-HWP at an angle of $(\alpha_{2}-\pi)/4$, given by
\begin{equation}
    U_{QHQ}=\ket{H}\bra{H}+e^{i\alpha_2}\ket{V}\bra{V}\text{,}
\end{equation}
where a global phase is ignored. After going through the second E-HWP with angle $\theta_2$ and the QHQ phase retarder, the two-qubit state of single photons can be prepared as
\begin{equation}
    \begin{split}
        \ket{\psi}\otimes\ket{\phi}=&\cos{2\theta_1}\cos{2\theta_2}\ket{s_0}\ket{H}+e^{i\alpha_2}\cos{2\theta_1}\sin{2\theta_2}\ket{s_0}\ket{V} \\
        & +e^{i\alpha_1}\sin{2\theta_1}\cos{2\theta_2}\ket{s_1}\ket{H}+e^{i(\alpha_1+\alpha_2)}\sin{2\theta_1}\sin{2\theta_2}\ket{s_1}\ket{V}\\
        =&(\cos2\theta_1\ket{s_0}+e^{i\alpha_1}\sin{2\theta_1}\ket{s_1})
        \otimes(\cos{2\theta_2}\ket{H}+e^{i\alpha_2}\sin{2\theta_2}\ket{V}).\\
    \end{split}
\end{equation}
\par
\textbf{Measurements.} For measurement modules in each strategy, photon detection is performed by multiple SPCMs. The detected counts are processed by a time tagger (Swabian, Ultra Performance), which outputs the measurement results. In the tomography module, a HWP and a QWP with three angle configurations \{HWP, QWP\}: $\{22.5^{\circ}, 0^{\circ}\}$, $\{0^{\circ}, -45^{\circ}\}$, $\{0^{\circ}, 0^{\circ}\}$, followed by a BD and two SPCMs, perform measurements of the three Pauli operators $\left(\hat{\sigma}_x,\hat{\sigma}_y,\hat{\sigma}_z\right)$, respectively. In the projection module, the LCPR and the E-HWP perform the inverse unitary $U^{\dagger}$, where $U^{\dagger}\ket{\tilde\phi} = \ket{0}$, to realize the projection onto $\ket{\tilde{\phi}}$. The successful projection is indicated by the detection of the photon in the horizontal polarization.
\par
In the SCM module, the Schur collective measurement based on the Schur transform consists of four projectors:
\begin{equation}
\begin{split}
    \hat{E}_1=\ket{00}\bra{00},\ \hat{E}_2=\ket{11}\bra{11},\ \hat{E}_+=\ket{\Psi_+}\bra{\Psi_+},\ \hat{E}_-=\ket{\Psi_-}\bra{\Psi_-}\text{,}
\end{split}
    \label{SCM_measure}
\end{equation}
where $\ket{00},\ \ket{11},\ \ket{\Psi_+} = \frac{1}{\sqrt{2}}(\ket{01}+\ket{10})$ are the triplet states and $\ket{\Psi_-} = \frac{1}{\sqrt{2}}(\ket{01}-\ket{10})$ is the singlet state. $\hat{E}_-$ is the projector onto the anti-symmetric subspace. The combination of the first three projectors, $\hat{E}_1+\hat{E}_2+\hat{E}_+$, forms a single measurement $\hat{E}_{sym}$, which projects onto the symmetric subspace. The probabilities of projecting a two-qubit state $\ket{\psi}\ket{\phi}$ onto the symmetric and anti-symmetric subspaces solely depend on the overlap $c$.  
\begin{figure}[!ht]
    \centering
    \includegraphics[width=0.6\textwidth]{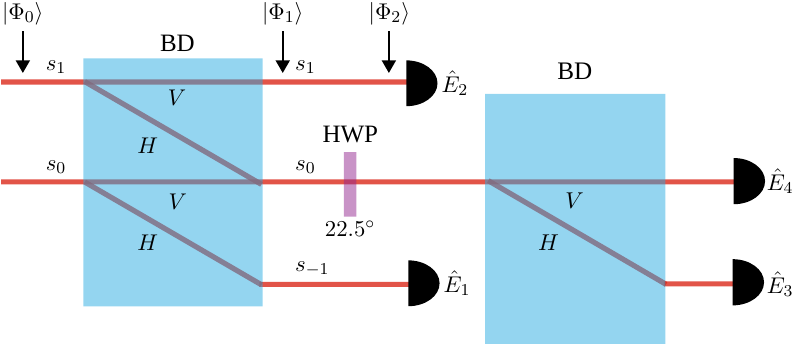}
    \caption{Realization of the Schur collective measurement. The input state $\ket{\Phi_0}$ is a two-qubit state, encoded using the polarization and path degrees of a single photon. The final detection of photons is achieved using four single photon counting modules (SPCMs). BD beam displacer, HWP half wave-plate.}
    \label{fig:SCM_shematic}
\end{figure}
\par
The measurement setup of SCM is illustrated in Fig.~\ref{fig:SCM_shematic}. The initial state is prepared as a general two-qubit state
\begin{equation}
    \ket{\Phi_0}= a_1\ket{s_0,H}+a_2\ket{s_0,V}+a_3\ket{s_1,H} + a_4\ket{s_1,V},\ |a_1|^2+|a_2|^2+|a_3|^2+|a_4|^2=1,
\end{equation}
where two-qubit state is encoded as $\ket{00}=\ket{s_0,H},\ket{01}=\ket{s_0,V},\ket{10}=\ket{s_1,H},\ket{11}=\ket{s_1,V}$. The first BD evolves the initial state as
\begin{equation}
    \ket{\Phi_1} = a_1\ket{s_{-1},H} + a_2\ket{s_{0},V} + a_3\ket{s_{0},H} + a_4\ket{s_{1},V},
\end{equation}
where $s_{-1}$ denotes the additional path mode introduced by the first BD. The state after HWP with angle $22.5^{\circ}$ on the path mode $s_0$ is
\begin{equation}
    \ket{\Phi_2} = a_1\ket{s_{-1},H} + \frac{(a_3-a_2)}{\sqrt{2}}\ket{s_{0},V} + \frac{(a_3+a_2)}{\sqrt{2}}\ket{s_{0},H} + a_4\ket{s_{1},V}.
\end{equation}
The two SPCMs following the first BD realize the first two projectors in Eq.~(\ref{SCM_measure}) with the detection probabilities as
\begin{equation}
\begin{split}
        p_1 & = |a_1|^2  = |\braket{s_{-1},H|\Phi_2}|^2 =\bra{\Phi_0}\hat{E}_1\ket{\Phi_0}, \\
        p_2 & = |a_4|^2 = |\braket{s_{1},V|\Phi_2}|^2 =\bra{\Phi_0}\hat{E}_2\ket{\Phi_0}. \\
\end{split}
\end{equation}
The second BD together with two SPCMs realize the last two projectors with the detection probabilities as
\begin{equation}
\begin{split}
        p_+ & = \left|\frac{(a_3+a_2)}{\sqrt{2}}\right|^2  = |\braket{s_{0},H|\Phi_2}|^2 =\bra{\Phi_0}\hat{E}_+\ket{\Phi_0}, \\
        p_- & = \left|\frac{(a_3-a_2)}{\sqrt{2}}\right|^2 = |\braket{s_{0},V|\Phi_2}|^2 =\bra{\Phi_0}\hat{E}_-\ket{\Phi_0}. \\
\end{split}
\end{equation}
When the initial state takes as a two-qubit product state as 
\begin{equation}
    \ket{\Phi_0} = \ket{\psi}\otimes\ket{\phi} = (b_1\ket{0}+b_2\ket{1})\otimes(d_1\ket{0}+d_2\ket{1}) = b_1d_1\ket{00}+b_1d_2\ket{01}+b_2d_1\ket{10}+b_2d_2\ket{11},
\end{equation}
the outcome probability of $\hat{E}_-$ is given by
\begin{equation}
    p_- = \left|\frac{b_2d_1-b_1d_2}{\sqrt{2}}\right|^2 = \frac{1}{2}(|b_2|^2|d_1|^2+|b_1|^2|d_2|^2- b_1^{*}b_2d_1d_2^{*}-b_1b_2^{*}d_1^{*}d_2) = \frac{1-|\braket{\psi|\phi}|^2}{2},
\end{equation}
here the overlap $c$ is given by $|\braket{\psi|\phi}|^2 = |b_1|^2|d_1|^2+|b_2|^2|d_2|^2+b_1^{*}b_2d_1d_2^{*}+b_1b_2^{*}d_1^{*}d_2$ with normalization conditions $|b_1|^2+|b_2|^2=1$ and $|d_1|^2+|d_2|^2=1$. The outcome probabilities of the other three projectors can be combined to a single one associated with the overlap as
\begin{equation}
    p_1+p_2+p_+ = 1-p_- = \frac{1+|\braket{\psi|\phi}|^2}{2}.
\end{equation}
We note that the SCM yields the same probability distribution as the ideal OST in Eq.~(\ref{eq-pf-iost}). In fact, the SCM strategy is equivalent to the Bell-basis algorithm proposed in \cite{Cincio_2018}, as an improved version of the swap test. 
\par
\section{Supplementary Results}
\textbf{Overlap estimation with a known state.} In this section, we discuss the overlap estimation when one of the two states, denoted as $\ket{\phi}$, is already known. The optimal strategy in this scenario involves projecting the unknown state $\ket{\psi}$ onto $\ket{\phi}$ using $N$ copies of $\ket{\psi}$. The number of successful projections $k$ follows a binomial distribution $\text{Bin}(k,N,c)$, and we estimate the overlap by the successful fraction $k/N$. The average variance through projection-based overlap estimation with a known state is given by $v_{proj} = c(1-c)/N$. 
\par
\begin{figure}
    \centering
    \includegraphics[width=0.55\textwidth]{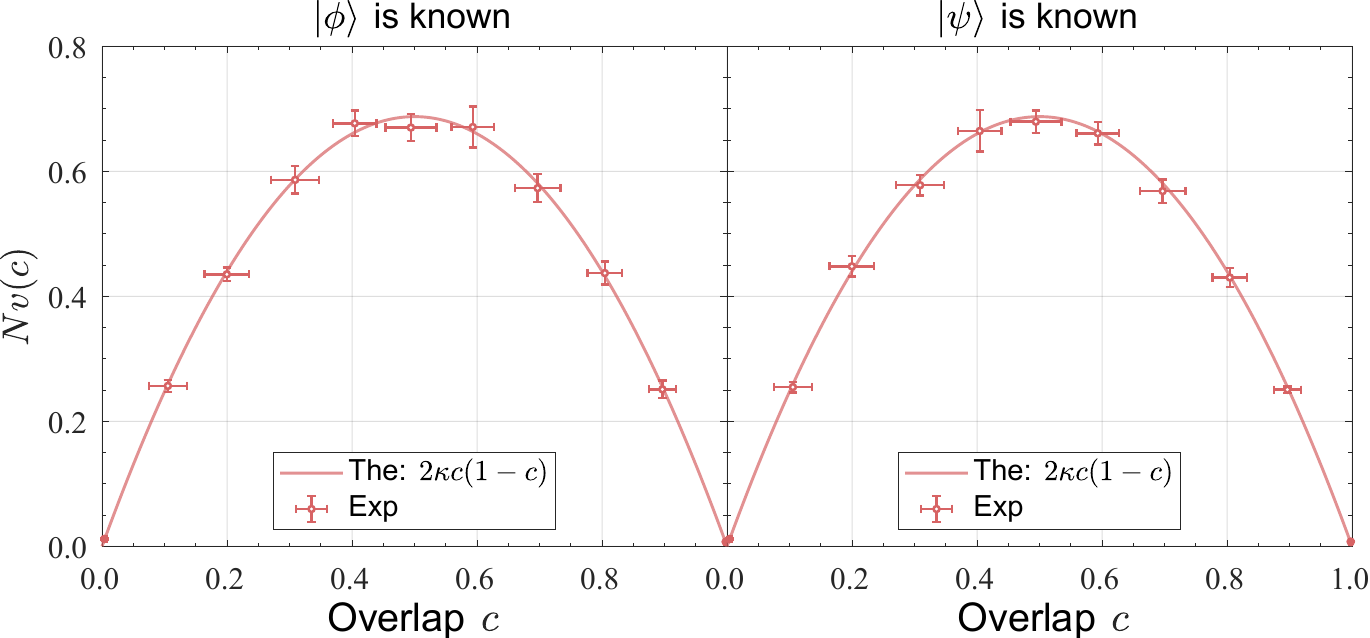}
    \caption{Experimental and theoretical results of scaled average variance $Nv(c)$ for tomography-based overlap estimation with a known state. The experimental results (markers) show good agreement with the theoretical results (solid lines).} 
    \label{fig:known_state_ove_v1}
\end{figure}
Additionally, we can reconstruct the unknown state $\ket{\psi}$ as $\ket{\tilde{\psi}}$ through quantum state tomography and calculate the estimated overlap $\tilde{c}_{tomo} = |\braket{\tilde{\psi}|\phi}|^2$. We derive $\tilde{c}_{tomo}$ from Eq.~(\ref{estimate_c}) by only considering the errors form $\ket{\tilde{\psi}}$ and approximating it as:
\begin{equation}
\begin{split}
        \tilde{c}_{tomo} & = \frac{1}{2}\left[ 1+(2c-1)\cos{\chi_1}+2\sqrt{c(1-c)}\cos{(\zeta_1-\varphi)}\sin{\chi_1} \right]\\
    & \approx c + \sqrt{c(1-c)}(t_1^c \cos{\varphi} + t_1^s \sin{\varphi})-\frac{2c-1}{4}\chi_1^2,
\end{split}
\end{equation}
where higher-order terms in the second equation are neglected. With the analysis in Section \ref{sec:tomo-error}, the average variance of tomography-based overlap estimation with a known state can be expressed as:
\begin{equation}
\begin{split}
        v_{tomo} = \overline{\left\langle (\tilde{c}_{tomo}-c)^2 \right\rangle} & = \frac{c(1-c)}{2}\left( \overline{\braket{(t_1^c)^2}}+\overline{\braket{(t_1^c)^2}} \right ) + O(\frac{1}{N^2})\\
        & = \frac{2\kappa c(1-c)}{N}+O(\frac{1}{N^2}).
\end{split}
\end{equation}
\par
To verify this theoretical result, we utilize the experimental data from the TT strategy to get experimental results for tomography-based overlap estimation with a known state. For the known state, we employ the complete measurement results (180,000 copies) to reconstruct the exact state. For the unknown state, we reconstruct the estimated state using $N = 900$ copies. This allows us to obtain two groups of average variance results, as depicted in Fig.~\ref{fig:known_state_ove_v1}. Consequently, we can attribute the errors in TT and TP strategy to the combinations of $v_{tomo}$ and $v_{proj}$, as mentioned in the main text.
\par
\textbf{Overlap estimation overhead.} Our analysis of various overlap estimation strategies shows that the average variance for all of them can be represented in a similar form: $v_s(c,N) = f_s(c)/N$, where $f_s(c)$ denotes the scaled average variance and $N$ is the number of copies. We can determine the overhead for overlap estimation from these variance results using Chebyshev’s inequality. Specifically, we can estimate the overlap $\tilde{c}_s$ with the probability
\begin{equation}
    \text{Pr}(|\tilde{c}_s-c|\geq\varepsilon) \leq \frac{v_s(c,N)}{\varepsilon^2} =\frac{f_s(c)}{N\epsilon^2} = \eta ,
\end{equation}
where $\varepsilon$ is the estimation error bound and $\eta$ is a threshold probability. Given a pair of $\varepsilon$ and $\eta$, the estimation error is less than $\varepsilon$ with a probability greater than $1-\eta$ by consuming $N\sim f_s(c)/\eta\varepsilon^2$ pairs of states, which represents the overhead for strategy $s$. From $\varepsilon = \sqrt{f_s(c)/N\eta}$, the estimation error $\varepsilon$ for all strategies scales as $O(1/\sqrt{N})$.
\par
\textbf{Extending the SCM strategy to qudits.} The SCM strategy can be naturally extended to higher dimensions, where $\ket{\psi}$ and $\ket{\phi}$ are defined in a $d$-dimensional Hilbert space:
\begin{equation}
\ket{\psi} = \sum_{i=0}^{d-1}\alpha_i\ket{i}, \quad \ket{\phi} = \sum_{j=0}^{d-1}\beta_j\ket{j},
\end{equation}
where ${\ket{i}}$ represents the basis for the qudits.  The SCM strategy for qudits utilizes a POVM with two elements:
\begin{equation}
    \hat{E}_{sym} = \hat{I} - \hat{E}_{ans},\ \hat{E}_{ans} = \sum_{i<j}\ket{\Pi_{ij}}\bra{\Pi_{ij}},
\end{equation}
where $\hat{E}_{ans}$ projects onto the anti-symmetric subspace, spanned by the vectors:
\begin{equation}
\ket{\Pi_{ij}} = \frac{1}{\sqrt{2}}\left(\ket{i}\ket{j}-\ket{j}\ket{i}\right), \quad i < j.
\end{equation}
The rank of $\hat{E}_{ans}$ is $d(d-1)/2$. The probability of obtaining the outcome associated with $\hat{E}_{ans}$, denoted as $p_{ans}$, is given by:
\begin{equation}
\begin{split}
    p_{ans} & = \bra{\psi}\bra{\phi}\hat{E}_{ans}\ket{\psi}\ket{\phi} \\
    & = \frac12\sum_{i<j}|\alpha_i\beta_j-\alpha_j\beta_i|^2 = \frac14\sum_{i\neq j}|\alpha_i\beta_j-\alpha_j\beta_i|^2 = \frac14\sum_{i,j=0}^{d-1}|\alpha_i\beta_j-\alpha_j\beta_i|^2 \\
    & = \frac{1}{2}\sum_{i=0,j=0}^{d-1}|\alpha_i|^2|\beta_j|^2-\frac{1}{2}(\sum_{i=0}^{d-1}\alpha_i\beta_i^{*})(\sum_{j=0}^{d-1}\alpha_j^{*}\beta_j) \\
    & = \frac{1}{2}(1-|\braket{\psi|\phi}|^2).
\end{split}
\end{equation}
Therefore, the overlap is given by $c = 1-2p_{ans}$. The estimation variance for high-dimensional SCM strategy is still $v_{scm}(c,N) = (1-c^2)/N$ and independent of $d$. This result highlights the scalability of the SCM strategy for overlap estimation in high-dimensional quantum systems.
\par
\textbf{Hong-Ou-Mandel interference.} As shown in Fig.\:\ref{fig:HOM_setup}, our experiments for the OST strategy involve two-mode Hong-Ou-Mandel interference (HOMI) between two input photons, with the delay between their arrival times at the interference NPBS set to zero. We use four SPCMs which are threshold photon detectors, to record the detection events of the 6 two-fold coincidence channels. We observe the HOMI varying with delay (path difference) for all 6 outcomes when two input photons are both in horizontal polarization mode, and the results are illustrated in Fig.\:\ref{fig:HOM_dip_peek}. The two coincidence channels between the SPCMs located on same sides of NPBS-1, as shown in Fig.~\hyperref[fig:HOM_dip_peek]{S11a}, correspond to the``pass" outcomes in OSTs, which manifest as peaks in HOMI pattern. The other four coincidence channels, corresponding to the ``fail" outcomes, manifest as dips in Fig.~\hyperref[fig:HOM_dip_peek]{S11b}. The high probability of ``pass" outcomes when the delay is zero is consistent with the overlap estimation results obtained through the OST strategy when the overlap is 1.
\begin{figure}[htbp]
    \centering
    \includegraphics[width=1\textwidth]{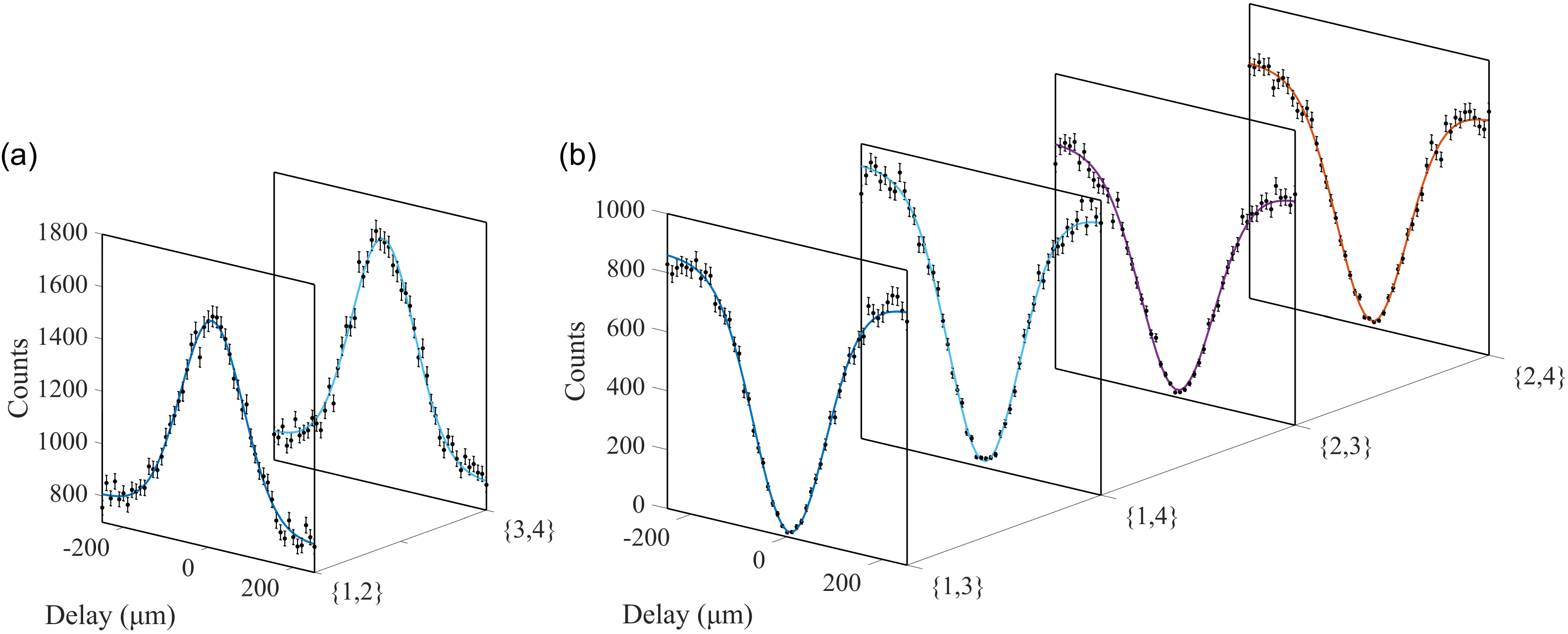}
    \caption{Hong-Ou-Mandel interference patterns for the 6 coincidence channels. \textbf{a} 2 one-side (of the NPBS-1 in Fig.\:\ref{fig:HOM_setup}) coincidence channels correspond to the ``pass" outcomes in optical swap tests. \textbf{b} 4 two-side coincidence channels correspond to the ``fail" outcomes. The colored solid lines represent curve fittings of the data (black dot) to a Gaussian function, while the error bars indicate uncertainties assuming Poisson count statistics.}
    \label{fig:HOM_dip_peek}
\end{figure}
\par
\textbf{Uncertainties of experimental measured $Nv(c)$.} As mentioned in the main text, the scaled average variance $Nv(c)$ measured in our experiments exhibits two types of uncertainties, represented by vertical and horizontal error bars in main text Fig.~2. Vertical errors indicate statistical uncertainties arising from the finite number of measuring the average variance, which can be reduced by increasing the number $n$ of experimental runs. 
\par
Horizontal errors reflect systematic uncertainties associated with our experimental setups for both state preparation and measurements. In state preparation, for each target overlap $c$, 100 qubit pairs are required to be prepared with the same overlap. However, due to experimental imperfections, the actual overlaps of different qubit pairs may deviate from the target value, leading to horizontal uncertainties in $Nv(c)$. Furthermore, systematic errors introduced by the measurement setups can result in biased estimations of the true overlaps, further contributing to the horizontal uncertainties. 
\par
These systematic errors can be attributed to imperfections in the optical elements used in our experimental setups. Specifically, the HWP and QWP suffer misalignment of the optics axis (typically $\sim 0.1$ degree), retardation errors (typically $\sim \lambda/300$ where $\lambda=830$nm) and inaccuracies in setting angles (typically $\sim 0.2$ degree). The LCPR is pre-calibrated by a co-linear inteferometer formed by an HWP-PR-HWP configuration and a BD, introducing some experimental errors. In the SCM and OST strategies, the interference visibility may experience slow drift and slight vibrations during the measurement process. In the SCM strategy, the average interference visibility between the two BDs is above 99.8\%, resulting in a minor influence on the precision of overlap estimation. Conversely, in the OST strategy, the average HOM interference visibility is approximately 96.5\%, significantly affecting the overlap estimation precision, as discussed in Section \ref{OST-spectral}. It is worth noting that the TT and TP strategies are subject to more systematic errors compared to SCM and OST. This is evident from the larger horizontal uncertainties observed in TT and TP, in contrast to OST and SCM. This phenomenon can be attributed to the fact that TT and TP  involve more measurement configurations, such as the three Pauli operators in the tomography process and projecting onto different states in the projection process. In contrast, the measurement setups in SCM and OST remain static, introducing fewer systematic errors and resulting in smaller horizontal uncertainties.

\end{widetext}

\end{document}